\documentclass[11pt]{article}

\usepackage[margin=1in,paperwidth=8.5in,paperheight=11in]{geometry}

\usepackage[utf8]{inputenc}
\usepackage[T1]{fontenc}

\usepackage{lmodern}  
\usepackage{textcomp}  

\usepackage{amsmath,amssymb,amsthm}
\usepackage{amsfonts}

\usepackage{graphicx}
\usepackage{booktabs,threeparttable,multirow,siunitx}
\usepackage{longtable}
\usepackage{caption}
\usepackage{array}
\usepackage{subcaption}   
\usepackage{tabularx}
\usepackage{float}

\usepackage{algorithm,algpseudocode}

\usepackage{adjustbox}
\usepackage{makecell}
\usepackage{setspace}
\usepackage{pdflscape}
\usepackage{rotating}
\usepackage[table]{xcolor}
\usepackage{csquotes}
\usepackage{microtype}
\usepackage[hyphens]{url}  

\usepackage{tikz}
\usetikzlibrary{arrows.meta,positioning,calc}

\usepackage[authoryear, round]{natbib}
\usepackage{etoolbox}
\AtBeginEnvironment{thebibliography}{%
    \setlength{\itemsep}{0pt}%
    \setlength{\parskip}{0pt}%
    \setlength{\topsep}{0pt}%
}

\usepackage[colorlinks=true, citecolor=blue, linkcolor=blue]{hyperref}

\usepackage[nameinlink,noabbrev]{cleveref}

\newtheorem{proposition}{Proposition}

\newtheorem{lemma}{Lemma}

\newtheorem*{theorem*}{Theorem}
\newtheorem*{lemma*}{Lemma}
\newtheorem*{corollary*}{Corollary}
\newtheorem*{proposition*}{Proposition}
\newtheorem*{remark*}{Remark}
\newtheorem*{definition*}{Definition}

\title{\textbf{Spurious Predictability in Financial Machine Learning}}

\author{%
    Sotirios D. Nikolopoulos\thanks{Email: \texttt{s.nikolopoulos@go.uop.gr}}\\%
    \textit{ Department of Accounting and Finance, University of Peloponnese,  Greece}
}

\date{}

\begin{document}
    \maketitle
    \thispagestyle{plain} 

    \begin{abstract}
        Adaptive specification search generates statistically significant backtests even under martingale-difference nulls. We introduce a falsification audit testing complete predictive workflows against synthetic reference classes, including zero-predictability environments and microstructure placebos. Workflows generating significant walk-forward evidence in these environments are falsified. For passing workflows, we quantify selection-induced performance inflation using an absolute magnitude gap linking optimized in-sample evidence to disjoint walk-forward realizations, adjusted for effective multiplicity. Simulations validate extreme-value scaling under correlated searches and demonstrate detection power under genuine structure. Empirical case studies confirm that many apparent findings represent methodological artifacts rather than genuine predictability.

        \vspace{5pt}
    \end{abstract}
    \section{Introduction}
    \label{sec:introduction}

    The integration of machine learning into financial econometrics has expanded the set of predictors, functional forms, and estimation strategies available in empirical finance. Methods such as random forests, gradient boosting, and neural networks are now common in applications spanning asset pricing and macroeconomic forecasting \citep{GuKellyXiu2020, Athey2019}. However, standard machine learning validation methods assume approximately i.i.d.\ data and can fail under the temporal dependence, non-stationarity, and causal ordering constraints of financial time series \citep{Varian2014, Diebold2015, Bergmeir2018}. As a result, empirical conclusions can be driven as much by the evaluation procedure as by the model class.

    Recent evidence highlights both the promise and fragility of ML-based findings in finance. Some studies report economically positive gains from ML-driven forecasting signals \citep{Binsbergen2023}, while subsequent work shows that these gains often disappear once data handling, timing, and bias controls respect the information available at each decision point \citep{Zhang2025}. Related evidence suggests that results can be highly specification-dependent, with many variants failing to outperform appropriate benchmarks once evaluation protocols are standardized \citep{Campbell2024}. These findings motivate evaluation procedures that explicitly account for adaptive search and temporal constraints. In particular, the contrast between the gains documented by \citet{Binsbergen2023} and the sensitivity to leakage and protocol details emphasized by \citet{Zhang2025} illustrates how workflow integrity can be a binding constraint for interpreting empirical performance.

    This problem is not unique to machine learning. Empirical finance has long faced challenges in out-of-sample stability and replication. \citet{HarveyLiuZhu2016} documented the extent of factor proliferation and emphasized the role of data mining, while \citet{HouXueZhang2020} showed that many published anomalies do not survive replication under more robust testing. Taken together, these literatures point to two interacting sources of distortion in predictive research: adaptive specification search and validation protocols that inadvertently leak future information.

    The identification problem is to separate genuine predictability—reflecting a stable relationship between available information and future outcomes—from spurious predictability generated by selection bias, information leakage, or violations of temporal ordering. This problem is acute in finance because signals are weak, dependence is persistent, and structural breaks are common \citep{Engle1982, Campbell1997}. Random cross-validation can break the temporal order of observations, inducing look-ahead bias \citep{Bergmeir2018}. More broadly, these issues relate to inference after model selection \citep{Belloni2014, Lee2016} and to the data-snooping literature \citep{White2000, Hansen2005, LopezDePrado2018}, which examines how repeated evaluation and selection inflate reported performance under the null.

    We propose an evaluation protocol that treats falsification as a prerequisite for interpreting backtest performance. The protocol evaluates the complete modeling workflow—including feature construction, tuning, selection, and portfolio mapping—on induced-null environments that preserve key features of financial data while removing any conditional mean predictability. Performance is assessed via strict walk-forward validation that enforces causal ordering. The observed walk-forward winner statistic is compared to its distribution under the null, estimated by Monte Carlo simulation. Systematic performance in these environments signals workflow invalidity—such as information leakage, selection distortion, or timing errors—rather than evidence of predictability.

    We also develop diagnostics to quantify the magnitude of selection-induced inflation. Under a global martingale-difference null, the optimized in-sample winner statistic grows as $\sqrt{\log K_{\mathrm{eff}}}$, where $K_{\mathrm{eff}}$ measures the effective number of independent strategies after accounting for correlation. In contrast, leakage-robust walk-forward evaluation remains stochastically bounded under the same null. Monte Carlo experiments quantify inflation across correlated strategy families, showing that effective multiplicity—not the nominal number of specifications—determines the expected in-sample maximum. To summarize inflation for the selected configuration, we report a stabilized Backtest Inflation Factor (BIF) and complementary gap-based diagnostics that remain interpretable when walk-forward evidence is weak.

    Empirical case studies illustrate recurrent confounds. Predictive models can reproduce known systematic risk premia, generating statistically significant gross returns while delivering zero alpha after factor adjustment \citep{FamaFrench1993}. Validation choices in volatility forecasting can mechanically improve reported performance under null settings when random cross-validation is used instead of strict walk-forward evaluation. These examples illustrate how the audit framework distinguishes methodological artifacts from genuine predictive content.

    We emphasize that our objective is aligned with the broader goal of promoting credible empirical finance. For example, \citet{GuKellyXiu2020} demonstrate the economic value of machine learning in asset pricing, while \citet{JensenKellyPedersen2023} conclude that factor research exhibits substantial internal and external validity once inference is grounded in economically appropriate benchmarks and multiplicity is handled in a dependence-aware manner, supported by transparent code and data. Our contribution is complementary: rather than re-litigating factor-level replicability, we provide a workflow-level diagnostic that screens for pipeline invalidity. By ``workflow'' we mean the full mapping from raw data to executed positions, encompassing preprocessing, tuning, and timing alignment. This diagnostic screens for upstream failures (e.g., leakage, temporal misalignment, and protocol artifacts) that can compromise any inferential framework—Bayesian or frequentist—if left unchecked.

    The remainder of the paper is organized as follows. \Cref{sec:mechanisms} formalizes mechanisms generating spurious predictability. \Cref{sec:theory} establishes the asymptotic motivation for the scaling behavior under global nulls and defines the inflation diagnostics. \Cref{sec:audit} defines the falsification audit and the induced-null environments used as reference nulls. \Cref{sec:simulation} validates the framework through Monte Carlo simulations. \Cref{sec:empirical_applications} presents empirical applications in returns and volatility forecasting. \Cref{sec:implications} discusses implications, and \Cref{sec:conclusion} concludes. The Appendices provide asymptotic derivations, complete specifications of the induced-null environments, and additional robustness analyses.

    \section{Mechanisms of Spuriousness in Financial Machine Learning}
    \label{sec:mechanisms}
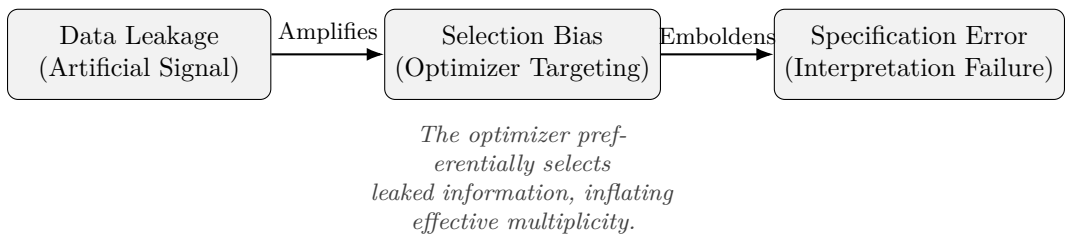
\begin{figure}[htbp]
        \centering
        \begin{tikzpicture}[
            node distance=1.5cm and 1.5cm,
            box/.style={draw, rounded corners, minimum width=3.5cm, minimum height=1.2cm, align=center, fill=gray!10, font=\small},
            arrow/.style={-Latex, thick}
            ]
            \node[box] (leakage) {Data Leakage\\(Artificial Signal)};
            \node[box, right=of leakage] (selection) {Selection Bias\\(Optimizer Targeting)};
            \node[box, right=of selection] (interp) {Specification Error\\(Interpretation Failure)};

            \draw[arrow] (leakage) -- node[above, font=\footnotesize] {Amplifies} (selection);
            \draw[arrow] (selection) -- node[above, font=\footnotesize] {Emboldens} (interp);

            \node[below=0.2cm of selection, font=\footnotesize, text width=4cm, align=center, color=darkgray] (note) {\textit{The optimizer preferentially selects \\ leaked information, inflating \\ effective multiplicity.}};
        \end{tikzpicture}
        \caption{The Cascade of Spuriousness. Data leakage creates artificial signal-to-noise ratios; adaptive selection algorithms preferentially target these artifacts, resulting in inflated test statistics that are subsequently misidentified as structural alpha.}
        \label{fig:mechanism_cascade}
    \end{figure}

    The core premise of this work is that statistical validation in financial machine learning constitutes a joint test of two hypotheses: the existence of a predictive signal and the integrity of the research workflow. Traditional backtesting conflates these, interpreting a high Sharpe ratio as evidence of the former while ignoring the latter. We organize our taxonomy of failure around the compounding cascade illustrated in Figure~\ref{fig:mechanism_cascade}. This cascade typically originates with data leakage, is amplified by selection bias, and is ultimately cemented by specification error.

    \subsection{The Global Null and Reference Classes}
    \label{subsec:null_def}
    To formalize these failures, we define the Global Null hypothesis ($H_0$) as the state where \textit{excess} asset returns $r_t$ follow a martingale difference sequence (MDS) with respect to the public information set $\mathcal{F}_{t-1}$, such that $\mathbb{E}[r_t \mid \mathcal{F}_{t-1}] = 0$ \citep{Campbell1997}. Under $H_0$, no algorithm should systematically extract excess returns. However, empirical financial data exhibits stylized facts—volatility clustering, heavy tails, and leverage effects—that can be exploited by overfitted models to generate spurious significance even when the conditional mean is zero \citep{Cont2001, Mandelbrot1963}.

    To rigorously test this, we define \textit{Induced-Null} environments. These include strict martingale difference reference classes and targeted placebos that preserve zero expected excess return in the latent efficient returns, yet may embed predictable transient structures (e.g., microstructure noise or bid--ask bounce) in observed prices. A workflow that generates statistically significant walk-forward performance on strict MDS reference classes is statistically falsified, demonstrating a capacity to hallucinate signal. Conversely, performance on targeted placebos (such as the bid-ask bounce environment in Section~\ref{sec:audit}) constitutes economic falsification, indicating the exploitation of non-investable artifacts rather than fundamental value.

    \subsection{The Cascade of Spuriousness}
    \label{subsec:cascade}
    The distinct mechanisms of spuriousness rarely occur in isolation; rather, they interact sequentially. The process often begins with data leakage, where model development conditions on information that is not measurable with respect to the filtration $\mathcal{F}_{t-1}$, so the feature set implicitly contains signals that would not have been available at the time of decision. While gross look-ahead bias is rare in mature workflows, subtler forms are endemic \citep{LopezDePrado2018}. These include normalization leakage, where features are scaled using statistics computed over the full sample rather than an expanding window; survivorship bias, where the investment universe is conditioned on constituents surviving to time $T$; and cross-validation contamination, where overlapping labeling horizons or preprocessing choices allow information to bleed across folds. Leakage lowers the effective noise floor and creates an artificial predictability that is indistinguishable from alpha during in-sample training.

    This artificial predictability amplifies selection bias. Selection bias occurs when a researcher or algorithm evaluates $K$ candidate specifications and reports only the specification $k^*$ that maximizes an absolute performance metric, typically the $t$-statistic, so that $Z_{IS}^* = \max_{k} |Z_{IS,k}|$. Under Gaussian statistics and independence, the expected maximum absolute statistic under the global null is approximately $\sqrt{2\ln(2K)}$ up to lower-order terms, so even a modest search (e.g., $K\approx 100$) can generate apparently significant in-sample results by chance \citep{HarveyLiuZhu2016}. When candidates are correlated, the relevant multiplicity is $K_{\mathrm{eff}}\le K$, and redundancy collapses $K_{\mathrm{eff}}$ toward unity, attenuating extreme-value inflation relative to the nominal count $K$. Even under a true signal, selection induces a Winner's Curse: the selected model's estimated effect size tends to exceed the population parameter  because selection preferentially retains realizations where estimation error aligns with the signal direction \citep{Bailey2014}.

    The cascade concludes with specification error, which we treat as an economic interpretation failure rather than a mere statistical misfit. Inflated statistics produced by leakage and selection encourage structural narratives that do not survive transport to the walk-forward distribution. A workflow may be statistically valid yet structurally spurious if it attributes predictive power to alpha that is in fact compensation for latent risk exposures such as momentum or liquidity, or if it fits transient regime relationships that are not stable out of sample \citep{FamaFrench1993, Carhart1997}. In our framework, these constitute specification errors because the model may describe the training set but fails to generalize to the target domain or is misidentified as an anomaly rather than a risk premium.

    \subsection{Relation to Data-Snooping Corrections}
    \label{subsec:snooping_relation}
    This framework complements but remains distinct from post-hoc corrections for data snooping, such as the Reality Check \citep{White2000} or the Superior Predictive Ability (SPA) test \citep{Hansen2005}. Those methods adjust the critical values of the test statistic to account for the size of the universe ($K$) after the fact. However, they rely on the assumption that the researcher knows the true value of $K$. In modern machine learning, $K$ is often implicit, effectively infinite, or obfuscated by the adaptive nature of the search. Practically, these tools occupy distinct stages in the quantitative lifecycle. The \textit{Falsification Audit} (Section \ref{sec:audit}) serves as a pre-commitment filter during workflow development, rejecting architectures prone to hallucination. The \textit{Inflation Diagnostic} (Section \ref{sec:theory}) is applied during internal validation to quantify the gap between in-sample optimization and walk-forward reality. In contrast, corrections such as the Reality Check are post-hoc inference tools appropriate for final external reporting.

    \section{Selection-Induced Backtest Inflation}
    \label{sec:theory}
    The falsification motivation in \Cref{sec:mechanisms} is operationalized in \Cref{sec:audit} as an induced-null audit that screens for workflow invalidity (leakage, temporal misalignment, microstructure exploitation, and protocol artifacts). Clearing that gate is necessary but not sufficient for economic signal: even under a martingale-difference global null, adaptive specification search can still produce apparently strong optimized in-sample backtests. This section formalizes the behavior of optimized backtests under the global null, motivates the $\Theta(\sqrt{\log K_{\mathrm{eff}}})$ scaling law for selection-induced inflation under correlated search, and defines inflation diagnostics that translate effective multiplicity into an economically interpretable correction to reported performance, centered on the absolute magnitude gap $\Delta Z$, with the stabilized Backtest Inflation Factor (BIF) used only as a supplementary within-study ranking heuristic. The objective is end-to-end pipeline diagnosis; classical data-snooping inference (e.g., Reality Check and SPA) remains the appropriate layer when the object of interest is a search-adjusted $p$-value for the best-performing specification.

    \subsection{Formalizing the Global Null Hypothesis}
    \label{subsec:global_null}

    Consider a time series of asset returns $\{r_t\}_{t=1}^T$. We define the global null hypothesis as conditional unpredictability with respect to the filtration $\mathcal{F}_{t-1}$ generated by all information available at time $t-1$:
    \begin{equation}
        \mathbb{E}[r_t \mid \mathcal{F}_{t-1}] = 0, \quad \forall t.
        \label{eq:null}
    \end{equation}
    This implies $\{r_t\}$ is a martingale difference sequence. A candidate strategy $k\in\{1,\dots,K\}$ produces a signal $x_{t,k}\in\mathcal{F}_{t-1}$ and the realized strategy return is $R_{t,k}=x_{t,k}r_t$. Under \Cref{eq:null}, $\mathbb{E}[R_{t,k}\mid\mathcal{F}_{t-1}]=0$, so $\{R_{t,k}\}$ is a martingale difference sequence for each fixed $k$.

    We distinguish two disjoint phases: an in-sample (selection) phase $\mathcal{T}_{\mathrm{IS}}$ and a walk-forward (evaluation) phase $\mathcal{T}_{\mathrm{WF}}$, with $\mathcal{T}_{\mathrm{IS}}\cap\mathcal{T}_{\mathrm{WF}}=\emptyset$. Disjointness requires that the entire mapping from data to the selected configuration---preprocessing, feature construction, tuning, and model choice---is measurable with respect to the in-sample $\sigma$-field only.

    Define the sample means $\bar{R}_{\mathrm{IS},k}$ and $\bar{R}_{\mathrm{WF},k}$ in the usual way. In finite samples, and in audit environments that intentionally embed microstructure frictions, strategy returns can display conditional heteroskedasticity and mild serial correlation. We therefore employ a HAC variance estimator with a Bartlett kernel \citep{NeweyWest1987}. We use the sample-size-based rule-of-thumb bandwidth $L_n = \lfloor 4(n/100)^{2/9} \rfloor$ proposed by \citet{NeweyWest1994} and widely adopted as a software default, where $n \in \{n_{\mathrm{IS}}, n_{\mathrm{WF}}\}$ denotes the sample size of the relevant phase, satisfying $L_n \to \infty$ and $L_n/n \to 0$ as $n \to \infty$. The studentized statistics are

    \begin{equation}
        Z_{\mathrm{IS},k} \equiv \frac{\bar{R}_{\mathrm{IS},k}}{\sqrt{\widehat{V}_{\mathrm{HAC}}(\bar{R}_{\mathrm{IS},k})}},
        \qquad
        Z_{\mathrm{WF},k} \equiv \frac{\bar{R}_{\mathrm{WF},k}}{\sqrt{\widehat{V}_{\mathrm{HAC}}(\bar{R}_{\mathrm{WF},k})}}.
        \label{eq:z_is_z_wf}
    \end{equation}
    Under \Cref{eq:null} and standard regularity conditions, for any fixed $k$, $Z_{\mathrm{IS},k}\Rightarrow\mathcal{N}(0,1)$ and $Z_{\mathrm{WF},k}\Rightarrow\mathcal{N}(0,1)$ as $T\to\infty$. However, we explicitly note that HAC $t$-statistics frequently exhibit severe size distortions in finite samples. For restricted walk-forward evaluation periods (e.g., $n_{\mathrm{WF}} = 252$ days), the empirical distribution of $Z_{\mathrm{WF}}$ develops heavier tails than the theoretical $\mathcal{N}(0,1)$ reference. While rule-of-thumb bandwidth selection mitigates first-order autocorrelation, it does not strictly eliminate finite-sample over-rejection under the null. Consequently, test statistics relying on asymptotic approximations are anti-conservative for size control in finite samples, tending to over-reject under the null. When used as an admission gate in the diagnostic audit, this bias makes the screen conservative with respect to pipeline acceptance.

    \subsection{The Multiplicity Problem and Effective Multiplicity}
    \label{subsec:keff}
    In the selection phase, the researcher evaluates $K$ candidate strategies and chooses the most extreme in-sample result:

    \begin{equation}
        Z_{\mathrm{IS}}^\star \equiv \max_{1\le k\le K} |Z_{\mathrm{IS},k}|,
        \qquad
        k^\star \equiv \operatorname*{arg\,max}_{1\le k\le K} |Z_{\mathrm{IS},k}|.
        \label{eq:z_is_star}
    \end{equation}

    When candidates are correlated, the nominal $K$ overstates the effective dimensionality of the search. Let $\mathbf{Z}_{\mathrm{IS}} \equiv (Z_{\mathrm{IS},1},\dots,Z_{\mathrm{IS},K})^\top$ and let $\Sigma$ denote the correlation matrix of $\mathbf{Z}_{\mathrm{IS}}$. We define effective multiplicity via the spectral participation ratio:
    \begin{equation}
        K_{\mathrm{eff}}
        = \frac{\left(\operatorname{tr}\Sigma\right)^2}{\|\Sigma\|_F^2}
        = \frac{\left(\sum_{i=1}^K \lambda_i\right)^2}{\sum_{i=1}^K \lambda_i^2},
        \label{eq:keff}
    \end{equation}
    where $\lambda_1,\dots,\lambda_K$ are the eigenvalues of $\Sigma$. Since $\Sigma$ is a correlation matrix, $\operatorname{tr}(\Sigma)=K$, so equivalently
    \begin{equation}
        K_{\mathrm{eff}} = \frac{K^2}{\|\Sigma\|_F^2}.
        \label{eq:keff_simplified}
    \end{equation}
    By construction, $K_{\mathrm{eff}}=K$ under independence and decreases toward $1$ as configurations become collinear. In high-dimensional regimes, we apply linear shrinkage toward the identity using the Schäfer--Strimmer estimator \citep{Schaefer2005} to ensure $\Sigma$ remains well-conditioned; details are in Appendix~\ref{app:keff_validation}.

    \subsection{The Scaling Law of Selection Bias}
    \label{subsec:scaling_law}

    Selection-induced inflation is governed by the extreme-value behavior of $\{Z_{\mathrm{IS},k}\}$ after optimization. Under two-sided selection, $Z_{\mathrm{IS}}^\star$ depends on the effective number of distinct candidates summarized by $K_{\mathrm{eff}}$.

    \begin{proposition}
        \label{thm:max_gaussian}
        Assume $\mathbf{Z}_{\mathrm{IS}}$ is approximately mean-zero multivariate Gaussian with correlation matrix $\Sigma$, and define $K_{\mathrm{eff}}$ by \Cref{eq:keff}. Suppose dependence is generated by a structured, high-correlation search geometry, for example an approximate-factor structure within families,
        $
        Z_{\mathrm{IS},k}=\sqrt{\rho}\,F_{c(k)}+\sqrt{1-\rho}\,\varepsilon_k
        $
        with $\rho\in(0,1)$, heterogeneous family sizes, and independent $(F_c,\varepsilon_k)$. Then, as $K_{\mathrm{eff}}\to\infty$,
        \[
        \mathbb{E}\!\left[Z_{\mathrm{IS}}^\star\right]
        = \Theta\!\left(\sqrt{\log K_{\mathrm{eff}}}\right).
        \]
    \end{proposition}

    While each studentized statistic is asymptotically Gaussian as $T\to\infty$, the finite-sample extreme-value approximation may underestimate the null benchmark for $Z_{\mathrm{IS}}^\star$---and hence the implied inflation penalty---under heavy-tailed return innovations, especially when $K$ is large relative to $T$. We therefore verify by simulation that, for the sample sizes and search dimensions studied here, the $\Theta(\sqrt{\log K_{\mathrm{eff}}})$ approximation remains accurate even under heavy-tailed Student-$t$ innovations (see the empirical robustness sweeps in Appendix~\ref{app:robustness_checks}).

    The $\Theta(\sqrt{\log K_{\mathrm{eff}}})$ statement should be read as a consequence of Gaussian maximal inequalities under search geometries whose canonical-metric packing complexity is of order $K_{\mathrm{eff}}$; Appendix~\ref{app:gaussian_max} formalizes this link (Lemma~\ref{lem:keff_entropy_bridge}) and clarifies that $K_{\mathrm{eff}}$ is not claimed to parameterize extremes for arbitrary $\Sigma$.

    In contrast, performance evaluated via a strict walk-forward protocol remains bounded in probability under the null, provided the evaluation is truly disjoint.

    \begin{lemma}
        \label{lem:wf_clt}
        Assume: (i) the selection rule $k^\star=\arg\max_k |Z_{\mathrm{IS},k}|$ is measurable with respect to the in-sample $\sigma$-field; (ii) walk-forward blocks are generated by a strictly causal evaluation protocol disjoint from selection; and (iii) standard HAC regularity conditions for \Cref{eq:z_is_z_wf} hold on the walk-forward blocks. Then, conditional on $\mathcal{T}_{\mathrm{IS}}$, the index $k^\star$ is fixed, and under the global null:
        \begin{equation}
            Z_{\mathrm{WF}}^\star \equiv |Z_{\mathrm{WF},k^\star}| = O_p(1).
        \end{equation}
    \end{lemma}
    The proof is given in Appendix~\ref{app:theory_proofs}.

    \subsection{Quantifying Inflation: The Absolute Gap and Stabilized BIF}
    \label{subsec:bif}

    To quantify the divergence between optimized in-sample evidence and realizable walk-forward evidence, the framework relies primarily on the absolute magnitude gap:
    \begin{equation}
        \Delta Z \equiv Z_{\mathrm{IS}}^\star - Z_{\mathrm{WF}}^\star,
        \label{eq:delta_z}
    \end{equation}
    where $Z_{\mathrm{IS}}^\star = \max_{k\le K} |Z_{\mathrm{IS},k}|$ and $Z_{\mathrm{WF}}^\star = |Z_{\mathrm{WF},k^\star}|$. Because strictly disjoint walk-forward evaluation remains bounded in probability under the null, $\Delta Z$ provides a linear, mathematically continuous measurement of the extreme-value bias induced by adaptive specification search. Unlike ratio-based formulations, $\Delta Z$ remains strictly well-defined and analytically consistent across the entire domain of out-of-sample realizations.

    To provide a supplementary relative measure of this degradation, we define the \emph{raw Backtest Inflation Factor} as the ratio of the optimized in‑sample winner to its walk‑forward counterpart:
    \begin{equation}
        \mathrm{BIF}^{\mathrm{raw}} \equiv \frac{Z_{\mathrm{IS}}^\star}{Z_{\mathrm{WF}}^\star}.
        \label{eq:bif_raw}
    \end{equation}
    Because this raw ratio is mathematically undefined when $Z_{\mathrm{WF}}^\star = 0$, we introduce the \emph{stabilized Backtest Inflation Factor} by enforcing a strict denominator floor:
    \begin{equation}
        \mathrm{BIF}^{\mathrm{stab}}_{\tau} \equiv \frac{Z_{\mathrm{IS}}^\star}{\max\{Z_{\mathrm{WF}}^\star,\tau\}}.
        \label{eq:bif_stab_def}
    \end{equation}
    Throughout this paper, we set the stability parameter to $\tau = 0.5$. While $\mathrm{BIF}^{\mathrm{stab}}_{\tau}$ coincides with $\mathrm{BIF}^{\mathrm{raw}}$ for values above this threshold, it saturates at $Z_{\mathrm{IS}}^\star/\tau$ for highly degenerate models. Consequently, $\mathrm{BIF}^{\mathrm{stab}}_{\tau}$ must be interpreted strictly as a numerically stabilized operational heuristic rather than a high-resolution degradation measure. Furthermore, because the absolute magnitude of $\mathrm{BIF}^{\mathrm{stab}}_{\tau}$ is heavily dictated by the choice of $\tau$, the metric lacks cross-study comparability. A backtest inflation factor generated under $\tau=0.5$ cannot be quantitatively compared to one derived under $\tau=0.3$. Therefore, $\mathrm{BIF}^{\mathrm{stab}}_{\tau}$ functions exclusively as an internal, intra-study ranking heuristic. Cross-study evaluations and formal benchmarking should therefore rely on the $\tau$-invariant magnitude gap $\Delta Z$, rather than on $\mathrm{BIF}^{\mathrm{stab}}_{\tau}$.

    \section{The Falsification Audit Framework}
    \label{sec:audit}
    The theoretical results in \Cref{sec:theory} establish that performance inflation is an inherent consequence of adaptive model selection under a global null. To operationalize these insights, we introduce a falsification-based audit framework that evaluates machine learning pipelines in controlled synthetic environments. The audit imposes a necessary condition for methodological validity: a pipeline must not generate systematic performance when the underlying data-generating process contains no exploitable mean signal.

    \subsection{Design Principles}
    The audit framework is grounded in three principles that align machine learning practice with econometric validation. First, falsifiability shifts the burden of proof: a pipeline is not validated by historical performance but is falsified if it extracts statistically significant signal from null data. This follows the logic of placebo tests in causal inference \citep{Imbens2015}. Second, the audit evaluates the pipeline end-to-end, including preprocessing, feature engineering, and execution timing. Spurious predictability typically arises from interactions across stages---such as look-ahead bias in normalization or misaligned execution lags---rather than from the learning algorithm in isolation. Third, the audit enforces causal validity by requiring that all signals and portfolio decisions at time $t$ be measurable with respect to $\mathcal{F}_{t-1}$ and executed in a manner consistent with the directional flow of time. Disjointness is interpreted operationally: no preprocessing, tuning, or model choice may use information from evaluation blocks, directly or indirectly.

    \subsection{Synthetic Null Environments}
    We construct five synthetic data-generating processes designed to target distinct failure modes. Environments A, B, D, and E satisfy the martingale-difference restriction $\mathbb{E}[r_t \mid \mathcal{F}_{t-1}] = 0$, while Environment C is a microstructure placebo that intentionally violates the martingale-difference restriction in \emph{observed} returns. Full mathematical specifications and default calibrations for all environments are detailed in Appendix~\ref{app:environments}.

    \paragraph{Environment A: White Noise.}
    Returns are i.i.d. Gaussian with zero mean. This environment isolates pure selection bias and provides the canonical setting for the $\Theta(\sqrt{\log K_{\mathrm{eff}}})$ scaling law.

    \paragraph{Environment B: Regime-Switching Volatility.}
    A two-state Markov chain governs conditional volatility while the conditional mean remains zero. This environment detects predictability arising from improper normalization or scale-based trading rules that "leak" future volatility states.

    \paragraph{Environment C: Friction Placebo (Bid--Ask Bounce).}
    This is not a ``no-predictability in observed returns'' null. It is a targeted placebo designed to detect timing, indexing, and execution-lag mistakes that can masquerade as alpha in real implementations when applied to microstructure-contaminated prices. We simulate a latent efficient return that satisfies the martingale-difference restriction, then generate an \emph{observed} return series with Roll-style bid--ask bounce (an MA(1) component with negative first-order autocorrelation). The intentional violation of the martingale-difference restriction in \emph{observed} returns is the diagnostic feature: a correctly implemented workflow should not convert this transient friction pattern into persistent walk-forward profitability once signals are aligned and executed properly.

    \paragraph{Environment D: Factor Null (Zero Alpha).}
    Returns are driven by a single mean-zero factor plus idiosyncratic noise. This environment isolates omitted-risk attribution: any apparent alpha must reflect systematic factor exposure being misread as skill.

    \paragraph{Environment E: GARCH(1,1) Volatility Clustering.}
    Returns exhibit persistent conditional heteroskedasticity while maintaining a zero conditional mean. This environment stresses the robustness of HAC-based inference and risk adjustment under realistic volatility dynamics.

    \subsection{The Two-Stage Audit Protocol}
    \label{subsec:two_stage_audit}

    The audit proceeds in two stages. Stage~1 falsifies pipelines that generate implausibly large walk-forward winner magnitudes under induced-null environments designed to preserve the pipeline’s mechanics while removing economic signal. Pipelines that are not falsified advance to Stage~2, where the identical end-to-end workflow is executed on the target dataset under a strict disjoint in-sample/walk-forward split, producing optimized in-sample and walk-forward winner magnitudes together with inflation diagnostics.

    \subsubsection{Stage 1: Falsification under induced nulls}
    \label{subsec:stage1_falsification}

    Stage~1 tests whether the full pipeline, held fixed, produces spuriously extreme walk-forward winners under induced-null environments and targeted placebos that preserve the mechanics of the workflow while stripping out genuine economic signal. The pipeline is executed across the induced-null environments in a Monte Carlo experiment. For each environment $e$ and replication $m=1,\ldots,M$, the workflow yields a walk-forward winner magnitude $Z^{\star}_{\mathrm{WF},e,m}$ drawn from the induced null of the adaptive selection procedure. Stage~1 is a conservative gate that controls false discovery across environments; the operational thresholds and decision rule are specified in Section~\ref{subsec:calibration}.

    \subsubsection{Stage 2: Target-data evaluation}
    \label{subsec:stage2_target}

    Stage~2 quantifies selection-induced inflation on the target dataset conditional on passing Stage~1. Pipelines that are not falsified are executed on the target sample under a strict disjoint in-sample/walk-forward split, with adaptive selection confined to the in-sample block and evaluation performed only on disjoint walk-forward blocks. This yields the optimized maximum $Z_{\mathrm{IS}}^\star$ and its walk-forward counterpart $Z_{\mathrm{WF}}^\star$, together with $\widehat{K}_{\mathrm{eff}}$ and the associated inflation diagnostics (including $\mathrm{BIF}^{\mathrm{stab}}_{\tau}$ and auxiliary measures). The null-calibrated thresholds and decision rules used to classify inflation are specified in Section~\ref{subsec:calibration}.

    We compute the test statistics utilizing HAC standard errors to account for residual dependence. However, we explicitly note that HAC $t$-statistics converge to the standard normal distribution asymptotically and frequently exhibit severe size distortions in finite samples. For restricted walk-forward evaluation periods (e.g., $T_{\mathrm{WF}} = 252$ days), the empirical distribution of $Z_{\mathrm{WF}}$ develops heavier tails than the theoretical $N(0,1)$ reference. While rule-of-thumb bandwidth selection (e.g., Newey-West) mitigates first-order autocorrelation, it does not strictly eliminate finite-sample over-rejection under the null. Consequently, the falsification thresholds derived from asymptotic approximations may be inherently conservative, classifying marginal noise as statistically significant out-of-sample evidence.

    \subsection{Operational Thresholds and Calibration}
    \label{subsec:calibration}
    Stage~1 (falsification gate) utilizes environment-specific critical values for the walk-forward winner magnitude. For each environment $e$, let $\zeta_{e,1-\alpha/5}$ denote the empirical $1-\alpha/5$ quantile of $\{Z^{\star}_{\mathrm{WF},e,m}\}_{m=1}^M$. The pipeline is falsified if $Z^{\star}_{\mathrm{WF},e,\mathrm{obs}}>\zeta_{e,1-\alpha/5}$ for any $e$, utilizing a Bonferroni correction across the multiple testing environments to control the family-wise error rate. We explicitly note that while this correction rigorously controls false discoveries under the null, it structurally reduces statistical power; pipelines may fail a specific environment due to finite-sample variation rather than systematic leakage.

    Stage~2 (inflation classification) applies null-calibrated thresholds strictly to the absolute magnitude gap ($\Delta Z$) computed after adaptive search. For a chosen quantile level $q$, the operational threshold is defined as:
    \begin{equation}
        \varepsilon_{q,\Delta} \equiv Q_q(\Delta Z).
        \label{eq:thresholds_final}
    \end{equation}
    We flag null-implausible inflation if $\Delta Z > \varepsilon_{0.99,\Delta}$. The recommended threshold values tabulated in Appendix~\ref{app:calib_thresholds} are strictly illustrative approximations based on idealized search geometries. In practice, researchers must simulate their own null distributions given their exact pipeline dimensions, correlation structures ($\widehat{K}_{\mathrm{eff}}$), and sample splits ($n_{\mathrm{IS}},n_{\mathrm{WF}}$) to generate valid reference cutoffs. When pipeline-specific null draws are unavailable, researchers must adopt a conservative worst-case calibration corresponding to maximal effective multiplicity.

    \subsection{Audit Robustness and Limitations}
    \label{subsec:limitations}
    The falsification audit provides necessary conditions for methodological validity but does not certify economic truth. In principle, adaptive tuning could be used to satisfy a fixed diagnostic without improving genuine predictive content, constituting a form of overfitting to the audit itself. The practical constraint is that the audit is intentionally multi-dimensional: passing requires robustness across five environments targeting distinct failure channels, including multiplicity, leakage, microstructure timing errors, factor confounding, and volatility-driven artifacts. Satisfying these constraints simultaneously reduces effective degrees of freedom and limits the flexibility that typically generates spurious in-sample success, rendering audit-aware tuning costly and fragile.

    A pipeline that passes the audit has demonstrated robustness to a specified set of structural artifacts. This does not guarantee that its predictive performance on real data reflects a genuine economic signal; it may still lack economic significance, exhibit poor concentration-adjusted performance, or fail to cover transaction costs. The audit is therefore a stress test, not a certification. To further limit local overfitting to the audit, we recommend following standard benchmark-design practices: fix the workflow specification prior to audit execution; randomize audit seeds and report seed sensitivity; reserve a holdout set of null environments (e.g., unseen parameterizations within the same null reference classes) for out-of-design validation; and publicly archive the audit code and seeds to enable independent verification.

    The current framework is developed for return prediction tasks with a well-defined trading lag.
    Extensions to classification settings (e.g., default prediction, regime labeling) and to non-trading contexts (e.g., macroeconomic forecasting) require adaptation of the performance metrics and null constructions, but the core logic of falsification remains transferable.

    \subsection{Defending Against Meta-Overfitting: The Blind-Parameter Protocol}
    \label{subsec:blind_protocol}
    The deployment of static falsification environments introduces a critical structural vulnerability: meta-overfitting. If the exact parameters of the null data-generating processes are globally observable, algorithmic search spaces can be adversarially optimized to bypass the diagnostic without capturing genuine economic structure. Consequently, passing a static falsification audit is strictly insufficient to prove out-of-sample validity.

    To mitigate this vulnerability, the audit must operate under a Blind-Parameter Protocol. Under this architecture, null environments are defined as continuous distributions over structural parameters rather than fixed point-estimates (e.g., for the MA(1) placebo draw $\theta \sim \mathrm{Uniform}[-0.8,-0.2]$ rather than fixing $\theta=-0.5$). The auditing entity provides a public set of parameter draws ($\Theta_{\mathrm{dev}}$) for pipeline calibration, while securing a disjoint, privately seeded set ($\Theta_{\mathrm{audit}}$) for the final falsification gate. This forces the workflow to demonstrate robustness across the parameter family, reducing the scope for adversarial optimization against the audit itself.

    \section{Simulation Evidence: Validating the Framework}
    \label{sec:simulation}
    We subject the theoretical framework to a comprehensive computational audit. The objective is twofold: first, to quantify the magnitude of spurious predictability generated by standard machine learning workflows under the global null hypothesis; and second, to illustrate how our diagnostics behave under known structural alternatives that mimic common backtesting pitfalls.

    All simulations are conducted with $N=1{,}000$ Monte Carlo replications per scenario using the \texttt{QuantAudit} \textsf{R} package. This library serves as the reference implementation for the framework, integrating modules for synthetic null generation, strict walk-forward validation, and the estimation of inflation diagnostics ($\Delta Z, \widehat{K}_{\mathrm{eff}}$). The \texttt{QuantAudit} R package and full replication scripts are prepared for public release and will be made available via GitHub and CRAN upon journal publication to facilitate independent replication and standardized auditing. Detailed uncertainty estimates and robustness checks are deferred to Appendix~\ref{app:sup_sim_results}.

    We evaluate representative simulation strategies categorized by their underlying mechanism. The \texttt{RandomBaseline} (White Noise Null) serves as a passive control, generating i.i.d.\ Gaussian signals under the global null. The \texttt{Lookahead} strategy (Oracle) has access to future returns $r_{t+1}$, representing the theoretical upper bound of forward-looking bias. \texttt{DataMiner} performs spurious selection by choosing the in-sample optimal configuration from $K = 100$ independent noise predictors, providing the primary test for selection-induced inflation.

    To evaluate structural failure modes, we define three strategies that interact with specific data-generating features: \texttt{Contrarian} implements a linear mean-reversion signal $S_t = -\theta r_{t-1}$; \texttt{RegimeDetector} applies threshold nonlinearity via state-dependent switching based on GARCH volatility estimates; and \texttt{FactorMimic} harvests a persistent structural risk premium to evaluate the conflation of systematic factor exposure with alpha.

    Finally, to investigate \emph{capacity-driven inflation}---the systematic in-sample performance amplification induced by adaptive search over high-capacity learners under the global null---and the impact of correlated hyperparameter search, we evaluate \texttt{FeatureMining} (orthogonal search) against \texttt{HyperparameterTuning} (correlated XGBoost grid search) and \texttt{TrendFollowing} (causal control).

    \subsection{The Scaling Law of Selection Bias}
    \label{subsec:scaling_law_sim}
\begin{figure}[htbp]
        \centering
        \includegraphics[width=0.8\textwidth]{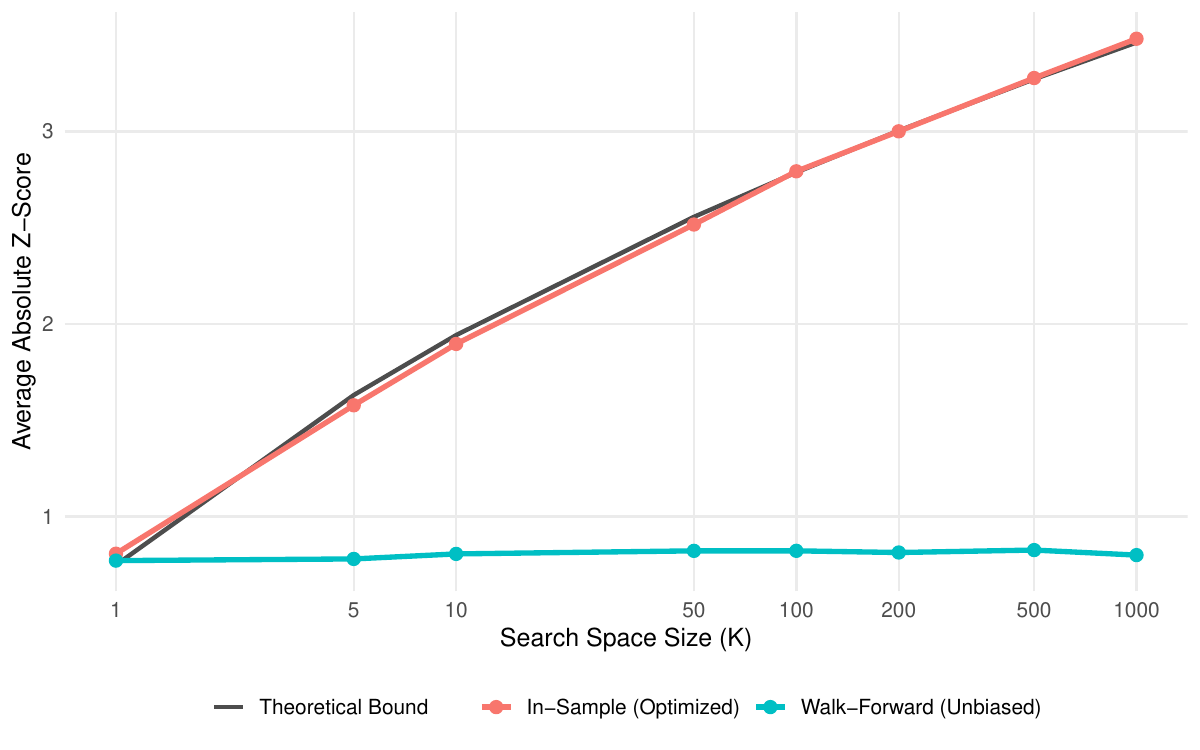}
        \caption{The scaling law of selection bias. The optimized in-sample winner magnitude grows on the $O(\sqrt{\log K})$ scale under the global null, while the walk-forward winner magnitude remains on the null scale when selection is performed on disjoint data.}
        \label{fig:scaling_law}
    \end{figure}

    We validate Proposition~\ref{thm:max_gaussian} by simulating a researcher who evaluates $K$ independent strategies in the White Noise environment (Environment~A). The researcher selects the configuration with the maximum absolute in-sample statistic, $Z^{\star}_{\mathrm{IS}}=\max_{k\le K}|Z_{\mathrm{IS},k}|$, and evaluates the selected index $k^{\star}$ on a disjoint walk-forward set, $Z^{\star}_{\mathrm{WF}}=|Z_{\mathrm{WF},k^{\star}}|$. Table~\ref{tab:scaling_law} reports the resulting dynamics, and Figure~\ref{fig:scaling_law} visualizes the scaling behavior.

    \begin{table}
        \caption{{The scaling law of selection bias (Environment A, $T=2{,}520$).} Winner FWP is the familywise false positive probability for the selected winner at the nominal threshold $|Z|>1.96$. $\mathrm{BIF}^{\mathrm{stab}}_{0.5}$ reports the median computed on the subset $\{|Z_{\mathrm{WF}}^\star| \ge 0.5\}$.}
        \label{tab:scaling_law}
        \centering
        \begin{tabular}[t]{lccccc}
            \toprule
            $K$ & Mean $Z_{\mathrm{IS}}^\star$ & Mean $Z_{\mathrm{WF}}^\star$ & $\Delta Z$ & $\mathrm{BIF}^{\mathrm{stab}}_{0.5}$ & Winner FWP (\%)\\
            \midrule
            1 & 0.81 & 0.77 & 0.04 & 0.65 & 5.3\\
            5 & 1.58 & 0.78 & 0.80 & 1.50 & 22.3\\
            10 & 1.90 & 0.81 & 1.09 & 1.80 & 40.2\\
            50 & 2.52 & 0.82 & 1.69 & 2.36 & 92.3\\
            100 & 2.79 & 0.82 & 1.97 & 2.60 & 99.9\\
            200 & 3.00 & 0.81 & 2.19 & 2.88 & 100.0\\
            500 & 3.28 & 0.83 & 2.45 & 3.18 & 100.0\\
            1000 & 3.48 & 0.80 & 2.68 & 3.33 & 100.0\\
            \bottomrule
        \end{tabular}
    \end{table}

    As predicted, optimized in-sample performance scales with $\sqrt{\log K}$. The expected maximum tracks the extreme-value benchmark for the maximum of $2K$ signed Gaussians, approximately $\sqrt{2\log(2K)}$. In finite samples, nominally significant in-sample discoveries become essentially certain as soon as the search budget is moderate. The winner false positive probability exhibits a sharp transition, rising quickly with $K$ and saturating near certainty for large search spaces despite the complete absence of signal.

    Crucially, the walk-forward winner magnitude remains on the null scale. Because selection is performed on disjoint in-sample data, $Z_{\mathrm{WF}}^\star$ behaves like the magnitude of a fixed configuration under the null, with $\mathbb{E}[Z_{\mathrm{WF}}^\star]$ remaining close to the half-normal expectation $\sqrt{2/\pi}\approx 0.80$ regardless of $K$. The divergence between in-sample and walk-forward evidence is therefore monotone in $K$ and is summarized by the Backtest Inflation Factor. At $K=1{,}000$, the median $\mathrm{BIF}_{0.5}$ on the reported subset is $5.12$, indicating that optimized in-sample evidence can exceed realized walk-forward evidence by roughly a factor of five solely due to multiplicity. Monte Carlo uncertainty summaries for the scaling curve are reported in Appendix~\ref{app:sup_sim_results}.

    \subsection{The Redundancy Law: Effective Multiplicity under Correlation}
    \label{subsec:redundancy_law_sim}
    Realistic strategy searches are rarely orthogonal, and they are rarely perfectly collinear. To test whether $\widehat{K}_{\mathrm{eff}}$ provides a usable proxy for effective search dimensionality under dependence, we simulate $K=500$ candidate strategies under the global null in two regimes. First, we use an idealized clustered construction that spans perfect redundancy ($\widehat{K}_{\mathrm{eff}}\approx 1$) to near-independence ($\widehat{K}_{\mathrm{eff}}=500$). Second, we verify robustness under imperfect intra-cluster correlation ($\rho<1$) with heterogeneous cluster sizes.

    \subsubsection{Clustered dependence benchmark}
    To test whether $\widehat{K}_{\mathrm{eff}}$ provides an operational adjustment for selection-driven inflation when candidates are redundant, we simulate clustered dependence spanning perfect redundancy to near-independence. Table~\ref{tab:redundancy_validation} compares observed in-sample inflation to extreme-value predictions computed using both nominal $K$ and $\widehat{K}_{\mathrm{eff}}$.

    \begin{table}[htbp]
        \centering
        \caption{Redundancy law validation ($K=500$). Strategies share $n_{\mathrm{clusters}}$ latent factors under the global null. $\widehat{K}_{\mathrm{eff}}$ is the shrinkage-corrected effective multiplicity. $\mathrm{EVT}(M)\approx a_M+\gamma b_M$ is the extreme-value expectation for $M$ standard normal variables, where $a_M=\sqrt{2\ln M}-\frac{\ln\ln M+\ln(4\pi)}{2\sqrt{2\ln M}}$, $b_M=\frac{1}{\sqrt{2\ln M}}$, and $\gamma\approx 0.5772$. The Ratio column compares observed inflation to the $\widehat{K}_{\mathrm{eff}}$-based prediction.}
        \label{tab:redundancy_validation}
        \begin{tabular}{lccccc}
            \toprule
            Dependence Regime & $\widehat{K}_{\mathrm{eff}}$ & $\mathbb{E}[|Z^{\star}_{\mathrm{IS}}|]$ & $\mathrm{EVT}(2K)$ & $\mathrm{EVT}(2\widehat{K}_{\mathrm{eff}})$ & Ratio \\
            \midrule
            Perfectly Correlated & 1.00 & 0.77 & 3.27 & 0.75 & 1.029 \\
            5 Clusters          & 5.03 & 1.57 & 3.27 & 1.63 & 0.963 \\
            10 Clusters         & 10.09 & 1.90 & 3.27 & 1.95 & 0.974 \\
            25 Clusters         & 25.46 & 2.26 & 3.27 & 2.31 & 0.978 \\
            50 Clusters         & 51.74 & 2.54 & 3.27 & 2.57 & 0.989 \\
            100 Clusters        & 106.66 & 2.78 & 3.27 & 2.81 & 0.990 \\
            Independent         & 500.00 & 3.26 & 3.27 & 3.27 & 0.998 \\
            \bottomrule
        \end{tabular}
    \end{table}

    The effective multiplicity estimator captures the compression of the search space induced by correlation. While nominal $K$ is fixed, the in-sample winner magnitude varies substantially with dependence. The observed winner statistic aligns closely with the $\mathrm{EVT}(2\widehat{K}_{\mathrm{eff}})$ benchmark, indicating that $\widehat{K}_{\mathrm{eff}}$ provides an operational adjustment for selection-driven inflation under the clustered dependence structures considered.

    The implications for inflation are summarized in Table~\ref{tab:redundancy_implications}. Median $\mathrm{BIF}^{\mathrm{stab}}_{0.5}$ rises from approximately $0.62$ under near-perfect redundancy to $3.23$ under independence, moving with the independence proxy $(\widehat{K}_{\mathrm{eff}}-1)/(K-1)$. In the limit $\widehat{K}_{\mathrm{eff}}\approx 1$, median $\mathrm{BIF}^{\mathrm{stab}}_{0.5}$ can fall below one due to finite-sample variation, consistent with the absence of systematic selection inflation. Full distributional summaries and uncertainty intervals are reported in Appendix~\ref{app:sup_sim_results}.

    \begin{table}[t]
        \centering
        \caption{Inflation implications under redundancy. The Independence Proxy is $(\widehat{K}_{\mathrm{eff}}-1)/(K-1)$. Median $\mathrm{BIF}^{\mathrm{stab}}_{0.5}$ is computed on the subset $\{|Z_{\mathrm{WF}}^\star| \ge 0.5\}$.}
        \label{tab:redundancy_implications}
        \begin{tabular}{lccc}
            \toprule
            Dependence Regime & Independence Proxy & $\mathrm{BIF}^{\mathrm{stab}}_{0.5}$ & $\mathbb{E}[Z^{\star}_{\mathrm{WF}}]$ \\
            \midrule
            Perfectly Correlated & 0.000 & 0.62 & 0.81 \\
            5 Clusters          & 0.008 & 1.44 & 0.84 \\
            10 Clusters         & 0.018 & 1.79 & 0.79 \\
            25 Clusters         & 0.049 & 2.22 & 0.80 \\
            50 Clusters         & 0.102 & 2.37 & 0.84 \\
            100 Clusters        & 0.212 & 2.69 & 0.82 \\
            Independent         & 1.000 & 3.23 & 0.76 \\
            \bottomrule
        \end{tabular}
    \end{table}

\begin{figure}[t]
        \centering
        \includegraphics[width=0.8\textwidth]{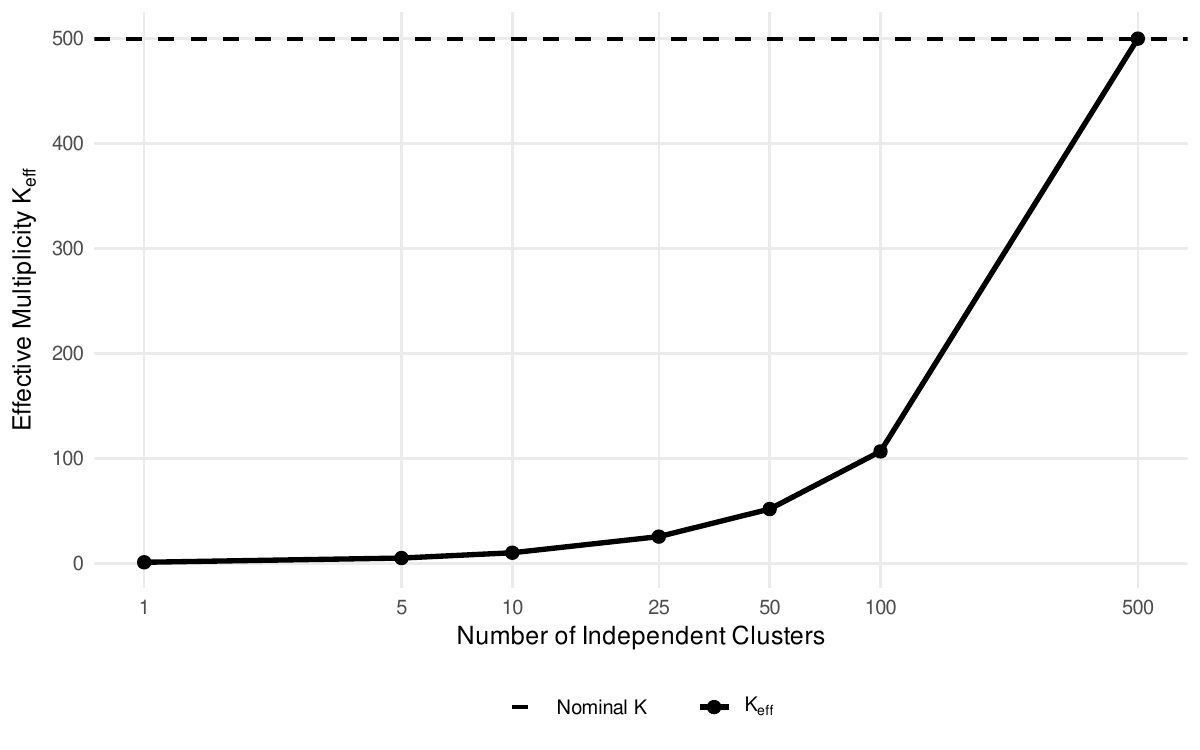}
        \caption{Effective multiplicity under correlation. The estimated $\widehat{K}_{\mathrm{eff}}$ tracks the actual number of independent clusters as redundancy increases. The horizontal line denotes the nominal search size $K=500$.}
        \label{fig:k_eff_vs_clusters}
    \end{figure}

    A complementary visualization of how redundancy collapses the search space is provided in Figure~\ref{fig:k_eff_vs_clusters}, which reports $\widehat{K}_{\mathrm{eff}}$ as a function of the number of latent clusters.

    \subsubsection{Robustness to imperfect intra-cluster correlation}
\begin{figure}[htbp]
        \centering
        \includegraphics[width=0.8\textwidth]{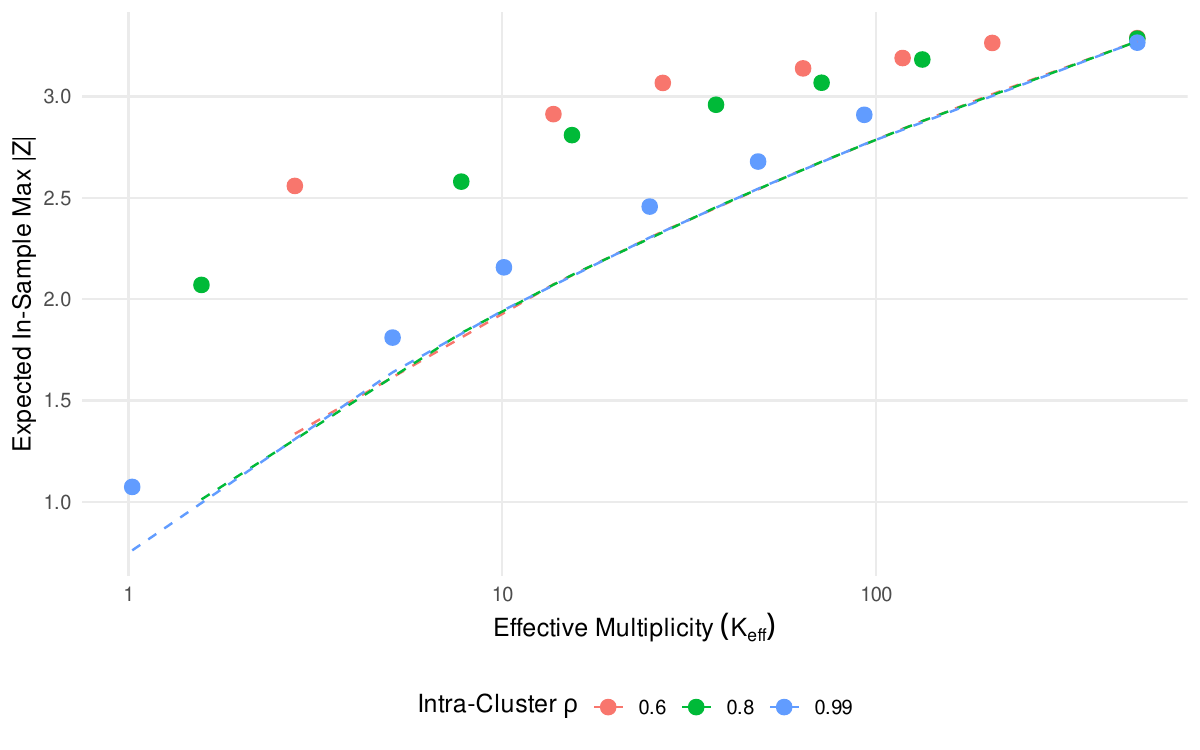}
        \caption{Validation of Redundancy Law under imperfect correlation. Observed in-sample winner magnitudes track the theoretical prediction based on $\widehat{K}_{\mathrm{eff}}$ across correlation regimes. Lower correlation increases effective multiplicity and inflates the winner statistic.}
        \label{fig:redundancy_law_new}
    \end{figure}

    To test whether the redundancy law is an artifact of idealized collinearity, we simulate $K=500$ candidate strategies with imperfect intra-cluster dependence. We vary the correlation $\rho \in \{0.6,0.8,0.99\}$ and assign strategies to clusters of heterogeneous sizes. Figure~\ref{fig:redundancy_law_new} visualizes that lower correlation increases $\widehat{K}_{\mathrm{eff}}$ and inflates the winner statistic in line with the $\sqrt{\log \widehat{K}_{\mathrm{eff}}}$ scaling. As shown in Table~\ref{tab:redundancy_validation_imperfect}, $\widehat{K}_{\mathrm{eff}}$ provides a robust calibration for the expected maximum, converging tightly to the EVT prediction as the number of independent clusters grows. The approximation naturally degrades only at the extreme boundary where the search space is approximately a single, highly noisy cluster.

    \begin{table}[htbp]
        \centering
        \caption{Redundancy law validation ($K=500$) with imperfect correlation. The $\widehat{K}_{\mathrm{eff}}$-based EVT benchmark calibrates the expected winner magnitude once the effective search space contains multiple independent clusters; in the boundary case where the universe is dominated by a single noisy cluster ($\widehat{K}_{\mathrm{eff}}$ near one), the EVT approximation can understate the observed maximum.}
        \label{tab:redundancy_validation_imperfect}
        \begin{tabular}{lrrrrr}
            \toprule
            $\rho$ & Clusters & $\widehat{K}_{\mathrm{eff}}$ & Obs. $\mathbb{E}[|Z^{\star}_{\mathrm{IS}}|]$ & Pred. (EVT) & Error \% \\
            \midrule
            0.60 & 1 & 2.8 & 2.56 & 1.34 & 91.5 \\
            & 5 & 13.7 & 2.91 & 2.07 & 40.5 \\
            & 10 & 26.9 & 3.07 & 2.33 & 31.4 \\
            & 25 & 63.8 & 3.14 & 2.64 & 18.9 \\
            & 50 & 117.8 & 3.19 & 2.84 & 12.3 \\
            & 100 & 204.7 & 3.26 & 3.01 & 8.4 \\
            & 500 & 500.0 & 3.29 & 3.27 & 0.5 \\
            \addlinespace
            0.80 & 1 & 1.6 & 2.07 & 1.01 & 104.5 \\
            & 5 & 7.8 & 2.58 & 1.83 & 40.7 \\
            & 10 & 15.4 & 2.81 & 2.12 & 32.6 \\
            & 25 & 37.3 & 2.96 & 2.45 & 20.6 \\
            & 50 & 71.5 & 3.07 & 2.68 & 14.6 \\
            & 100 & 133.0 & 3.18 & 2.88 & 10.6 \\
            & 500 & 500.0 & 3.28 & 3.27 & 0.4 \\
            \addlinespace
            0.99 & 1 & 1.0 & 1.07 & 0.76 & 41.1 \\
            & 5 & 5.1 & 1.81 & 1.64 & 10.5 \\
            & 10 & 10.1 & 2.16 & 1.95 & 10.8 \\
            & 25 & 24.8 & 2.46 & 2.30 & 6.6 \\
            & 50 & 48.4 & 2.68 & 2.54 & 5.3 \\
            & 100 & 93.0 & 2.91 & 2.76 & 5.3 \\
            & 500 & 500.0 & 3.27 & 3.27 & -0.2 \\
            \bottomrule
        \end{tabular}
    \end{table}

    \subsection{Capacity-Driven Inflation: Saturation vs.\ Inflation}
    \label{subsec:boosting_effect_sim}
\begin{figure}[htbp]
        \centering
        \includegraphics[width=0.8\textwidth]{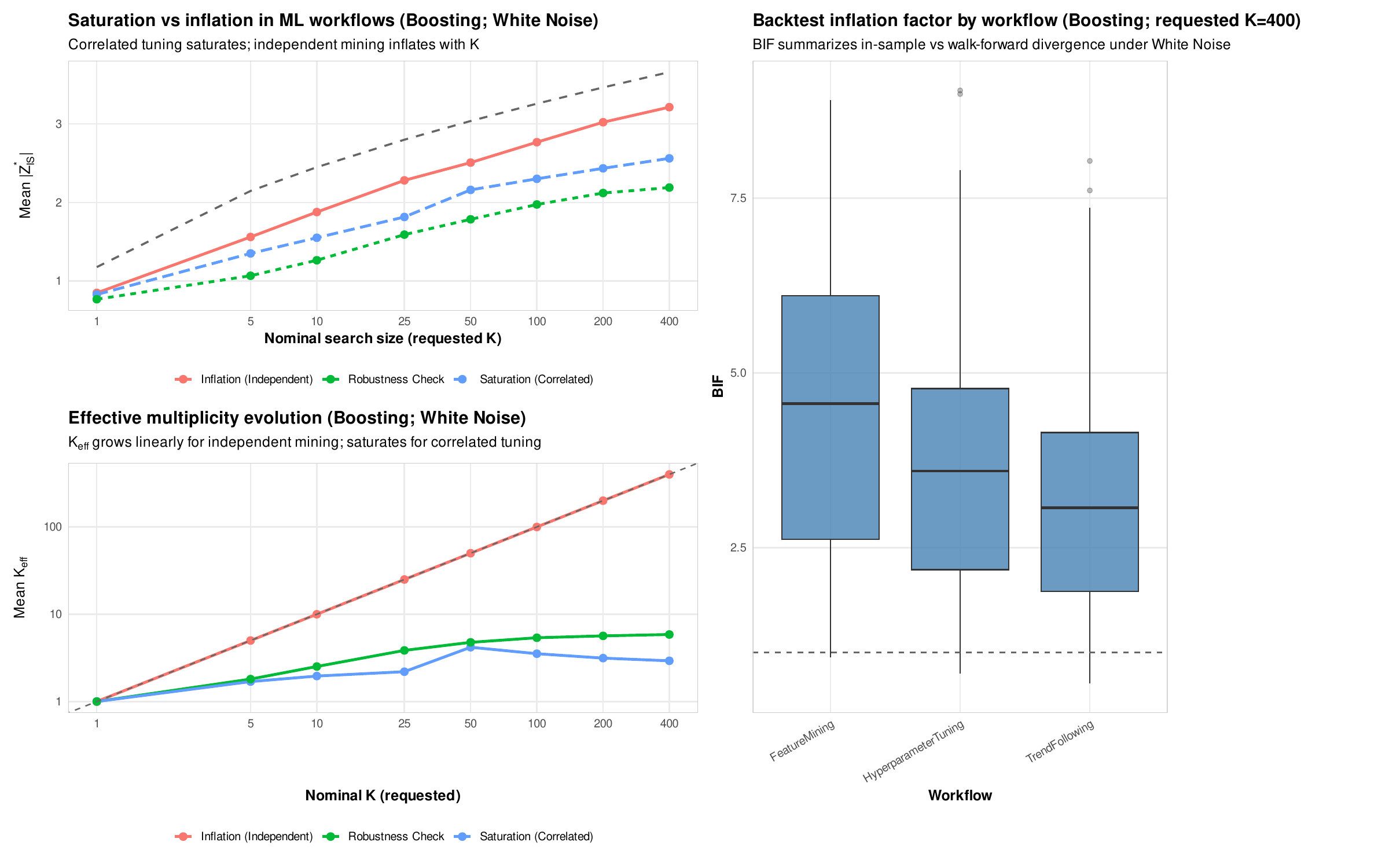}
        \caption{Saturation versus inflation in machine learning workflows under a global null. The left panels summarize in-sample extremeness and the evolution of $\widehat{K}_{\mathrm{eff}}$ with nominal $K$ under White Noise. The right panel summarizes the distribution of BIF at $K=400$ under White Noise; the dashed horizontal line marks the neutral benchmark $\mathrm{BIF}=1$.}
        \label{fig:capacity_driven_plots}
    \end{figure}

    In this section, we compare three machine learning workflows under the global null: \texttt{FeatureMining} (orthogonal search over independent candidates), \texttt{HyperparameterTuning} (XGBoost tuning over a correlated configuration family), and \texttt{TrendFollowing} (causal control with a small parameter grid). Table~\ref{tab:exp4b_compare_kmax} reports results at $K=400$ under the White Noise null environment, and Figure~\ref{fig:capacity_driven_plots} summarizes how in-sample extremeness and $\widehat{K}_{\mathrm{eff}}$ evolve with nominal search size.

    \begin{table}[htbp]
        \centering
        \small
        \caption{Workflow comparison under White Noise at $K=400$ (Boosting). Means over $N=1{,}000$ simulations; $\mathrm{BIF}^{\mathrm{stab}}_{0.5}$ is the median computed on the subset $\{|Z_{\mathrm{WF}}^\star| \ge 0.5\}$.}
        \label{tab:exp4b_compare_kmax}
        \resizebox{\textwidth}{!}{%
            \begin{tabular}{@{}lccccccc@{}}
                \toprule
                Workflow & $\widehat{K}_{\mathrm{eff}}$ & Mean $Z_{\mathrm{IS}}^\star$ & Mean $Z_{\mathrm{WF}}^\star$ & $\Delta Z$ & $\mathrm{BIF}^{\mathrm{stab}}_{0.5}$ & IS Fail (\%) & WF Fail (\%) \\
                \midrule
                FeatureMining & 400.00 & 3.21 & 0.84 & 2.37 & 2.91 & 100.0 & 6.6 \\
                \addlinespace
                \makecell[l]{Hyperparameter\\Tuning} & 2.93 & 2.56 & 0.79 & 1.77 & 2.49 & 88.2 & 4.7 \\
                \addlinespace
                TrendFollowing & 5.86 & 2.19 & 0.78 & 1.41 & 2.12 & 64.2 & 5.0 \\
                \bottomrule
            \end{tabular}%
        }
    \end{table}

    \texttt{FeatureMining} exhibits the expected inflation regime: $\widehat{K}_{\mathrm{eff}}\approx K$ and in-sample extremeness grows mechanically with the effective number of independent trials, while walk-forward evidence remains null-like. \texttt{HyperparameterTuning} exhibits capacity-driven inflation: despite low $\widehat{K}_{\mathrm{eff}}$, it produces materially inflated in-sample winners and high in-sample failure rates. The results indicate that saturation of a correlation-based redundancy measure does not imply immunity to selection inflation when the learner can fit sample idiosyncrasies on a correlated manifold. \texttt{TrendFollowing} lies between these regimes, but its effective search dimension is no longer negligible under causal execution, reaching $\widehat{K}_{\mathrm{eff}}\approx 5.9$ at $K=400$. Supplementary tables that separate saturation ratios and in-sample extremeness trajectories by workflow are reported in Appendix~\ref{app:sup_sim_results}.

    \subsection{Mechanism: Threshold Amplification}
    \label{subsec:threshold_amp_sim}
    To explain why a low $\widehat{K}_{\mathrm{eff}}$ on raw predictions can coincide with substantial in-sample extremeness, we isolate the role of non-linear signal extraction. We simulate highly correlated prediction scores ($\rho=0.9$) and compare the effective multiplicity of the raw scores to the effective multiplicity of thresholded trading signals. Table~\ref{tab:threshold_amp_kmax} reports results at $K=1{,}000$.

\begin{figure}[htbp]
        \centering
        \includegraphics[width=0.8\textwidth]{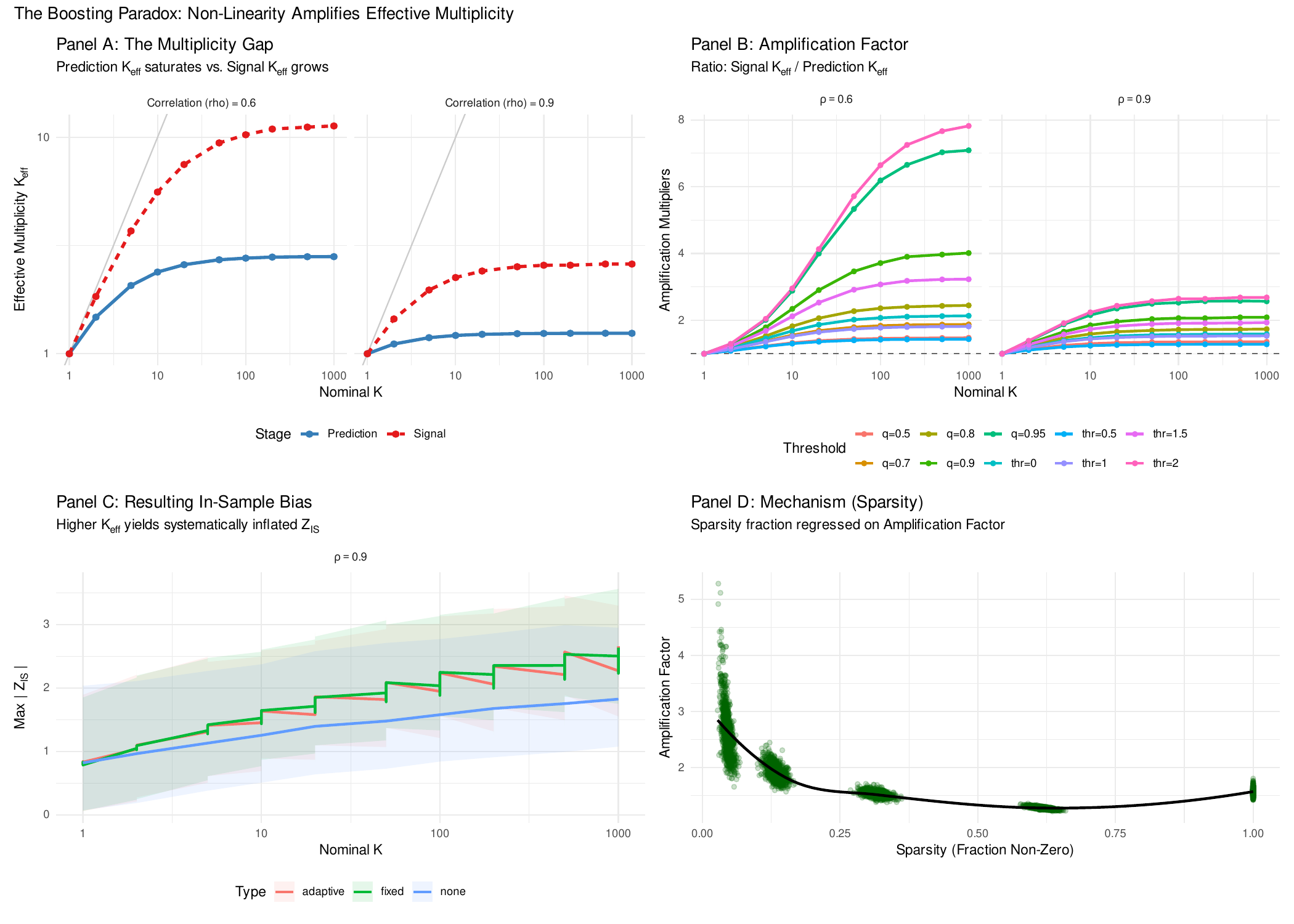}
        \caption{Threshold amplification and spurious orthogonality. Panel A: Effective multiplicity ($K_{\mathrm{eff}}$) of raw predictions and realized thresholded signals versus nominal $K$ for $\rho\in\{0.6,0.9\}$. Panel B: Amplification factor $K_{\mathrm{eff}}(\mathrm{Signal})/K_{\mathrm{eff}}(\mathrm{Pred})$ across threshold stringencies for $\rho\in\{0.6,0.9\}$. Panel C: Optimized in-sample extremum $Z_{\mathrm{IS}}^\star=\max_k |Z_{\mathrm{IS},k}|$ (mean with 5th--95th percentile band) at $\rho=0.9$. Panel D: Amplification versus signal sparsity (fraction non-zero) for fixed thresholds at $K=100$, $\rho=0.9$, showing amplification even when dense.}
        \label{fig:threshold_amp}
    \end{figure}

    The structural mechanism driving the divergence between raw predictor redundancy and resultant signal diversity is defined here as spurious orthogonality. As demonstrated in Figure~\ref{fig:threshold_amp}, the application of non-linear filtering mechanisms induces two distinct decorrelation dynamics. First, converting continuous predictions to binary states via sign extraction alters the correlation structure independently of sparsity. Second, and more critically, stringent thresholding induces sparsity within the signal vectors. This sparsity systematically decorrelates the resulting signal matrix. Consequently, the effective multiplicity ($K_{\mathrm{eff}}$) of the thresholded signals is substantially amplified relative to the highly correlated raw predictions (Panel A and Panel B). Panel C shows that thresholding is associated with an upward shift in the optimized in-sample extremum: in the $\rho=0.9$ regime plotted, both the mean of $\max\lvert Z_{\mathrm{IS}}^\star\rvert$ and its upper quantiles (as summarized by the 5th--95th percentile band) increase relative to the unthresholded baseline as $K$ grows, consistent with the higher $\widehat{K}_{\mathrm{eff}}$ induced by non-linear signal extraction; the panel is descriptive of in-sample optimization and does not by itself establish any corresponding walk-forward improvement. Panel D shows that sparsity is a dominant predictor of amplification within the fixed-threshold family at $K=100$ and $\rho=0.9$: the amplification factor $\widehat{K}_{\mathrm{eff}}(\mathrm{Signal})/\widehat{K}_{\mathrm{eff}}(\mathrm{Pred})$ increases as the fraction of non-zero signal entries falls, while remaining above one even when the fraction non-zero is near one, reflecting a discretization effect from mapping continuous scores to dense sign signals (e.g., $\mathrm{thr}=0$) that operates even in the absence of sparsity.

    \begin{table}[htbp]
        \centering
        \caption{Threshold amplification at $K=1{,}000$ ($N=1{,}000$). $\widehat{K}_{\mathrm{eff}}$ (Pred) is computed on raw scores; $\widehat{K}_{\mathrm{eff}}$ (Signal) is computed on realized trading signals. Amp.\ Factor is $\widehat{K}_{\mathrm{eff}}(\text{Signal})/\widehat{K}_{\mathrm{eff}}(\text{Pred})$.}
        \label{tab:threshold_amp_kmax}
        \begin{tabular}{llcccc}
            \toprule
            Threshold Type & Level & $\widehat{K}_{\mathrm{eff}}$ (Pred) & $\widehat{K}_{\mathrm{eff}}$ (Signal) & Amp.\ Factor & Mean $|Z_{\mathrm{IS}}^\star|$ \\
            \midrule
            None          & --      & 1.24 & 1.24 & 1.00 & 1.82 \\
            Fixed         & thr=1.0 & 1.24 & 1.91 & 1.54 & 2.42 \\
            Fixed         & thr=2.0 & 1.24 & 3.34 & 2.68 & 2.63 \\
            Adaptive      & q=0.50  & 1.24 & 1.69 & 1.36 & 2.27 \\
            Adaptive      & q=0.90  & 1.24 & 2.60 & 2.09 & 2.62 \\
            Adaptive      & q=0.95  & 1.24 & 3.20 & 2.57 & 2.66 \\
            \bottomrule
        \end{tabular}
    \end{table}

    Although the raw predictions remain nearly one-dimensional in this regime, thresholding materially increases the effective dimensionality of the realized signals by reducing tail co-movement. At $K=1{,}000$ and $\rho=0.9$, $\widehat{K}_{\mathrm{eff}}(\mathrm{Pred})$ remains fixed at 1.24 across specifications, whereas $\widehat{K}_{\mathrm{eff}}(\mathrm{Signal})$ rises from 1.24 under no thresholding to values between 1.69 and 3.34 under thresholded rules. The corresponding mean optimized in-sample extremum rises from 1.82 to values between 2.27 and 2.66. The strongest amplification in the reported grid occurs under the fixed-threshold specification $\mathrm{thr}=2.0$, with an amplification factor of 2.68, followed closely by adaptive thresholding at $q=0.95$, with an amplification factor of 2.57; adaptive thresholding at $q=0.90$ remains substantial, with an amplification factor of 2.09, but is not the maximal case. The mechanism is therefore broader than adaptive thresholding alone: both discretization and sparsity can transform a highly redundant prediction manifold into a materially less redundant signal family, thereby inflating optimized in-sample evidence without producing corresponding walk-forward gains. Full uncertainty summaries for the threshold-amplification sweep are provided in Appendix~\ref{app:sup_sim_results}.

    \subsection{The Linear Control: Regularization-Induced Redundancy}
    \label{subsec:linear_control_sim}
    To test whether the saturation pattern under correlated tuning is specific to non-linear learners, we replicate the null-workflow audit with a regularized linear model (Elastic Net via \texttt{glmnet}). The tuning workflow searches over $\alpha\in\{0,0.5,1\}$ and a dense regularization path in $\lambda$, subsampled to a nominal budget of $K$ configurations. Table~\ref{tab:exp4a_white_keff} reports the effective multiplicity trajectory under the White Noise null.

    \begin{table}[htbp]
        \centering
        \caption{Effective multiplicity under White Noise (Elastic Net). Mean $\widehat{K}_{\mathrm{eff}}$ across workflows and nominal search size $K$.}
        \label{tab:exp4a_white_keff}
        \begin{tabular}{lcccccccc}
            \toprule
            Workflow & K=1 & K=5 & K=10 & K=25 & K=50 & K=100 & K=200 & K=400 \\
            \midrule
            Elastic Net Tuning & 1 & 1.91 & 2.00 & 1.83 & 1.74 & 1.72 & 1.70 & 1.73 \\
            FeatureMining & 1 & 5.00 & 10.00 & 25.00 & 50.00 & 100.00 & 200.00 & 400.00 \\
            TrendFollowing & 1 & 1.81 & 2.52 & 3.85 & 4.76 & 5.38 & 5.65 & 5.86 \\
            \bottomrule
        \end{tabular}
    \end{table}

    Correlated regularization-path tuning exhibits substantial redundancy, with the effective search space compressed far below the nominal grid size. By $K=400$, Elastic Net Tuning stabilizes at $\widehat{K}_{\mathrm{eff}}=1.73$, indicating that hundreds of nominal evaluations compress to fewer than two effectively independent trials. Redundancy limits, but does not eliminate, selection inflation. Table~\ref{tab:exp4a_compare_kmax} summarizes cross-workflow outcomes at $K=400$ and shows that walk-forward evidence remains null-like while in-sample optimization remains materially inflated.

    \begin{table}[htbp]
        \centering
         \caption{Cross-workflow summary at $K=400$ (White Noise, $N=1{,}000$). Results correspond to the glmnet-based linear-control experiment. Reported values are Monte Carlo means, except $\mathrm{BIF}^{\mathrm{stab}}_{0.5}$, which reports the median on the subset $\{|Z_{\mathrm{WF}}^\star| \ge 0.5\}$.}
        \label{tab:exp4a_compare_kmax}
        \resizebox{\textwidth}{!}{%
            \begin{tabular}{lccccccc}
                \toprule
                Workflow & $\widehat{K}_{\mathrm{eff}}$ & Mean $|Z_{\mathrm{IS}}^\star|$ & Mean $|Z_{\mathrm{WF}}^\star|$ & Mean $\Delta Z$ & $\mathrm{BIF}^{\mathrm{stab}}_{0.5}$ & Fail IS (\%) & Fail WF (\%) \\
                \midrule
                FeatureMining      & 400.000 & 3.211 & 0.837 & 2.374 & 2.911 & 100.0 & 6.6 \\
                Elastic Net Tuning & 1.727 & 1.817 & 0.808 & 1.009 & 1.668 & 35.8 & 5.4 \\
                TrendFollowing     & 5.859 & 2.188 & 0.781 & 1.407 & 2.117 & 64.2 & 5.0 \\
                \bottomrule
            \end{tabular}%
        }
    \end{table}

    Confidence intervals for these estimates, reported in Appendix Table~\ref{tab:exp4a_unc_kmax}, confirm the precision of the point estimates and the robustness of the saturation pattern.

    The implication is structural: the multiplicity penalty is governed by the effective number of independent trials rather than the nominal grid size. Correlated tuning families compress the search volume and cap $\widehat{K}_{\mathrm{eff}}$, whereas orthogonal searches preserve $\widehat{K}_{\mathrm{eff}}\approx K$ and mechanically amplify optimized in-sample evidence under the null. The full $K$-sweep demonstrating saturation of $\widehat{K}_{\mathrm{eff}}$ and the corresponding plateau in in-sample extremeness is provided in Appendix~\ref{app:sup_sim_results}.

    \subsection{Positive Control: The Detection Frontier}
    \label{subsec:positive_control}

    A falsification framework must reject workflows under the global null while retaining power when predictability is real but weak. We therefore introduce a positive control that maps the detection frontier of the audit as a function of signal strength and event scarcity.

    We simulate a threshold autoregressive process designed to capture the dynamics targeted by breakout and trend-following rules:
    \begin{equation}
        r_t = \phi r_{t-1} \mathbb{I}(|r_{t-1}| > \theta \sigma) + \epsilon_t, \quad \epsilon_t \sim \mathcal{N}(0, \sigma^2),
        \label{eq:tar_dgp}
    \end{equation}
    where $\phi$ governs autocorrelation during active regimes and $\theta$ governs activation rarity (in standard deviations). Annualized volatility is fixed at $\sigma_{\text{ann}} = 15\%$, and the DGP is strictly causal, embedding genuine predictability when $\phi \neq 0$.

    We vary $\phi \in \{0.05, 0.10, 0.15, 0.20, 0.25\}$ and $\theta \in \{1.0, 1.5, 2.0, 2.5, 3.0\}$ over $N=1{,}000$ Monte Carlo replications per cell with $T=2{,}520$ trading days and a 60/40 in-sample/walk-forward split. The adaptive workflow selects the breakout threshold $k^\star$ from $\{0.5, 0.7, \dots, 3.5\}$ by maximizing the absolute in-sample t-statistic, subject to a minimum of 10 trades in-sample. Walk-forward inference uses HAC standard errors, and a strategy is classified as detected if $|Z_{\mathrm{WF}}| > 1.96$.

\begin{figure}[t]
        \centering
        \includegraphics[width=0.8\textwidth]{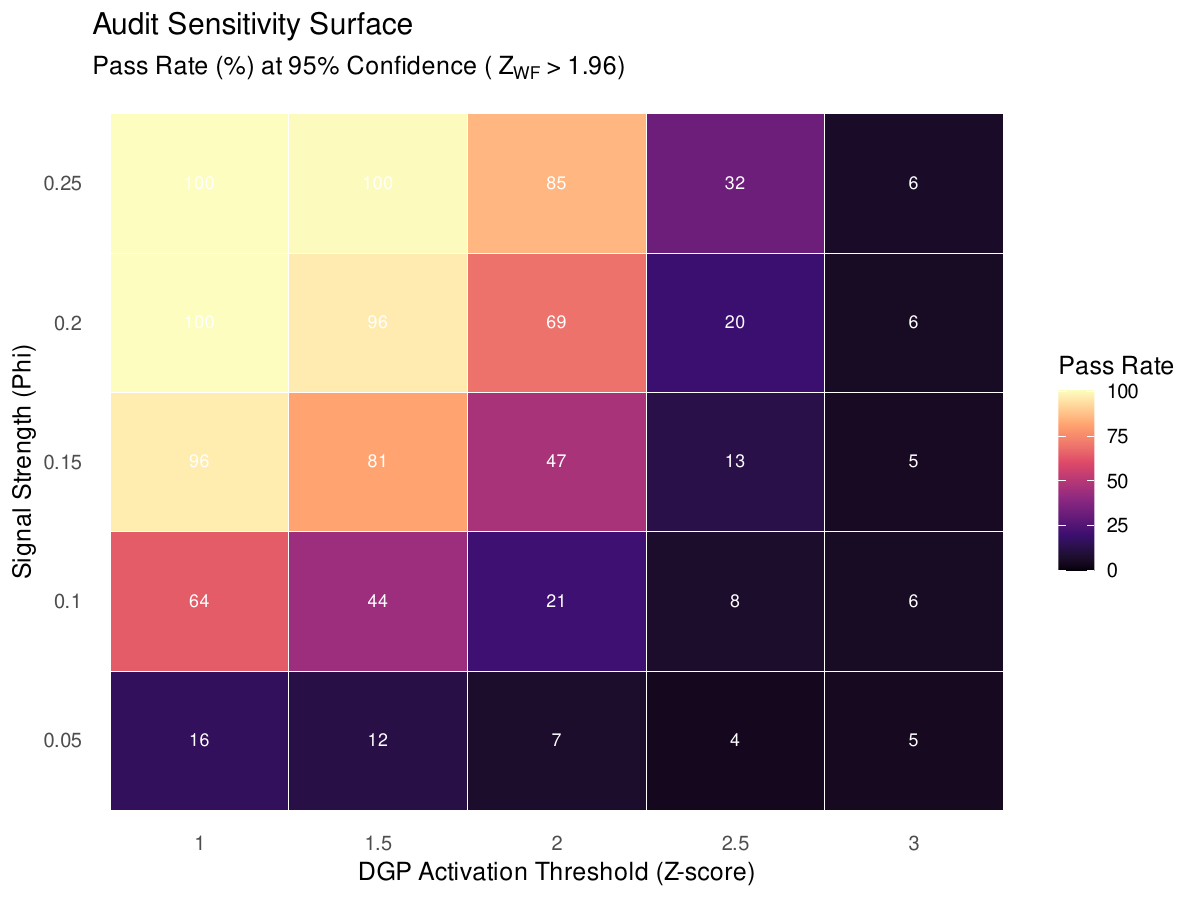}
        \caption{Detection frontier of the falsification audit. The heatmap shows pass rates (\% of simulations with $|Z_{\mathrm{WF}}| > 1.96$) across the parameter space.}
        \label{fig:detection_frontier}
    \end{figure}

    Table~\ref{tab:detection_limits} reports detection rates across the grid, and Figure~\ref{fig:detection_frontier} visualizes the detection surface.

    \begin{table}[htbp]
        \centering
        \caption{Detection limits: Pass rate (\%) of the falsification audit across signal strength ($\phi$) and activation threshold ($\theta$). Values are based on $N=1,\!000$ simulations per cell with HAC-adjusted walk-forward inference.}
        \label{tab:detection_limits}
        \begin{tabular}{lccccc}
            \toprule
            \multirow{2}{*}{Signal $\phi$} & \multicolumn{5}{c}{Activation threshold $\theta$ (standard deviations)} \\
            \cmidrule(lr){2-6}
            & 1.0 & 1.5 & 2.0 & 2.5 & 3.0 \\
            \midrule
            0.05 & 16.2 & 12.3 & 7.1 & 4.4 & 4.8 \\
            0.10 & 64.0 & 43.6 & 20.6 & 7.5 & 5.6 \\
            0.15 & 96.3 & 81.1 & 46.6 & 12.9 & 5.4 \\
            0.20 & 100.0 & 96.2 & 68.8 & 20.0 & 5.6 \\
            0.25 & 100.0 & 99.6 & 84.9 & 32.0 & 6.5 \\
            \bottomrule
        \end{tabular}
    \end{table}

    Detectability increases primarily with $\phi$ and declines with $\theta$ through a mechanically smaller effective sample size in the walk-forward window.
    For moderate-to-strong signals combined with non-scarce activation, the audit exhibits high power. However, the simulation demonstrates a severe power collapse under sparse signal regimes. Specifically, for weak and rare activation parameters ($\phi=0.05$, $\theta=3.0$), the detection rate falls to 4.8\%, operating strictly at or below the nominal 5\% size of the test. This confirms that the falsification framework possesses effectively zero statistical power to detect highly sparse, intermittent causal structures. The diagnostic architecture structurally prioritizes strict Type I error control (rejecting spurious pipelines) over Type II error minimization. Researchers must concede that while the audit reliably flags structural artifacts, it will systematically fail to validate genuine predictability if the underlying economic mechanism activates too infrequently within the walk-forward sample.

    Relative to the global-null simulations in \Cref{subsec:scaling_law_sim}, the positive control confirms asymmetric diagnostic behavior: under the null, BIF rises systematically with search size because optimized in-sample extremeness grows while walk-forward evidence remains null-like; under detectable alternatives, inflation diagnostics remain materially closer to unity because out-of-sample evidence rises with genuine structure. This establishes that the audit is not a universal rejection device and that its stringency concentrates on spurious predictability rather than suppressing real, causally valid signal.

    \subsection{Summary of Simulation Evidence}
    \label{subsec:simulation_summary}

    The simulations support the framework on five dimensions.
    Optimized in-sample winners grow on the $\Theta(\sqrt{\log K_{\mathrm{eff}}})$ scale under the null, while disjoint walk-forward winners remain null-like.
    Effective multiplicity $\widehat{K}_{\mathrm{eff}}$ provides an operational proxy for search dimensionality under clustered dependence and aligns observed inflation with extreme-value predictions.
    The falsification environments separate leakage and microstructure exploitation from workflows that are merely selection-inflated, while remaining near-nominal when the null is correctly specified.
    Correlated search families can exhibit severe redundancy in $\widehat{K}_{\mathrm{eff}}$ and still generate material selection inflation, with non-linear workflows particularly susceptible through threshold amplification that increases effective dimensionality during signal extraction.
    Finally, the positive control establishes that the audit's stringency is correctly targeted: it successfully filters out spurious selection bias without destroying statistical power for detecting genuine, causally valid economic signals.

    \section{Empirical Applications: Case Studies}
    \label{sec:empirical_applications}
    This section applies the two-stage audit protocol of \Cref{sec:audit} to market data to illustrate empirically relevant failure modes. Stage~1 runs each end-to-end pipeline on the induced-null environments and records pass/fail using the calibrated thresholds in \Cref{subsec:calibration}. Specifications that survive are then evaluated in Stage~2 on the historical sample under a strict disjoint in-sample/walk-forward design, with all adaptive selection confined to the in-sample block. We report the resulting in-sample and walk-forward statistics, the effective multiplicity estimate, the stabilized inflation diagnostics, and the null-calibrated inflation class. The objective is diagnostic rather than competitive: to determine whether apparent performance is consistent with robustness to structural nulls, coherent attribution after controlling for systematic exposures, and strictly time-respecting evaluation.

    These case studies are demonstrations aligned with specific channels targeted by the audit rather than an estimate of the population prevalence of spurious findings. Evidence on power and false-positive control is therefore deferred to the simulation section, which maps the detection frontier under controlled alternatives and null environments.

    \subsection{Mechanism Diagnosis: Deep Learning on SPY}
    \label{subsec:app1_spy}
    When evaluated on SPY (S\&P 500 ETF proxy) under strict temporal segregation with an internal pre-2016 validation split and without class rebalancing, the deep learning pipeline exhibits no out-of-sample evidence of directional predictability. On the post-2015 test window, the annualized alpha relative to buy-and-hold is $0.39\%$ with $p=0.931$, implying that any apparent in-sample edge does not survive a clean holdout evaluation.

    The dominant empirical pattern is threshold-induced activity clustering. Using decision thresholds calibrated exclusively on the training sample, the position trigger rate increases from $69.4\%$ in low-volatility states to $88.7\%$ in high-volatility states (Table \ref{tab:app_01_mechanism}), while the out-of-sample score distribution exhibits a material reduction in probability mass in a neighborhood of the decision boundary. Let $s_t$ denote the model score and let $s_{\mathrm{med}}$ be the training-sample median score used to anchor the symmetric threshold rule with margin $\delta$; the corresponding boundary-mass statistic $\Pr!\left(\bigl||s_t-s_{\mathrm{med}}|-\delta\bigr|\le \varepsilon\right)$ declines across volatility regimes, consistent with high-variance periods reallocating probability mass away from the decision margin and mechanically increasing trade activation.

    \begin{table}[t]
        \centering
        \small
        \caption{{Mechanism Diagnosis (SPY).} Activity increases in high volatility as score dispersion pushes mass away from the decision boundary.}
        \label{tab:app_01_mechanism}
        \begin{tabular}{lcc}
            \toprule
            Metric & Calm & High Vol \\
            \midrule
            Trigger Rate (Activity) & 69.4\% & 88.7\% \\
            Boundary Mass ($\Pr\!\left(\bigl||s_t-s_{\mathrm{med}}|-\delta\bigr|\le \varepsilon\right)$) & 0.174 & 0.066 \\
            \bottomrule
            \multicolumn{3}{l}{\footnotesize \textit{Note:} $\delta$ and $s_{\mathrm{med}}$ calibrated on training data; regime cutoff is the training 80th percentile} \\
            \multicolumn{3}{l}{\footnotesize of lagged volatility; $\varepsilon=0.01$. Alpha vs Buy\&Hold: 0.39\% ($p=0.931$). Returns are gross of} \\
            \multicolumn{3}{l}{\footnotesize transaction costs.}
        \end{tabular}
    \end{table}

    This reduction indicates that high-volatility periods reallocate probability mass away from the decision margin and into regions beyond the fixed thresholds, mechanically increasing trade activation.

    A Monte Carlo falsification on a zero-mean GARCH(1,1) null confirms that this mechanism does not require any conditional-mean structure. Across $N=30$ independent null paths, the identical pipeline reproduces regime-dependent activity, with a mean calm-regime trigger rate of $68.9\%$ and a mean high-volatility trigger rate of $79.1\%$ (Table \ref{tab:app_01_garch_mc}).

    \begin{table}[htbp]
        \centering
        \small
        \caption{{GARCH(1,1) Falsification (Monte Carlo).} Mean statistics across $N=30$ independent generator paths.}
        \label{tab:app_01_garch_mc}
        \begin{tabular}{lc}
            \toprule
            Metric & Mean Value \\
            \midrule
            Mean Gross Z (Model) & -0.014 \\
            Mean Alpha vs B\&H (Ann) & -0.17\% \\
            Mean Trigger Rate (Calm) & 68.9\% \\
            Mean Trigger Rate (High Vol) & 79.1\% \\
            \bottomrule
        \end{tabular}
    \end{table}

    In contrast, null performance remains centered near zero (mean annualized alpha $-0.17\%$; mean gross $Z=-0.014$), while the empirical distribution exhibits a non-negligible right tail (Figure \ref{fig:app_01_combined}, Panel B).

    \begin{figure}[htbp]
        \centering
        \includegraphics[width=0.8\textwidth]{app_01_combined_corrected.pdf}
        \caption{{Empirical vs. Falsification Performance.} Panel A shows the cumulative return of the deep learning model versus buy-and-hold on the SPY out-of-sample data. Panel B displays the distribution of annualized alpha generated by the identical pipeline on zero-mean GARCH(1,1) null paths.}
        \label{fig:app_01_combined}
    \end{figure}

    Taken together, the evidence supports a strictly mechanical interpretation: the workflow functions primarily as a nonlinear volatility-responsive activation rule. Heteroskedastic feature realizations widen the score distribution relative to fixed thresholds calibrated in calmer training conditions, raising turnover without generating a robust directional signal. Because results are reported gross of transaction costs, the economic impact of this structural failure is understated.

    \subsection{Cross-Sectional Machine Learning and the Beta Confound}
    \label{subsec:app2_cross_sectional}
    Cross-sectional machine learning pipelines can generate apparently predictive returns by reconstructing static exposure to priced risk premia rather than producing abnormal performance. This beta confound is diagnosed when gross long--short performance is statistically significant but the factor-adjusted intercept is indistinguishable from zero.

    To evaluate this failure channel, we apply an expanding-window ridge regression ($\lambda=0.01$) to the Fama--French 25 Portfolios sorted by Size and Book-to-Market. Each month, the model produces out-of-sample expected returns which are used to construct an extremal long--short portfolio that goes long the top three and short the bottom three portfolios. Following strict data integrity protocols, Fama--French sentinel missing values ($-99.99$) are scrubbed prior to estimation to avoid injecting artificial outliers into the ridge coefficients.

    The audit applies a Newey--West HAC intercept test in the FF3 regression to separate systematic exposure from residual performance. As documented in Table~\ref{tab:beta_trap_results}, the gross long--short strategy yields a statistically significant Bartlett-HAC Z-score of 3.153 ($p = 0.002$). In this sample, the beta-confound pattern does not obtain: when subjected to the FF3 audit regression, the strategy retains an economically and statistically significant intercept ($Z_{\alpha} = 2.708$, $p = 0.007$). Accordingly, the performance is not mechanically attributable to standard market, size, and value exposures, and the systematic-risk replication diagnostic does not trigger.

\begin{figure}[htbp]
        \centering
        \includegraphics[width=0.8\textwidth]{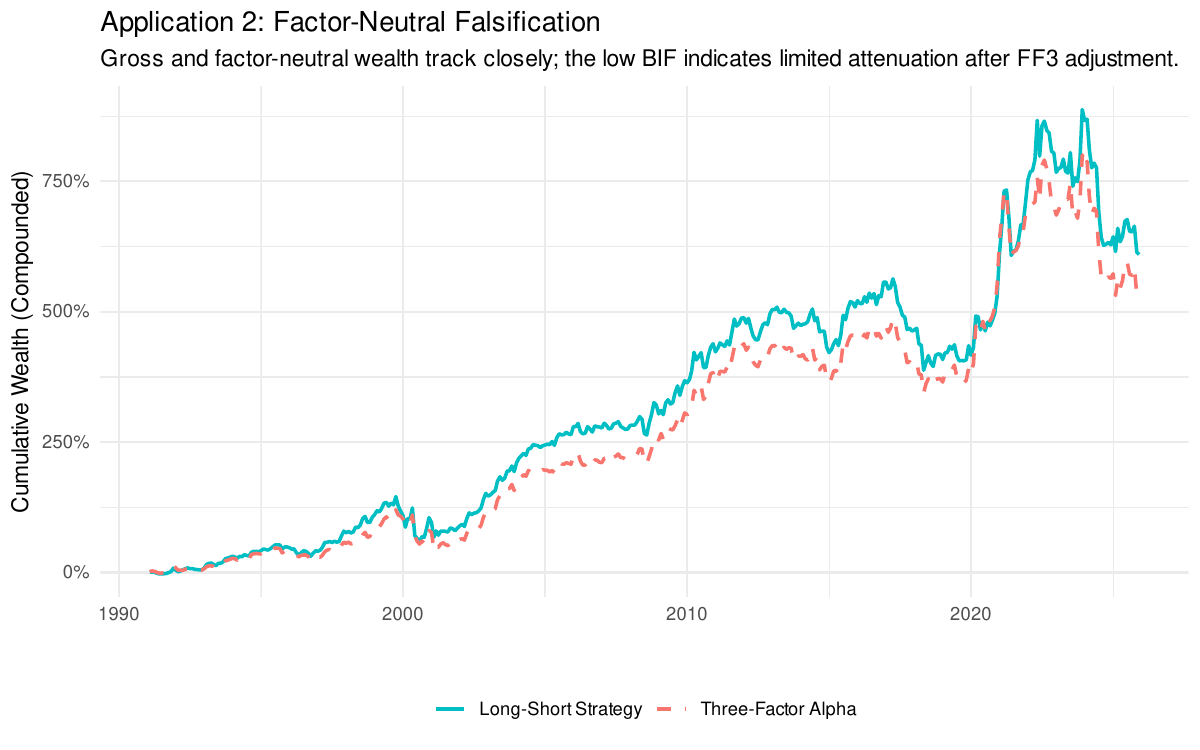}
        \caption{Cross-sectional ridge regression performance. Cumulative wealth of the out-of-sample long--short strategy that goes long the top three and short the bottom three FF25 portfolios each month (1991--2025). The dashed line plots the factor-neutral return series obtained by subtracting the fitted FF3 factor component from the strategy return. The limited divergence between the two series is consistent with the low Backtest Inflation Factor ($\mathrm{BIF}=1.16$).}
        \label{fig:beta_trap_wealth}
    \end{figure}

    The tau-floored Backtest Inflation Factor, $\mathrm{BIF}=|Z_{\mathrm{gross}}|/\max(|Z_{\alpha}|,\tau)$ with $\tau=0.5$, evaluates to $1.16$. This tight diagnostic spread indicates limited attenuation when moving from gross performance to factor-neutral performance under FF3. Figure \ref{fig:beta_trap_wealth} provides a visual diagnostic: the factor-neutral series closely tracks the gross series over the full out-of-sample window.

    \begin{table}[htbp]
        \centering
        \caption{\textbf{Cross-sectional systematic-risk replication diagnostics.} Ridge regression on FF25 out-of-sample performance (1991--2025).}
        \label{tab:beta_trap_results}
        \begin{tabular}{lc}
            \toprule
            Metric & Value \\
            \midrule
            Gross Z-Score & 3.153 \\
            Gross $p$-value & 0.002 \\
            Alpha Z-Score (FF3) & 2.708 \\
            Alpha $p$-value (FF3) & 0.007 \\
            BIF (Gross / Alpha) & 1.16 \\
            OOS Months & 419 \\
            \bottomrule
        \end{tabular}

        \vspace{1mm}
        {\footnotesize \textit{Note:} Alpha statistics computed via Newey-West HAC standard errors.}
    \end{table}

    \subsection{Validation Inflation in Volatility Forecasting}
    \label{subsec:app3_volatility}
    Financial time series exhibiting strong serial dependence, such as realized volatility, can be vulnerable to validation inflation when evaluated using machine learning protocols that ignore temporal ordering. In principle, random $k$-fold cross-validation may mix contiguous, highly correlated observations across training and validation partitions, allowing the model-selection process to exploit local regime persistence. Whether this distortion is material, however, is ultimately an empirical question.

    To assess this issue, we evaluate a daily volatility forecasting task on SPY (log realized volatility). The feature set includes standard Heterogeneous Autoregressive (HAR) lags (daily, weekly, and monthly) alongside 20 days of lagged returns. We contrast three specifications: a baseline linear HAR model, a Lasso regression utilizing standard random $k$-fold cross-validation, and a strictly causal expanding-window walk-forward Lasso. To eliminate temporal leakage during walk-forward refits, internal hyperparameter tuning ($\lambda$) is restricted to chronological block cross-validation.

    As documented in Table \ref{tab:section6_volatility_results}, the baseline HAR model establishes a genuine predictive signal, achieving an out-of-sample $R^2$ of $23.8\%$ ($0.238$).

    \begin{table}[htbp]
        \centering
        \small
        \caption{{Volatility prediction performance.} HAR and Walk-Forward metrics reflect true chronological holdout performance. The Random CV metrics reflect the internal cross-validated estimate, allowing comparison between time-inappropriate and time-respecting validation.}
        \label{tab:section6_volatility_results}
        \begin{tabular}{lcc}
            \toprule
            Model & OOS $R^2$ & RMSE \\
            \midrule
            HAR (Baseline) & 0.238 & 0.000584 \\
            Lasso + Random CV & 0.212 & 0.000446 \\
            Lasso + Walk-Forward (refit=5d) & 0.211 & 0.000594 \\
            \bottomrule
        \end{tabular}
    \end{table}

    \begin{table}[htbp]
        \centering
        \small
        \caption{{Validation inflation diagnostics.} VIR is defined as $R^2_{\mathrm{RandomCV}}/R^2_{\mathrm{WF}}$. $\Delta R^2$ is the absolute divergence between the internal estimate and the chronological realization.}
        \label{tab:section6_volatility_vir}
        \begin{tabular}{lc}
            \toprule
            Metric & Value \\
            \midrule
            Validation Inflation Ratio (VIR) & 1.006 \\
            Delta $R^2$ (RandomCV - Walk-Forward) & 0.001 \\
            \bottomrule
        \end{tabular}
    \end{table}

    Table \ref{tab:section6_volatility_vir} summarizes the comparison. The ratio-based diagnostic is now informative, and it indicates near-equivalence rather than inflation: $\mathrm{VIR}=1.006$ and $\Delta R^2 = R^2_{\mathrm{RandomCV}}-R^2_{\mathrm{WF}}=0.001$. In other words, the random-fold and walk-forward evaluations deliver almost the same estimated performance for the Lasso specification in this sample.

    The substantive conclusion is therefore different from a volatility-trap narrative. The higher-dimensional Lasso model does not materially outperform under random cross-validation, nor does it materially deteriorate under strict chronological evaluation; instead, both validation schemes yield nearly identical results, while the simpler HAR benchmark remains modestly stronger. The broader methodological lesson remains that temporal causality should be respected in tuning and evaluation for serially dependent targets, but the empirical magnitude of validation inflation must be established case by case rather than presumed.

    \subsection{Unified Interpretation Across Applications}
    \label{subsec:synthesis}

    Across the applications, apparent predictive success is shown to be compatible with multiple non-signal channels that become empirically distinguishable only when the full workflow is treated as the object under test. The case studies span threshold--volatility interactions that inflate activity without robust directional content, cross-sectional performance driven by systematic risk replication versus factor-neutral intercepts, and validation inflation from time-inappropriate fold construction under volatility clustering.

    These cases motivate a unified auditing discipline. Each pipeline is first subjected to end-to-end falsification under economically structured null environments, then decomposed via factor attribution, and finally evaluated under strictly chronological out-of-sample partitions. The simulation positive control in \Cref{subsec:positive_control} confirms that these gates do not sterilize genuine structure: when causal predictability is present, nontrivial walk-forward power remains detectable.

    \section{Implications for Research and Practice}
    \label{sec:implications}

    The falsification audit framework has direct implications for three stakeholder groups in the financial machine learning ecosystem: practitioners, academic researchers, and supervisors/policymakers. The unifying message is practical: in weak-signal, temporally ordered data, reported performance is not credible unless it survives (i) temporally causal validation, (ii) falsification under structured nulls, and (iii) explicit adjustment for adaptive specification search. These requirements operationalize long-standing concerns about data mining and non-replication in empirical finance \citep{HarveyLiuZhu2016,HouXueZhang2020} and align with placebo-based falsification principles in econometrics \citep{Imbens2015}.

    \subsection{Implications for Practitioners}

    For asset managers and quantitative research teams, the falsification audit is best treated as a pre-deployment gate that sits between research backtesting and any form of capital allocation. The objective is not to suppress exploration, but to prevent investment in strategies whose apparent edge is mechanically produced by validation artifacts, adaptive selection, or microstructure sensitivity rather than stable conditional mean structure. In weak-signal, temporally ordered data, the baseline evidentiary standard is procedural: a claim is not credible unless the entire end-to-end pipeline respects causal information flow and remains null-like under structured induced-null environments.

    Operationally, strict walk-forward evaluation should be enforced as a workflow invariant, not as a final reporting choice. Random cross-validation breaks time ordering and can import future regimes through repeated re-use of observations, which can inflate apparent predictive performance even when the model class is well-specified; this risk applies to preprocessing, feature construction, hyperparameter tuning, and model selection, not only to the final fit \citep{Bergmeir2018}. The audit then provides a direct robustness stress test: running the unchanged pipeline across martingale-difference reference classes and targeted placebos identifies whether performance survives when economic predictability is removed by design, thereby diagnosing leakage, timing misalignment, and protocol-driven uplift.

    Finally, adaptive search should be priced explicitly as model risk. Because selection-induced inflation is intrinsic to modern ML workflows, practitioners should record inflation diagnostics alongside headline performance, using the absolute magnitude gap $\Delta Z$ as the primary, well-defined measure and the stabilized $\mathrm{BIF}$ together with $\widehat{K}_{\mathrm{eff}}$ as an internal ranking and governance summary. This produces a defensible research record in which backtest strength, falsification outcomes, and search-induced inflation are jointly visible, enabling deployment decisions to be made on realized evidence rather than optimized in-sample extremeness.

    \section{Conclusion}
    \label{sec:conclusion}

    This paper formalizes a falsification framework to diagnose spurious predictability in financial machine learning pipelines. By evaluating complete predictive workflows against synthetic reference classes—including strict martingale-difference environments and targeted microstructure placebos—the audit systematically filters out performance driven by data leakage, indexing errors, and selection bias.

    For workflows that survive the falsification gates, the framework quantifies the multiplicity penalty using the absolute magnitude gap ($\Delta Z$). Because disjoint walk-forward evaluation remains bounded in probability under the null, $\Delta Z$ provides a mathematically continuous, unconfounded measurement of selection-induced decay. Crucially, we restrict the interpretation of the ratio-based Backtest Inflation Factor ($\mathrm{BIF}^{\mathrm{stab}}_{\tau}$) to a bounded statistical heuristic, explicitly acknowledging that it cannot serve as a strictly linear economic adjustment when out-of-sample evidence collapses toward zero.

    The empirical applications demonstrate the audit in action by separating mechanical success from genuine structure across representative failure modes.

    While this framework establishes rigorous bounds for extreme-value bias under clustered search geometries, significant extensions remain. Currently, the inflation metrics operate as descriptive statistics evaluated against simulated null thresholds. Future research must formalize the statistical inference surrounding these measurements. Developing a robust hypothesis testing framework—such as a pipeline-specific bootstrap test for the magnitude gap—will transition the quantification of backtest decay from a descriptive diagnostic into a formal inferential standard. Ultimately, shifting the burden of proof from historical backtest performance to algorithmic robustness against strictly defined nulls is essential for the maturation of financial machine learning.

    \setlength{\bibhang}{0pt}
    \bibliographystyle{abbrvnat}
    \bibliography{references}

\begin{thebibliography}{30}
\providecommand{\natexlab}[1]{#1}
\providecommand{\url}[1]{\texttt{#1}}
\expandafter\ifx\csname urlstyle\endcsname\relax
  \providecommand{\doi}[1]{doi: #1}\else
  \providecommand{\doi}{doi: \begingroup \urlstyle{rm}\Url}\fi

\bibitem[Athey(2019)]{Athey2019}
S.~Athey.
\newblock The impact of machine learning on economics.
\newblock In \emph{The Economics of Artificial Intelligence: An Agenda}, pages
  507--547. University of Chicago Press, 2019.

\bibitem[Bailey and L{\'o}pez~de Prado(2014)]{Bailey2014}
D.~H. Bailey and M.~L{\'o}pez~de Prado.
\newblock The deflated sharpe ratio: Correcting for selection bias, backtest
  overfitting, and non-normality.
\newblock \emph{The Journal of Portfolio Management}, 40\penalty0 (5):\penalty0
  94--107, 2014.

\bibitem[Belloni et~al.(2014)Belloni, Chernozhukov, and Hansen]{Belloni2014}
A.~Belloni, V.~Chernozhukov, and C.~Hansen.
\newblock Inference on treatment effects after selection among high-dimensional
  controls.
\newblock \emph{The Review of Economic Studies}, 81\penalty0 (2):\penalty0
  608--650, 2014.

\bibitem[Bergmeir et~al.(2018)Bergmeir, Hyndman, and Koo]{Bergmeir2018}
C.~Bergmeir, R.~J. Hyndman, and B.~Koo.
\newblock A note on the validity of cross-validation for evaluating
  autoregressive time series prediction.
\newblock \emph{Computational Statistics \& Data Analysis}, 120:\penalty0
  70--83, 2018.

\bibitem[Campbell et~al.(2024)Campbell, Ham, Lu, and Wood]{Campbell2024}
J.~L. Campbell, H.~Ham, Z.~Lu, and K.~Wood.
\newblock Expectations matter: When (not) to use machine learning earnings
  forecasts.
\newblock \emph{SSRN Electronic Journal}, 2024.
\newblock URL
  \url{https://papers.ssrn.com/sol3/papers.cfm?abstract_id=4495297}.
\newblock Working Paper.

\bibitem[Campbell et~al.(1997)Campbell, Lo, and MacKinlay]{Campbell1997}
J.~Y. Campbell, A.~W. Lo, and A.~C. MacKinlay.
\newblock \emph{The Econometrics of Financial Markets}.
\newblock Princeton University Press, 1997.

\bibitem[Carhart(1997)]{Carhart1997}
M.~M. Carhart.
\newblock On persistence in mutual fund performance.
\newblock \emph{The Journal of Finance}, 52\penalty0 (1):\penalty0 57--82,
  1997.

\bibitem[Cont(2001)]{Cont2001}
R.~Cont.
\newblock Empirical properties of asset returns: stylized facts and statistical
  issues.
\newblock \emph{Quantitative Finance}, 1\penalty0 (2):\penalty0 223--236, 2001.

\bibitem[Diebold(2015)]{Diebold2015}
F.~X. Diebold.
\newblock Comparing predictive accuracy, twenty years later: A personal
  perspective on the use and abuse of diebold-mariano tests.
\newblock \emph{Journal of Business \& Economic Statistics}, 33\penalty0
  (1):\penalty0 1--1, 2015.
\newblock \doi{10.1080/07350015.2014.983236}.

\bibitem[Engle(1982)]{Engle1982}
R.~F. Engle.
\newblock Autoregressive conditional heteroscedasticity with estimates of the
  variance of united kingdom inflation.
\newblock \emph{Econometrica}, 50\penalty0 (4):\penalty0 987--1007, 1982.

\bibitem[Fama and French(1993)]{FamaFrench1993}
E.~F. Fama and K.~R. French.
\newblock Common risk factors in the returns on stocks and bonds.
\newblock \emph{Journal of Financial Economics}, 33\penalty0 (1):\penalty0
  3--56, 1993.
\newblock \doi{10.1016/0304-405X(93)90023-5}.

\bibitem[Giné and Nickl(2016)]{GineNickl2016}
E.~Giné and R.~Nickl.
\newblock \emph{Mathematical Foundations of Infinite-Dimensional Statistical
  Models}.
\newblock Cambridge Series in Statistical and Probabilistic Mathematics.
  Cambridge University Press, Cambridge, 2016.

\bibitem[Gu et~al.(2020)Gu, Kelly, and Xiu]{GuKellyXiu2020}
S.~Gu, B.~Kelly, and D.~Xiu.
\newblock Empirical asset pricing via machine learning.
\newblock \emph{The Review of Financial Studies}, 33\penalty0 (5):\penalty0
  2223--2273, 2020.

\bibitem[Hansen(2005)]{Hansen2005}
P.~R. Hansen.
\newblock A test for superior predictive ability.
\newblock \emph{Journal of Business \& Economic Statistics}, 23\penalty0
  (4):\penalty0 365--380, 2005.

\bibitem[Harvey et~al.(2016)Harvey, Liu, and Zhu]{HarveyLiuZhu2016}
C.~R. Harvey, Y.~Liu, and H.~Zhu.
\newblock ...and the cross-section of expected returns.
\newblock \emph{The Review of Financial Studies}, 29\penalty0 (1):\penalty0
  5--68, 2016.

\bibitem[Hou et~al.(2020)Hou, Xue, and Zhang]{HouXueZhang2020}
K.~Hou, C.~Xue, and L.~Zhang.
\newblock Replicating anomalies.
\newblock \emph{The Review of Financial Studies}, 33\penalty0 (5):\penalty0
  2019--2133, 2020.
\newblock \doi{10.1093/rfs/hhy131}.

\bibitem[Imbens and Rubin(2015)]{Imbens2015}
G.~W. Imbens and D.~B. Rubin.
\newblock \emph{Causal Inference in Statistics, Social, and Biomedical
  Sciences}.
\newblock Cambridge University Press, 2015.

\bibitem[Jensen et~al.(2023)Jensen, Kelly, and
  Pedersen]{JensenKellyPedersen2023}
T.~I. Jensen, B.~Kelly, and L.~H. Pedersen.
\newblock Is there a replication crisis in finance?
\newblock \emph{The Journal of Finance}, 78\penalty0 (5):\penalty0 2465--2518,
  2023.

\bibitem[Lee et~al.(2016)Lee, Sun, Sun, and Taylor]{Lee2016}
J.~D. Lee, D.~L. Sun, Y.~Sun, and J.~E. Taylor.
\newblock Exact post-selection inference, with application to the lasso.
\newblock \emph{The Annals of Statistics}, 44\penalty0 (3):\penalty0 907--927,
  2016.

\bibitem[Lopez~de Prado(2018)]{LopezDePrado2018}
M.~Lopez~de Prado.
\newblock \emph{Advances in Financial Machine Learning}.
\newblock John Wiley \& Sons, 2018.

\bibitem[Mandelbrot(1963)]{Mandelbrot1963}
B.~Mandelbrot.
\newblock The variation of certain speculative prices.
\newblock \emph{The Journal of Business}, 36\penalty0 (4):\penalty0 394--419,
  1963.

\bibitem[Newey and West(1987)]{NeweyWest1987}
W.~K. Newey and K.~D. West.
\newblock A simple, positive semi-definite, heteroskedasticity and
  autocorrelation consistent covariance matrix.
\newblock \emph{Econometrica}, 55\penalty0 (3):\penalty0 703--708, 1987.

\bibitem[Newey and West(1994)]{NeweyWest1994}
W.~K. Newey and K.~D. West.
\newblock Automatic lag selection in covariance matrix estimation.
\newblock \emph{The Review of Economic Studies}, 61\penalty0 (4):\penalty0
  631--653, 1994.

\bibitem[Sch{\"a}fer and Strimmer(2005)]{Schaefer2005}
J.~Sch{\"a}fer and K.~Strimmer.
\newblock A shrinkage approach to large-scale covariance matrix estimation and
  implications for functional genomics.
\newblock \emph{Statistical Applications in Genetics and Molecular Biology},
  4\penalty0 (1):\penalty0 Article 32, 2005.

\bibitem[Shao(2010)]{Shao2010}
X.~Shao.
\newblock The dependent wild bootstrap.
\newblock \emph{Journal of the American Statistical Association}, 105\penalty0
  (489):\penalty0 218--235, 2010.

\bibitem[van Binsbergen et~al.(2023)van Binsbergen, Han, and
  Lopez-Lira]{Binsbergen2023}
J.~H. van Binsbergen, X.~Han, and A.~Lopez-Lira.
\newblock Man versus machine learning: The term structure of earnings
  expectations and conditional biases.
\newblock \emph{The Review of Financial Studies}, 36\penalty0 (6):\penalty0
  2361--2396, 2023.

\bibitem[Varian(2014)]{Varian2014}
H.~R. Varian.
\newblock Big data: New tricks for econometrics.
\newblock \emph{Journal of Economic Perspectives}, 28\penalty0 (2):\penalty0
  3--28, 2014.

\bibitem[Wainwright(2019)]{Wainwright2019}
M.~J. Wainwright.
\newblock \emph{High-Dimensional Statistics: A Non-Asymptotic Viewpoint}.
\newblock Cambridge Series in Statistical and Probabilistic Mathematics.
  Cambridge University Press, Cambridge, 2019.
\newblock ISBN 978-1108498029.

\bibitem[White(2000)]{White2000}
H.~White.
\newblock A reality check for data snooping.
\newblock \emph{Econometrica}, 68\penalty0 (5):\penalty0 1097--1126, 2000.

\bibitem[Zhang et~al.(2025)Zhang, Zhu, and Linnainmaa]{Zhang2025}
Y.~Zhang, Y.~Zhu, and J.~T. Linnainmaa.
\newblock Man versus machine learning revisited.
\newblock \emph{The Review of Financial Studies}, 38\penalty0 (12):\penalty0
  3768--3790, 2025.

\end{thebibliography}

    \clearpage
    \appendix
    \section*{Appendices}
    \addcontentsline{toc}{section}{Appendices}
    \renewcommand{\thesection}{\Alph{section}}
    \renewcommand{\thefigure}{\thesection.\arabic{figure}}
    \renewcommand{\thetable}{\thesection.\arabic{table}}
    \makeatletter
    \@addtoreset{equation}{section}
    \makeatother
    \renewcommand{\theequation}{\thesection.\arabic{equation}}

    \section{Asymptotic Motivation for Selection-Induced Inflation}
    \label{app:theory_proofs}

    This appendix provides the asymptotic motivation for the scaling behavior of in-sample maximization under the global null and outlines the finite-sample variance dynamics of walk-forward performance. The mathematical derivations herein represent heuristic limiting cases (where the hypothesis space $K$ is fixed and $T \to \infty$) intended to illustrate the base mechanisms of selection bias. In the high-dimensional finite-sample environments typical of machine learning ($K \ge T$), the uniform Gaussian approximation is not guaranteed, and we do not rely on it. Consequently, the scaling law $\Theta(\sqrt{\ln K_{\mathrm{eff}}})$ is empirically validated by the simulation suite in \Cref{sec:simulation}, rather than by non-asymptotic mathematical proof. Throughout, $K$ denotes the number of evaluated strategies and $T$ the time-series length used to construct the studentized statistics.

    \subsection{Gaussian maximal inequalities and effective multiplicity}
    \label{app:gaussian_max}

    Let $\mathbf{Z}=(Z_1,\dots,Z_K)$ be a vector of centered, unit-variance statistics. As an asymptotic limit, assume that as $T\to\infty$,
    \[
    \mathbf{Z}\ \Rightarrow\ \mathcal{N}(0,\Sigma),
    \]
    for some correlation matrix $\Sigma$. Define the canonical metric on the index set $\{1,\dots,K\}$ by
    \[
    d(i,j)\ =\ \sqrt{\mathbb{E}(Z_i-Z_j)^2}\ =\ \sqrt{2(1-\Sigma_{ij})}.
    \]
    Let $N(\varepsilon)$ denote the $\varepsilon$-covering number and $M(\varepsilon)$ the $\varepsilon$-packing number of $\{1,\dots,K\}$ under $d$. Define the effective multiplicity (stable rank proxy) by
    \[
    K_{\mathrm{eff}}\ =\ \frac{(\mathrm{tr}\,\Sigma)^2}{\|\Sigma\|_F^2}.
    \]

    \begin{lemma}[Gaussian maxima are governed by metric entropy]
        \label{lem:gaussian_entropy_bounds}
        There exist universal constants $0<c<C<\infty$ such that for any centered Gaussian vector $\mathbf{Z}\sim\mathcal{N}(0,\Sigma)$ with unit variances,
        \[
        \mathbb{E}\Big[\max_{1\le k\le K} Z_k\Big]\ \le\ C\int_{0}^{\mathrm{diam}}\sqrt{\ln N(\varepsilon)}\,d\varepsilon,
        \qquad
        \mathbb{E}\Big[\max_{1\le k\le K} Z_k\Big]\ \ge\ c\sup_{\varepsilon>0}\ \varepsilon\sqrt{\ln M(\varepsilon)},
        \]
        where $\mathrm{diam}=\max_{i,j} d(i,j)\le 2$. The same bounds hold, up to universal constants, with $\max_k Z_k$ replaced by $\max_k |Z_k|$.
    \end{lemma}

    \vspace{0.5em}
    \noindent\textbf{Derivation.} The upper bound is Dudley’s entropy integral inequality for Gaussian processes \citep{Wainwright2019}. The lower bound follows from Sudakov’s minoration theorem for Gaussian vectors \citep{GineNickl2016}: for any $\varepsilon>0$, $\varepsilon\sqrt{2\ln M(\varepsilon)} \le 2\mathbb{E}[\max_{1\le k\le K} Z_k]$. For the extension to $\max_k |Z_k|$, consider the Gaussian process indexed by $\{(k,+),(k,-)\}$ with values $Z_{(k,+)}=Z_k$ and $Z_{(k,-)}=-Z_k$; applying the same bounds to $\max \widetilde{Z}$ yields the desired result, because $\max_k |Z_k| = \max\{\max_k Z_k,\max_k(-Z_k)\}$. \hfill $\blacksquare$

    \begin{lemma}[Block-redundancy geometry links entropy and $K_{\mathrm{eff}}$]
        \label{lem:keff_entropy_bridge}
        Let $\mathbf{Z}\sim\mathcal{N}(0,\Sigma)$ with unit variances. Suppose $\Sigma$ is block diagonal with $m$ independent clusters of sizes $s_1,\dots,s_m$ and within-block equicorrelation $\rho\in(0,1)$. Let $K=\sum_{c=1}^m s_c$, and $s_{\max}=\max_c s_c$.

        \begin{enumerate}
            \item[(i)] The Frobenius norm satisfies $\|\Sigma\|_F^2=\sum_{c=1}^m\big(s_c+\rho^2 s_c(s_c-1)\big)$, yielding
            $K_{\mathrm{eff}}=K^2 / \sum_{c=1}^m\big(s_c+\rho^2 s_c(s_c-1)\big)$.
            In the strong-redundancy limit $\rho\uparrow 1$, $K_{\mathrm{eff}}\to m$ under balanced clusters ($s_c=K/m$).
            \item[(ii)] Under the canonical metric $d$, choosing one representative index per block yields a packing of size $m$ at separation $\delta=\sqrt{2}$. Hence $M(\delta)\ge m$. Moreover, each block is contained in a ball of radius $\eta=\sqrt{2(1-\rho)}$ centered at its representative, so $N(\varepsilon)\le m$ for any $\varepsilon\ge\eta$.
        \end{enumerate}
        Consequently, by Lemma~\ref{lem:gaussian_entropy_bounds}, the bounds are $\mathbb{E}[\max_{k\le K}|Z_k|] \le C\big(\eta\sqrt{\ln K} + 2\sqrt{\ln m}\big)$ and $\mathbb{E}[\max_{k\le K}|Z_k|] \ge c\sqrt{2}\sqrt{\ln m}$.
    \end{lemma}

    \vspace{0.5em}
    \noindent\textbf{Derivation.} Part (i) follows by direct trace calculation. For part (ii), the distance between distinct block representatives is $\sqrt{2}$ due to orthogonal independence, forming a $\sqrt{2}$-packing. For any index $k$ in block $c$, $d(k,i_c)=\sqrt{2(1-\rho)}=\eta$, establishing the covering bound. Applying Dudley's integral split at $\eta$ and Sudakov's minoration at $\delta=\sqrt{2}$ recovers the specified bounds. The term $\eta\sqrt{\ln K}$ captures the potential inflation from within-cluster noise; when $\eta$ is small (i.e., $\rho \to 1$), this term is dominated, and the expected maximum scales proportionally to $\sqrt{\ln m}$. In the balanced-cluster regime $s_c\equiv K/m$, as $\rho \to 1$, $K_{\mathrm{eff}}\asymp m$, motivating the $\Theta(\sqrt{\log K_{\mathrm{eff}}})$ expectation. \hfill $\blacksquare$

    \subsection{Finite-Sample Walk-Forward Dynamics and Conditional Persistence}
    \label{app:wf_stability}

    The evaluation of the selected walk-forward sequence requires explicit differentiation between asymptotic limits and finite-sample conditional mechanics.

    Let the in-sample period end at time $T_{\mathrm{IS}}$ and the walk-forward period be $\mathcal{T}_{\mathrm{WF}}=\{T_{\mathrm{IS}}+1,\dots,T\}$. Let $k^\star = \text{argmax}_{k} |Z_{\mathrm{IS},k}|$ be the index selected strictly using in-sample data. Define walk-forward P\&L increments by $Y_t=x_{t,k^\star}r_t$.

    If the underlying returns $r_t$ represent a strictly independent and identically distributed (i.i.d.) martingale difference sequence, then conditional on the selection event, $Y_t$ trivially inherits the unconditional moments. As $|\mathcal{T}_{\mathrm{WF}}| \to \infty$, standard Central Limit Theorem conditions apply, and a HAC variance estimator $\widehat{V}_n$ converges, yielding $Z^\star_{\mathrm{WF}} \xrightarrow{d} \mathcal{N}(0,1)$.

    However, financial returns systematically violate strict i.i.d. assumptions, frequently exhibiting volatility clustering (e.g., GARCH dynamics). Under conditional heteroskedasticity, the maximization function $k^\star$ can interact with finite-sample variance trajectories. The algorithm evaluates candidate strategies $x_{t,k}$ whose specific market exposures may have aligned with elevated variance states preceding $T_{\mathrm{IS}}$.

    Consequently, the initial state of the walk-forward sequence $Y_{T_{\mathrm{IS}}+1}$ can be conditioned on an elevated terminal variance realization. Due to the persistence of conditional variance, this state can carry over into the finite walk-forward period. This post-selection conditioning creates a documented risk of finite-sample size distortion for standard HAC estimators. While HAC estimators are asymptotically consistent for unconditional long-run variance, they can understate variance in finite samples skewed by transient conditional shocks, leading to artificial inflation of the test statistic $Z^\star_{\mathrm{WF}}$ and degraded Type I error control.

    \vspace{0.5em}
    \noindent\textbf{Operational Inference and Robustness.} Because finite-sample size distortions can compromise asymptotic HAC estimators under conditional heteroskedasticity, standard test statistics should be interpreted with caution. In this paper, our default operational inference relies directly on the induced-null simulation reference classes (\Cref{sec:audit}), which bypass asymptotic approximations by empirically calibrating the exact finite-sample distribution under the null. For practitioners who require standalone inference without simulating a full suite of reference environments, we recommend incorporating a dependent wild bootstrap or block bootstrap \citep{Shao2010} as a robustness layer. This resampling procedure approximates the finite-sample distribution of $Z^\star_{\mathrm{WF}}$ while preserving conditional heteroskedasticity and weak dependence, without relying on the potentially distorted finite-sample moment assumptions of standard asymptotic estimators.

    \section{Synthetic Reference Environments: Full Specifications}
    \label{app:environments}

    This section provides the complete mathematical specifications and default parameter values for the five canonical reference environments used in the falsification audit (\Cref{sec:audit}). Unless stated otherwise, let $\mathcal{F}^{(r)}_{t}=\sigma(r_1,\dots,r_t)$ denote the natural filtration generated by the observed return process. Environments A, B, D, and E are strict martingale-difference nulls with respect to $\mathcal{F}^{(r)}_{t-1}$, satisfying
    \[
    \mathbb{E}[r_t \mid \mathcal{F}^{(r)}_{t-1}] = 0,
    \]
    so that no exploitable conditional mean dynamics are present by construction. Environment C is a targeted microstructure placebo: it is deliberately constructed so that the observed return process fails the martingale-difference restriction, serving as a normative disqualifier for workflows that inadvertently monetize execution, indexing, or alignment artifacts.

    \subsection{Environment A: White Noise (Pure Selection Bias)}
    \label{app:envA}

    \paragraph{Data generating process.}
    \[
    r_t = \sigma_{\text{daily}}\,\epsilon_t,
    \qquad
    \epsilon_t \stackrel{\text{i.i.d.}}{\sim}\mathcal{N}(0,1).
    \]
    Then $\mathbb{E}[r_t\mid\mathcal{F}^{(r)}_{t-1}]=0$ and $\operatorname{Cov}(r_t,r_{t-\ell})=0$ for all $\ell\ge 1$.

    \paragraph{Default calibration.}
    Annualized volatility target $\sigma^{\text{ann}} = 0.20$, giving
    \[
    \sigma_{\text{daily}}=\frac{0.20}{\sqrt{252}}.
    \]

    \paragraph{Targeted failure mode.}
    Isolates pure selection bias. Under this environment, any statistically significant performance must arise from adaptive specification search rather than genuine predictability.

    \subsection{Environment B: Regime-Switching Volatility (Markov, Zero Mean)}
    \label{app:envB}

    This environment introduces persistent volatility regimes while maintaining a zero conditional mean. Let $s_t\in\{1,2\}$ be an unobserved two-state Markov chain. Even though $\mathcal{F}^{(r)}_{t-1}$ is informative about $s_t$ through volatility clustering, the conditional mean remains zero because the shock has mean zero and is independent of the past.

    \paragraph{Data generating process.}
    Let $s_t \in \{1,2\}$ be a two-state Markov chain with transition matrix
    \[
    P=\begin{pmatrix}
        p_{11} & 1-p_{11} \\
        1-p_{22} & p_{22}
    \end{pmatrix}.
    \]
    Returns are generated as
    \[
    r_t = \sigma_{s_t}\,\epsilon_t,
    \qquad
    \epsilon_t \stackrel{\text{i.i.d.}}{\sim}\mathcal{N}(0,1),
    \]
    with $\epsilon_t$ independent of $\sigma(s_u,u\le t)$ and hence independent of $\mathcal{F}^{(r)}_{t-1}$. Therefore $\mathbb{E}[r_t\mid\mathcal{F}^{(r)}_{t-1}]=0$.

    \paragraph{Default calibration.}
    Low-regime annualized volatility $\sigma^{\text{ann}}=0.10$, high-regime volatility multiplier $m=3.0$, regime persistence $p_{11}=p_{22}=0.98$. Hence
    \[
    \sigma_{1}=\frac{0.10}{\sqrt{252}},
    \qquad
    \sigma_{2}=\frac{0.10\,m}{\sqrt{252}}=\frac{0.30}{\sqrt{252}}.
    \]

    \paragraph{Targeted failure mode.}
    Detects predictability arising from improper normalization, regime leakage, or scale-based trading rules that inadvertently exploit volatility regimes.

    \subsection{Environment C: Friction Placebo (Bid--Ask Bounce / Roll MA(1))}
    \label{app:envC}

    This environment is a targeted microstructure placebo. A latent efficient return satisfies the martingale-difference restriction, but the observed return process is constructed to exhibit bid--ask bounce and hence to violate the martingale-difference restriction under $\mathcal{F}^{(r)}_{t-1}$.

    \paragraph{Data generating process.}
    Let the latent efficient return be
    \[
    u_t = \sigma_\epsilon\,\epsilon_t,
    \qquad
    \epsilon_t \stackrel{\text{i.i.d.}}{\sim}\mathcal{N}(0,1),
    \]
    and define the latent filtration $\mathcal{F}^{(u)}_{t}=\sigma(u_1,\dots,u_t)$. Then $\mathbb{E}[u_t\mid\mathcal{F}^{(u)}_{t-1}]=0$. Observed returns are generated as a Roll-style bounce reduced form
    \[
    r_t = u_t + \theta\,u_{t-1},
    \qquad
    |\theta|<1,
    \]
    which is an MA(1) in the latent shocks. The martingale-difference restriction fails for the observed returns whenever $\theta\neq 0$ and $u_{t-1}\in\mathcal{F}^{(r)}_{t-1}$ is predictable in $L^2$ from past observed returns, which holds for invertible MA(1) processes. A simple, filtration-free diagnostic is the nonzero lag-one autocovariance:
    \[
    \operatorname{Cov}(r_t,r_{t-1})
    =
    \operatorname{Cov}(u_t+\theta u_{t-1},\,u_{t-1}+\theta u_{t-2})
    =
    \theta\,\operatorname{Var}(u_{t-1})
    =
    \theta\,\sigma_\epsilon^2.
    \]
    If $\{r_t\}$ were a martingale-difference sequence with respect to $\mathcal{F}^{(r)}_{t-1}$, then $\mathbb{E}[r_t r_{t-1}]=0$ would follow from $\mathbb{E}[r_t\mid\mathcal{F}^{(r)}_{t-1}]=0$; the above autocovariance therefore certifies the intentional violation when $\theta\neq 0$.

    Innovations are scaled to match a target annualised volatility $\sigma^{\text{ann}}$. Let
    \[
    \sigma_{\text{daily}}=\frac{\sigma^{\text{ann}}}{\sqrt{252}},
    \qquad
    \sigma_\epsilon=\frac{\sigma_{\text{daily}}}{\sqrt{1+\theta^2}},
    \]
    so that $\operatorname{Var}(r_t)=\sigma_{\text{daily}}^2$.

    \paragraph{Default calibration.}
    Annualised volatility target $\sigma^{\text{ann}}=0.20$, MA(1) coefficient $\theta=-0.5$.

    \paragraph{Targeted failure mode.}
    Targets timing, indexing, alignment, and execution-lag errors that can spuriously monetize transient microstructure patterns.

    \subsection{Environment D: Factor Null (One-Factor, Zero Alpha)}
    \label{app:envD}

    This environment contains systematic risk exposure but no abnormal performance by construction.

    \paragraph{Data generating process.}
    \[
    r_t = \beta f_t + e_t,
    \]
    with
    \[
    f_t \stackrel{\text{i.i.d.}}{\sim}\mathcal{N}(0,\sigma_f^2),
    \qquad
    e_t \stackrel{\text{i.i.d.}}{\sim}\mathcal{N}(0,\sigma_e^2),
    \qquad
    f_t \perp e_t,
    \]
    and $(f_t,e_t)$ independent of $\mathcal{F}^{(r)}_{t-1}$. Hence $\mathbb{E}[r_t\mid\mathcal{F}^{(r)}_{t-1}]=0$.

    \paragraph{Default calibration.}
    $\beta=1.0$ and daily volatilities
    \[
    \sigma_f=\frac{0.20}{\sqrt{252}},
    \qquad
    \sigma_e=\frac{0.10}{\sqrt{252}}.
    \]

    \paragraph{Targeted failure mode.}
    Isolates omitted-risk attribution. Any apparent ``alpha'' must reflect systematic exposure being misread as skill rather than genuine abnormal performance.

    \subsection{Environment E: Volatility Clustering Null (GARCH(1,1), Zero Mean)}
    \label{app:envE}

    This environment produces volatility clustering while preserving a zero conditional mean.

    \paragraph{Data generating process.}
    \[
    r_t=\sqrt{h_t}\,z_t,
    \qquad
    z_t \stackrel{\text{i.i.d.}}{\sim}\mathcal{N}(0,1),
    \]
    \[
    h_t=\omega+\alpha r_{t-1}^2+\beta h_{t-1}.
    \]
    Conditional on $\mathcal{F}^{(r)}_{t-1}$, $h_t$ is measurable and $\mathbb{E}[z_t]=0$, hence $\mathbb{E}[r_t\mid\mathcal{F}^{(r)}_{t-1}]=0$.

    \paragraph{Default calibration.}
    Set $\alpha=0.10$, $\beta=0.85$ (so $\alpha+\beta=0.95<1$). Choose $\omega$ to match the target daily variance. Let $\sigma_{\mathrm{daily}}=\sigma^{\mathrm{ann}}/\sqrt{252}$ with $\sigma^{\mathrm{ann}}=0.20$. Then
    \[
    \omega=(1-\alpha-\beta)\,\sigma_{\mathrm{daily}}^2.
    \]

    \paragraph{Targeted failure mode.}
    Stresses risk adjustment and HAC-based inference under conditional heteroskedasticity.

    \subsection{Extended Microstructure Environment (Optional Robustness Check)}
    \label{app:envCstar}
    In addition to the canonical microstructure placebo (Environment C), we retain an optional martingale-bar stress test for robustness. Synthetic intraday price bars are generated from a driftless process and aggregated to bar returns. This extended environment is reported separately when used and does not alter the canonical five-environment falsification suite. Full specifications are available in the \texttt{QuantAudit} package documentation.

    \section{Supplementary Simulation Results and Robustness}
    \label{app:sup_sim_results}
    This appendix provides comprehensive supplementary results for all simulation experiments reported in \Cref{sec:simulation}. We report Monte Carlo uncertainty estimates, complete tables for the falsification diagnostic matrix, detailed evolution of effective multiplicity and in-sample inflation across search sizes, exhaustive threshold amplification sweeps, and a battery of robustness checks that confirm the stability and generality of our findings. All results are based on $N = 1,\!000$ Monte Carlo replications per configuration unless otherwise noted.

    \subsection{Monte Carlo Uncertainty for Scaling and Redundancy Laws}
    \label{app:monte_carlo_uncertainty}

    \subsubsection{Scaling law uncertainty}
    Table~\ref{tab:scaling_uncertainty} reports $95\%$ confidence intervals for the key statistics underlying the scaling-law experiment (\Cref{subsec:scaling_law_sim}). Estimates are precise throughout the range of $K$, with confidence-interval widths for mean $|Z_{\mathrm{IS}}|$ consistently below $0.04$ absolute units. The winner familywise false positive probability becomes essentially certain for $K \ge 200$, as indicated by confidence bounds rounding to $100\%$.

    \begin{table}
        \caption{{Monte Carlo uncertainty for the scaling-law experiment.} 95\% intervals for key simulation estimates. Winner FWP denotes the familywise false positive probability for the selected winner at the nominal threshold $|Z|>1.96$. BIF Median is computed on the subset $\{|Z_{\mathrm{WF}}^\star| \ge 0.5\}$.}
        \label{tab:scaling_uncertainty}
        \resizebox{\textwidth}{!}{%
            \centering
            \begin{tabular}[t]{lccccccccc}
                \toprule
                K & n & IS Mean & IS CI & WF Mean & WF CI & Winner FWP (\%) & Winner FWP CI & BIF Median & BIF CI\\
                \midrule
                1 & 1000 & 0.81 & {}[0.77, 0.85] & 0.77 & {}[0.73, 0.81] & 5.3 & {}[3.9, 6.7] & 0.65 & {}[0.03, 2.90]\\
                5 & 1000 & 1.58 & {}[1.54, 1.61] & 0.78 & {}[0.74, 0.82] & 22.3 & {}[19.7, 24.9] & 1.50 & {}[0.45, 3.71]\\
                10 & 1000 & 1.90 & {}[1.86, 1.93] & 0.81 & {}[0.77, 0.85] & 40.2 & {}[37.2, 43.2] & 1.80 & {}[0.62, 4.54]\\
                50 & 1000 & 2.52 & {}[2.49, 2.54] & 0.82 & {}[0.78, 0.86] & 92.3 & {}[90.6, 94.0] & 2.36 & {}[1.02, 5.06]\\
                100 & 1000 & 2.79 & {}[2.77, 2.82] & 0.82 & {}[0.78, 0.86] & 99.9 & {}[99.7, 100.1] & 2.60 & {}[1.09, 5.56]\\
                200 & 1000 & 3.00 & {}[2.98, 3.03] & 0.81 & {}[0.78, 0.85] & 100.0 & {}[100.0, 100.0] & 2.88 & {}[1.17, 5.78]\\
                500 & 1000 & 3.28 & {}[3.25, 3.30] & 0.83 & {}[0.79, 0.86] & 100.0 & {}[100.0, 100.0] & 3.18 & {}[1.26, 6.70]\\
                1000 & 1000 & 3.48 & {}[3.46, 3.50] & 0.80 & {}[0.76, 0.84] & 100.0 & {}[100.0, 100.0] & 3.33 & {}[1.38, 6.76]\\
                \bottomrule
            \end{tabular}
        }
    \end{table}

    \subsubsection{Redundancy law uncertainty}
    Table~\ref{tab:redundancy_uncertainty} provides confidence intervals for the redundancy law experiment (\Cref{subsec:redundancy_law_sim}). The effective multiplicity estimator exhibits remarkable precision, with interval widths less than $0.2$ even in the $100$-cluster case. In-sample winner statistics are estimated with similarly high precision.

    \begin{table}[htbp]
        \centering
        \caption{Monte Carlo uncertainty for the redundancy law experiment. For each dependence regime, we report the point estimate and 95\% confidence interval for effective multiplicity $\widehat{K}_{\mathrm{eff}}$, the in-sample winner magnitude $|Z_{\mathrm{IS}}^\star|$, the walk-forward winner magnitude $|Z_{\mathrm{WF}}^\star|$, and the stabilized Backtest Inflation Factor $\mathrm{BIF}^{\mathrm{stab}}_{0.5}$.}
        \label{tab:redundancy_uncertainty}
        \resizebox{\textwidth}{!}{%
            \begin{tabular}{lcccccccc}
                \toprule
                Regime & Keff & Keff CI & IS Mean & IS CI & WF Mean & WF CI & BIF Median & BIF CI \\
                \midrule
                Perfectly Correlated & 1.00 & [1.00, 1.00] & 0.77 & [0.74, 0.81] & 0.81 & [0.77, 0.85] & 0.62 & [0.02, 2.53] \\
                5 Clusters          & 5.03 & [5.01, 5.04] & 1.57 & [1.54, 1.61] & 0.84 & [0.81, 0.88] & 1.44 & [0.48, 4.02] \\
                10 Clusters         & 10.09 & [10.06, 10.11] & 1.90 & [1.86, 1.93] & 0.79 & [0.75, 0.83] & 1.79 & [0.67, 4.23] \\
                25 Clusters         & 25.46 & [25.39, 25.52] & 2.26 & [2.23, 2.29] & 0.80 & [0.76, 0.84] & 2.22 & [0.81, 4.98] \\
                50 Clusters         & 51.74 & [51.60, 51.87] & 2.54 & [2.52, 2.57] & 0.84 & [0.80, 0.88] & 2.37 & [0.99, 5.83] \\
                100 Clusters        & 106.66 & [106.39, 106.94] & 2.78 & [2.75, 2.80] & 0.82 & [0.78, 0.85] & 2.69 & [1.10, 5.59] \\
                Independent         & 500.00 & [499.99, 500.00] & 3.26 & [3.24, 3.29] & 0.76 & [0.72, 0.79] & 3.23 & [1.38, 6.35] \\
                \bottomrule
            \end{tabular}
        }
    \end{table}

    \subsubsection{\texorpdfstring{Workflow comparison uncertainty at $K=400$ (Elastic Net)}{Workflow comparison uncertainty at K=400 (Elastic Net)}}
    To formalize the dispersion of performance metrics under the global null, Table~\ref{tab:exp4a_unc_kmax} reports the 95\% Monte Carlo uncertainty intervals for the linear-control experiment under the White Noise null at the maximum nominal search size ($K=400$). The empirical bounds confirm the structural divergence between independent and correlated search topologies. For the orthogonal \texttt{FeatureMining} workflow, the effective dimensionality remains strictly at $400$, driving the in-sample rejection rate to 100\% with a median $\mathrm{BIF}^{\mathrm{stab}}_{0.5}$ of 2.91. Conversely, \texttt{Elastic Net Tuning} induces strong algorithmic shrinkage, compressing the effective multiplicity to $\widehat{K}_{\mathrm{eff}} \approx 1.73$. Its in-sample extremeness is therefore materially attenuated relative to the independent search benchmark, although its in-sample failure rate remains 35.8\%. \texttt{TrendFollowing} occupies an intermediate regime, with $\widehat{K}_{\mathrm{eff}} \approx 5.86$ and an in-sample failure rate of 64.2\%. Across all three workflows, the walk-forward failure rates remain near nominal (about 5--7\%), confirming that walk-forward validation filters the spurious in-sample alpha.

    \begin{table}[htbp]
        \centering
        \caption{Monte Carlo uncertainty for the linear-control experiment under White Noise at $K=400$ ($N=1{,}000$ simulations per workflow). For each metric, we report the mean across simulations (median for $\mathrm{BIF}^{\mathrm{stab}}_{0.5}$), together with 95\% uncertainty intervals. For continuous statistics, the intervals are the empirical 2.5\% and 97.5\% quantiles of the simulation distribution. For failure rates (in percentage), the intervals are approximate 95\% Wald intervals for the proportion. The BIF summaries are computed on the stability subset $\{|Z_{\mathrm{WF}}^\star| \ge 0.5\}$.}
        \label{tab:exp4a_unc_kmax}
        \resizebox{\textwidth}{!}{%
            \begin{tabular}{lcccccccccccccccccc}
                \toprule
                Workflow
                & $\widehat{K}_{\mathrm{eff}}$ & $\widehat{K}_{\mathrm{eff}}$ Lo & $\widehat{K}_{\mathrm{eff}}$ Hi
                & $|Z_{\mathrm{IS}}^\star|$ & IS Lo & IS Hi
                & $|Z_{\mathrm{WF}}^\star|$ & WF Lo & WF Hi
                & Fail IS (\%) & FIS Lo & FIS Hi
                & Fail WF (\%) & FWF Lo & FWF Hi
                & $\mathrm{BIF}^{\mathrm{stab}}_{0.5}$ & BIF Lo & BIF Hi \\
                \midrule
                FeatureMining & 400.00 & 400.00 & 400.00 & 3.21 & 2.62 & 4.03 & 0.84 & 0.04 & 2.41 & 100.0 & 100.00 & 100.00 & 6.6 & 5.06 & 8.14 & 2.91 & 1.23 & 6.24 \\
                Elastic Net Tuning & 1.73 & 1.39 & 1.99 & 1.82 & 0.90 & 3.28 & 0.81 & 0.04 & 2.32 & 35.8 & 32.83 & 38.77 & 5.4 & 4.00 & 6.80 & 1.67 & 0.54 & 4.28 \\
                TrendFollowing & 5.86 & 4.10 & 8.13 & 2.19 & 1.44 & 3.31 & 0.78 & 0.02 & 2.21 & 64.2 & 61.23 & 67.17 & 5.0 & 3.65 & 6.35 & 2.12 & 0.82 & 4.67 \\
                \bottomrule
            \end{tabular}%
        }
    \end{table}

    \subsubsection{Workflow comparison uncertainty}
    \label{subsubsec:workflow_uncertainty}

    Table~\ref{tab:exp4b_unc_kmax} reports $95\%$ confidence intervals for the key statistics of the capacity-driven inflation experiment under White Noise at $K=400$ (\Cref{subsec:boosting_effect_sim}), based on $N=1{,}000$ Monte Carlo replications. For continuous metrics ($\widehat{K}_{\mathrm{eff}}$, $|Z_{\mathrm{IS}}|$, $|Z_{\mathrm{WF}}|$, and $\mathrm{BIF}^{\mathrm{stab}}_{0.5}$), the intervals are empirical quantiles; for failure rates, they are asymptotic Wald intervals. Feature Mining exhibits zero uncertainty in $\widehat{K}_{\mathrm{eff}}$ by construction, as the independent strategies guarantee $\widehat{K}_{\mathrm{eff}}=K$. Hyperparameter Tuning and Trend Following show moderate sampling variability, with $\widehat{K}_{\mathrm{eff}}$ intervals spanning approximately $[2.5,3.5]$ and $[4.1,8.1]$, respectively.

    \begin{table}[htbp]
        \centering
        \caption{Monte Carlo uncertainty for workflow comparison under White Noise at $K=400$. 95\% empirical bounds for continuous metrics; asymptotic binomial bounds for failure probabilities.}
        \label{tab:exp4b_unc_kmax}
        \resizebox{\textwidth}{!}{%
            \begin{tabular}{lcccccccccccccccccc}
                \toprule
                Workflow & $\widehat{K}_{\mathrm{eff}}$ & $\widehat{K}_{\mathrm{eff}}$ Lo & $\widehat{K}_{\mathrm{eff}}$ Hi
                & $|Z_{\mathrm{IS}}^\star|$ & IS Lo & IS Hi   & $|Z_{\mathrm{WF}}^\star|$ & WF Lo & WF Hi
                & Fail IS (\%) & FIS Lo & FIS Hi  & Fail WF (\%) & FWF Lo & FWF Hi
                & $\mathrm{BIF}^{\mathrm{stab}}_{0.5}$ & BIF Lo & BIF Hi\\
                \midrule
                FeatureMining & 400.00 & 400.00 & 400.00 & 3.21 & 2.62 & 4.03 & 0.84 & 0.04 & 2.41 & 100.0 & 100.00 & 100.00 & 6.6 & 5.06 & 8.14 & 2.91 & 1.23 & 6.24\\
                \addlinespace
                HyperparameterTuning & 2.93 & 2.50 & 3.50 & 2.56 & 1.72 & 3.80 & 0.79 & 0.04 & 2.21 & 88.2 & 86.20 & 90.20 & 4.7 & 3.39 & 6.01 & 2.49 & 0.93 & 5.52\\
                \addlinespace
                TrendFollowing & 5.86 & 4.10 & 8.13 & 2.19 & 1.44 & 3.31 & 0.78 & 0.02 & 2.21 & 64.2 & 61.23 & 67.17 & 5.0 & 3.65 & 6.35 & 2.12 & 0.82 & 4.67\\
                \bottomrule
            \end{tabular}%
        }
    \end{table}

    \subsection{Complete Falsification Diagnostic Matrix}
    \label{app:sim_stats}

    \subsubsection{In-sample failure rates}
    Table~\ref{tab:falsification_failure_is} reports in-sample failure rates (percentage of simulations with $|Z_{\mathrm{IS}}| > 1.96$) for the six canonical pipelines. Two patterns stand out. First, DataMiner and Lookahead exhibit near-certain in-sample rejection across all environments. Second, Contrarian remains near the nominal level in the strict null environments but fails in-sample with certainty in the Microstructure placebo. Comparison with walk-forward failure rates (Table~\ref{tab:falsification_failure_wf}) distinguishes pure selection bias from genuine workflow invalidity.

    \begin{table}[htbp]
        \centering
        \caption{{In-sample failure rates.} Percent of simulations with $|Z_{\mathrm{IS}}| > 1.96$ (diagnostic for selection bias).}
        \label{tab:falsification_failure_is}
        \begin{tabular}{lcccc}
            \toprule
            Pipeline & WhiteNoise & RegimeSwitch & Microstructure & FactorNull \\
            \midrule
            RandomBaseline & 6.0 & 5.1 & 4.6 & 4.9 \\
            DataMiner      & 99.3 & 99.2 & 99.8 & 99.5 \\
            Contrarian     & 5.6 & 5.1 & 100.0 & 4.9 \\
            Lookahead      & 100.0 & 100.0 & 100.0 & 100.0 \\
            RegimeDetector & 4.9 & 4.1 & 2.0 & 4.3 \\
            FactorMimic    & 4.6 & 3.4 & 2.8 & 5.1 \\
            \bottomrule
        \end{tabular}
    \end{table}

    \begin{table}[htbp]
        \centering
        \caption{{Walk-forward failure rates.} Percent of simulations with $|Z_{\mathrm{WF}}| > 1.96$ (expected $\approx 5\%$ under the null).}
        \label{tab:falsification_failure_wf}
        \begin{tabular}{lcccc}
            \toprule
            Pipeline & WhiteNoise & RegimeSwitch & Microstructure & FactorNull \\
            \midrule
            RandomBaseline & 4.6 & 4.8 & 6.9 & 5.3 \\
            DataMiner      & 5.5 & 5.3 & 6.3 & 6.2 \\
            Contrarian     & 4.1 & 5.5 & 100.0 & 6.5 \\
            Lookahead      & 100.0 & 100.0 & 100.0 & 100.0 \\
            RegimeDetector & 5.1 & 5.0 & 2.5 & 3.7 \\
            FactorMimic    & 4.3 & 4.5 & 2.1 & 5.7 \\
            \bottomrule
        \end{tabular}
    \end{table}

    \subsubsection{Pipeline classification}
    Table~\ref{tab:falsification_pipeline_classification} provides a rule-based diagnosis of each pipeline based on its walk-forward failure pattern. The classification now identifies four distinct outcomes: severe methodological error in Contrarian, selection bias in DataMiner, universal leakage in Lookahead, and robust operation in RandomBaseline, RegimeDetector, and FactorMimic.

    \begin{table}[htbp]
        \centering
        \caption{{Pipeline classification.} Rule-based diagnosis using environment-specific 95th percentiles from RandomBaseline as falsification thresholds.}
        \label{tab:falsification_pipeline_classification}
        \small
        \setlength{\tabcolsep}{4pt}
        \begin{tabular}{l c c c l}
            \toprule
            Pipeline & WF Fail (Mean) & IS Fail (Mean) & Max |Z| (WF) & Diagnosis \\
            \midrule
            Contrarian     & 29.23 & 28.90 & 10.72 & Severe Methodological Error \\
            DataMiner      & 5.60 & 99.45 & 0.82 & Selection Bias (IS-only) \\
            FactorMimic    & 3.93 & 3.98 & 0.81 & Robust (Passes WF) \\
            Lookahead      & 100.00 & 100.00 & 42.43 & Universal Failure (WF) \\
            RandomBaseline & 5.00 & 5.15 & 0.82 & Robust (Passes WF) \\
            RegimeDetector & 3.85 & 3.83 & 0.80 & Robust (Passes WF) \\
            \bottomrule
        \end{tabular}
    \end{table}

    \subsubsection{Per-cell confidence intervals}
    Table~\ref{tab:falsification_uncertainty_by_cell} reports $95\%$ confidence intervals for walk-forward statistics and failure rates for every pipeline--environment pair. The intervals confirm the classification logic. DataMiner exhibits near-null walk-forward failure rates across environments despite extreme in-sample rejection, Contrarian displays a decisive breakdown in the Microstructure placebo, and Lookahead remains universally invalid with failure-rate intervals concentrated at $100\%$. Robust pipelines maintain walk-forward failure-rate intervals that remain close to the nominal level.

    \begin{table}[htbp]
        \centering
        \caption{{Monte Carlo uncertainty for falsification diagnostics.} 95\% confidence intervals for walk-forward and in-sample statistics, walk-forward failure rates, and the stabilized Backtest Inflation Factor ($\mathrm{BIF}^{\mathrm{stab}}_{0.5}$).}
        \label{tab:falsification_uncertainty_by_cell}
        \resizebox{\textwidth}{!}{%
            \begin{tabular}{llccccccccccccc}
                \toprule
                Pipeline & Environment &
                \multicolumn{3}{c}{Walk-forward $|Z|$} & \multicolumn{3}{c}{WF Fail \%} &
                \multicolumn{3}{c}{In-sample $|Z|$} & \multicolumn{3}{c}{$\mathrm{BIF}^{\mathrm{stab}}_{0.5}$} \\
                \cmidrule(lr){3-5} \cmidrule(lr){6-8} \cmidrule(lr){9-11} \cmidrule(lr){12-14}
                & & Mean & Low & High & Rate & Low & High & Mean & Low & High & Median & Low & High \\
                \midrule
                Contrarian     & FactorNull     & 0.85 & 0.81 & 0.88 & 6.5 & 5.10 & 8.20 & 0.80 & 0.76 & 0.84 & 0.60 & 0.03 & 2.58 \\
                Contrarian     & Microstructure & 10.72 & 10.67 & 10.77 & 100.0 & 99.60 & 100.00 & 13.09 & 13.04 & 13.14 & 1.22 & 1.01 & 1.50 \\
                Contrarian     & RegimeSwitch   & 0.81 & 0.77 & 0.85 & 5.5 & 4.20 & 7.10 & 0.79 & 0.76 & 0.83 & 0.60 & 0.03 & 2.82 \\
                Contrarian     & WhiteNoise     & 0.80 & 0.76 & 0.83 & 4.1 & 3.00 & 5.50 & 0.82 & 0.78 & 0.86 & 0.65 & 0.02 & 2.95 \\
                \addlinespace
                DataMiner      & FactorNull     & 0.81 & 0.77 & 0.85 & 6.2 & 4.90 & 7.90 & 2.77 & 2.75 & 2.79 & 2.67 & 1.02 & 5.55 \\
                DataMiner      & Microstructure & 0.81 & 0.77 & 0.85 & 6.3 & 5.00 & 8.00 & 2.79 & 2.76 & 2.81 & 2.76 & 1.08 & 5.55 \\
                DataMiner      & RegimeSwitch   & 0.82 & 0.78 & 0.85 & 5.3 & 4.10 & 6.90 & 2.76 & 2.73 & 2.79 & 2.77 & 1.12 & 5.57 \\
                DataMiner      & WhiteNoise     & 0.80 & 0.77 & 0.84 & 5.5 & 4.20 & 7.10 & 2.77 & 2.74 & 2.79 & 2.69 & 1.08 & 5.85 \\
                \addlinespace
                FactorMimic    & FactorNull     & 0.79 & 0.76 & 0.83 & 5.7 & 4.40 & 7.30 & 0.82 & 0.78 & 0.86 & 0.62 & 0.03 & 2.92 \\
                FactorMimic    & Microstructure & 0.69 & 0.66 & 0.73 & 2.1 & 1.40 & 3.20 & 0.71 & 0.68 & 0.74 & 0.63 & 0.03 & 2.64 \\
                FactorMimic    & RegimeSwitch   & 0.81 & 0.77 & 0.85 & 4.5 & 3.40 & 6.00 & 0.74 & 0.71 & 0.78 & 0.54 & 0.04 & 2.63 \\
                FactorMimic    & WhiteNoise     & 0.80 & 0.77 & 0.84 & 4.3 & 3.20 & 5.70 & 0.79 & 0.76 & 0.83 & 0.64 & 0.03 & 2.74 \\
                \addlinespace
                Lookahead      & FactorNull     & 42.43 & 42.30 & 42.56 & 100.0 & 99.60 & 100.00 & 51.65 & 51.52 & 51.78 & 1.22 & 1.08 & 1.37 \\
                Lookahead      & Microstructure & 37.99 & 37.88 & 38.11 & 100.0 & 99.60 & 100.00 & 46.23 & 46.10 & 46.35 & 1.21 & 1.07 & 1.40 \\
                Lookahead      & RegimeSwitch   & 21.32 & 21.20 & 21.44 & 100.0 & 99.60 & 100.00 & 24.71 & 24.59 & 24.82 & 1.16 & 0.92 & 1.43 \\
                Lookahead      & WhiteNoise     & 42.36 & 42.23 & 42.49 & 100.0 & 99.60 & 100.00 & 51.72 & 51.59 & 51.86 & 1.22 & 1.07 & 1.38 \\
                \addlinespace
                RandomBaseline & FactorNull     & 0.79 & 0.75 & 0.83 & 5.3 & 4.10 & 6.90 & 0.81 & 0.77 & 0.85 & 0.66 & 0.04 & 3.00 \\
                RandomBaseline & Microstructure & 0.82 & 0.78 & 0.86 & 6.9 & 5.50 & 8.60 & 0.80 & 0.76 & 0.83 & 0.62 & 0.04 & 2.79 \\
                RandomBaseline & RegimeSwitch   & 0.79 & 0.75 & 0.83 & 4.8 & 3.60 & 6.30 & 0.82 & 0.78 & 0.85 & 0.65 & 0.04 & 2.99 \\
                RandomBaseline & WhiteNoise     & 0.79 & 0.75 & 0.82 & 4.6 & 3.50 & 6.10 & 0.83 & 0.79 & 0.87 & 0.65 & 0.04 & 3.17 \\
                \addlinespace
                RegimeDetector & FactorNull     & 0.79 & 0.76 & 0.83 & 3.7 & 2.70 & 5.10 & 0.77 & 0.73 & 0.80 & 0.65 & 0.02 & 2.67 \\
                RegimeDetector & Microstructure & 0.65 & 0.62 & 0.68 & 2.5 & 1.70 & 3.70 & 0.67 & 0.63 & 0.70 & 0.60 & 0.03 & 2.48 \\
                RegimeDetector & RegimeSwitch   & 0.80 & 0.76 & 0.83 & 5.0 & 3.80 & 6.50 & 0.81 & 0.77 & 0.84 & 0.60 & 0.03 & 2.78 \\
                RegimeDetector & WhiteNoise     & 0.79 & 0.75 & 0.83 & 5.1 & 3.90 & 6.60 & 0.80 & 0.76 & 0.84 & 0.65 & 0.04 & 2.73 \\
                \bottomrule
            \end{tabular}%
        }
    \end{table}

    \subsection{Detailed Results: Capacity-Driven Inflation}
    \label{app:boosting_paradox_details}

    \subsubsection{Evolution of effective multiplicity across \texorpdfstring{$K$}{K}}
    Table~\ref{tab:exp4b_white_keff} reports the evolution of $\widehat{K}_{\mathrm{eff}}$ with nominal search size $K$ for the three workflows under White Noise. Feature Mining maintains perfect tracking ($\widehat{K}_{\mathrm{eff}}=K$) by construction. Under Hyperparameter Tuning, $\widehat{K}_{\mathrm{eff}}$ remains in the low single digits and settles around $\approx 3$ by $K=400$; the mild non-monotonicity across intermediate $K$ reflects Monte Carlo variability rather than an expansion of independent trials. Trend Following exhibits intermediate but non-negligible effective search dimensionality, rising to $\widehat{K}_{\mathrm{eff}}\approx 5.9$ by $K=400$.

    \begin{table}[htbp]
        \caption{Effective multiplicity under White Noise (Boosting). Mean $\widehat{K}_{\mathrm{eff}}$ across workflows.}
        \label{tab:exp4b_white_keff}
        \centering
        \resizebox{\textwidth}{!}{%
            \begin{tabular}[t]{lcccccccc}
                \toprule
                Workflow & $K=1$ & $K=5$ & $K=10$ & $K=25$ & $K=50$ & $K=100$ & $K=200$ & $K=400$ \\
                \midrule
                Saturation (Correlated) & 1 & 1.69 & 1.96 & 2.20 & 4.19 & 3.53 & 3.14 & 2.93\\
                Inflation (Independent) & 1 & 5.00 & 10.00 & 25.00 & 50.00 & 100.00 & 200.00 & 400.00\\
                Robustness Check & 1 & 1.81 & 2.52 & 3.85 & 4.76 & 5.38 & 5.65 & 5.86\\
                \bottomrule
        \end{tabular}}
    \end{table}

    \subsubsection{Saturation ratio}
    Table~\ref{tab:exp4b_white_satratio} reports the saturation ratio $\widehat{K}_{\mathrm{eff}} / K$. For Hyperparameter Tuning, the ratio declines from $0.34$ at $K=5$ to $0.01$ at $K=400$, indicating extreme redundancy. Trend Following also displays declining saturation ratios, though from a less compressed effective search space. Feature Mining maintains a ratio of $1.00$ by construction.

    \begin{table}[htbp]
        \caption{Saturation ratio under White Noise (Boosting). Mean ratio $\widehat{K}_{\mathrm{eff}}/K$ by workflow.}
        \centering
        \label{tab:exp4b_white_satratio}
        \resizebox{\textwidth}{!}{%
            \begin{tabular}[t]{lcccccccc}
                \toprule
                Workflow & $K=1$ & $K=5$ & $K=10$ & $K=25$ & $K=50$ & $K=100$ & $K=200$ & $K=400$ \\
                \midrule
                Saturation (Correlated) & 1 & 0.34 & 0.20 & 0.09 & 0.08 & 0.04 & 0.02 & 0.01\\
                Inflation (Independent) & 1 & 1.00 & 1.00 & 1.00 & 1.00 & 1.00 & 1.00 & 1.00\\
                Robustness Check & 1 & 0.36 & 0.25 & 0.15 & 0.10 & 0.05 & 0.03 & 0.01\\
                \bottomrule
        \end{tabular}}
    \end{table}

    \subsubsection{In-sample inflation across \texorpdfstring{$K$}{K}}
    Table~\ref{tab:exp4b_white_zis} documents the growth of $\mathbb{E}[|Z_{\mathrm{IS}}^\star|]$ with $K$. Feature Mining follows the $\sqrt{\log K}$ scaling predicted by theory, reaching $3.21$ at $K=400$. Hyperparameter Tuning shows slower but still substantial growth, from $1.35$ at $K=5$ to $2.56$ at $K=400$. Trend Following also exhibits substantial inflation, rising from $1.07$ at $K=5$ to $2.19$ at $K=400$.

    \begin{table}[htbp]
        \centering
        \caption{In-sample inflation under White Noise (Boosting). Mean absolute in-sample statistic $\mathbb{E}[|Z^{\star}_{\mathrm{IS}}|]$ by workflow and $K$.}
        \label{tab:exp4b_white_zis}
        \resizebox{\textwidth}{!}{%
            \begin{tabular}[t]{lcccccccc}
                \toprule
                Workflow & $K=1$ & $K=5$ & $K=10$ & $K=25$ & $K=50$ & $K=100$ & $K=200$ & $K=400$ \\
                \midrule
                Saturation (Correlated) & 0.83 & 1.35 & 1.55 & 1.81 & 2.16 & 2.30 & 2.43 & 2.56 \\
                Inflation (Independent) & 0.85 & 1.56 & 1.88 & 2.28 & 2.51 & 2.77 & 3.02 & 3.21 \\
                Robustness Check & 0.77 & 1.07 & 1.26 & 1.59 & 1.78 & 1.97 & 2.12 & 2.19 \\
                \bottomrule
        \end{tabular}}
    \end{table}

    \subsection{Threshold Amplification: Full Sweeps}
    \label{app:threshold_amp_sweeps}

    \subsubsection{Amplification across threshold rules and correlation levels}
    Table~\ref{tab:threshold_amp_full} provides a comprehensive view of threshold amplification, including $5\%$--$95\%$ envelopes for the in-sample statistic. Walk-forward performance remains bounded near the null expectation regardless of thresholding and is reported here only through the mean $|Z_{\mathrm{WF}}|$, while in-sample extremeness increases systematically with threshold aggressiveness.

    \begin{table}[htbp]
        \centering
        \caption{Threshold amplification at $K=1{,}000$ under high correlation ($\rho=0.9$). $\widehat{K}_{\mathrm{eff}}$ (Pred) is computed on raw continuous scores, and $\widehat{K}_{\mathrm{eff}}$ (Signal) is computed on realized thresholded trading signals. Amp.\ Factor is $\widehat{K}_{\mathrm{eff}}(\mathrm{Signal})/\widehat{K}_{\mathrm{eff}}(\mathrm{Pred})$. Results are based on $N=1{,}000$ independent simulations per cell.}

        \label{tab:threshold_amp_full}
        \resizebox{\textwidth}{!}{%
            \begin{tabular}{lccccc}
                \toprule
                Threshold Rule & $\widehat{K}_{\mathrm{eff}}$ (Pred) & $\widehat{K}_{\mathrm{eff}}$ (Signal) & Amp. & $|Z_{\mathrm{IS}}|$ Envelope & $|Z_{\mathrm{WF}}|$ Mean \\
                \midrule
                None & 1.24 & 1.24 & 1.00 & 1.82 (1.07--2.95) & 0.80 \\
                Fixed (thr=0.0) & 1.24 & 1.97 & 1.59 & 2.50 (1.75--3.56) & 0.79 \\
                Fixed (thr=0.5) & 1.24 & 1.60 & 1.28 & 2.23 (1.51--3.39) & 0.79 \\
                Fixed (thr=1.0) & 1.24 & 1.91 & 1.54 & 2.42 (1.69--3.52) & 0.79 \\
                Fixed (thr=1.5) & 1.24 & 2.41 & 1.93 & 2.59 (1.88--3.57) & 0.80 \\
                Fixed (thr=2.0) & 1.24 & 3.34 & 2.68 & 2.63 (1.98--3.45) & 0.86 \\
                Adaptive (q=0.50) & 1.24 & 1.69 & 1.36 & 2.27 (1.55--3.30) & 0.80 \\
                Adaptive (q=0.70) & 1.24 & 1.94 & 1.56 & 2.42 (1.71--3.58) & 0.79 \\
                Adaptive (q=0.80) & 1.24 & 2.16 & 1.74 & 2.56 (1.82--3.64) & 0.77 \\
                Adaptive (q=0.90) & 1.24 & 2.60 & 2.09 & 2.62 (1.92--3.64) & 0.80 \\
                Adaptive (q=0.95) & 1.24 & 3.20 & 2.57 & 2.66 (1.99--3.48) & 0.82 \\
                \bottomrule
        \end{tabular}}
    \end{table}

    \subsubsection{Sweep over correlation and \texorpdfstring{$K$}{K}}
    Tables~\ref{tab:threshold_amp_zis_sweep} and \ref{tab:threshold_amp_zwf_sweep} report in-sample selection bias and walk-forward performance across a grid of $\rho \in \{0, 0.9\}$ and $K$ ranging from $1$ to $1000$. The incremental effect of thresholding on in-sample extremeness is strongest at high correlation: when $\rho=0.9$, both fixed and adaptive thresholding shift $\mathbb{E}[\max_k |Z_{\mathrm{IS},k}|]$ materially above the no-threshold baseline, whereas under independence ($\rho=0$) the three regimes remain close across the full $K$ sweep. By contrast, walk-forward magnitudes remain stable near the null benchmark across both $\rho$ and $K$, indicating that the threshold-induced increase is an in-sample selection effect rather than a corresponding out-of-sample improvement.

    \begin{table}[htbp]
        \centering
        \caption{In-sample selection bias sweep across nominal $K$. Entries report $\mathbb{E}[\max_k |Z_{\mathrm{IS},k}|]$ under the null.}
        \label{tab:threshold_amp_zis_sweep}
        \resizebox{\textwidth}{!}{%
            \begin{tabular}{llcccccccccc}
                \toprule
                $\rho$ & Regime & $K=1$ & $K=2$ & $K=5$ & $K=10$ & $K=20$ & $K=50$ & $K=100$ & $K=200$ & $K=500$ & $K=1000$ \\
                \midrule
                0.0 & None          & 0.79 & 1.18 & 1.57 & 1.91 & 2.19 & 2.55 & 2.79 & 3.01 & 3.29 & 3.51 \\
                0.0 & Fixed (1)     & 0.81 & 1.17 & 1.60 & 1.92 & 2.21 & 2.53 & 2.79 & 3.01 & 3.30 & 3.51 \\
                0.0 & Adaptive (0.9)& 0.83 & 1.17 & 1.57 & 1.88 & 2.18 & 2.51 & 2.73 & 2.96 & 3.23 & 3.40 \\
                \midrule
                0.9 & None          & 0.82 & 0.96 & 1.13 & 1.25 & 1.39 & 1.48 & 1.58 & 1.67 & 1.75 & 1.82 \\
                0.9 & Fixed (1)     & 0.79 & 1.03 & 1.32 & 1.51 & 1.72 & 1.88 & 2.01 & 2.16 & 2.31 & 2.42 \\
                0.9 & Adaptive (0.9)& 0.81 & 1.07 & 1.40 & 1.62 & 1.81 & 2.04 & 2.18 & 2.34 & 2.52 & 2.62 \\
                \bottomrule
        \end{tabular}}
    \end{table}

    \begin{table}[htbp]
        \centering
        \caption{Walk-forward performance of the configuration selected in-sample. Entries report $\mathbb{E}[|Z_{\mathrm{WF}}(k^\star)|]$.}
        \label{tab:threshold_amp_zwf_sweep}
        \resizebox{\textwidth}{!}{%
            \begin{tabular}{llcccccccccc}
                \toprule
                $\rho$ & Regime & $K=1$ & $K=2$ & $K=5$ & $K=10$ & $K=20$ & $K=50$ & $K=100$ & $K=200$ & $K=500$ & $K=1000$ \\
                \midrule
                0.0 & None          & 0.79 & 0.82 & 0.83 & 0.84 & 0.83 & 0.80 & 0.80 & 0.82 & 0.77 & 0.80 \\
                0.0 & Fixed (1)     & 0.80 & 0.80 & 0.84 & 0.82 & 0.80 & 0.82 & 0.85 & 0.82 & 0.81 & 0.80 \\
                0.0 & Adaptive (0.9)& 0.79 & 0.76 & 0.80 & 0.83 & 0.80 & 0.83 & 0.82 & 0.82 & 0.81 & 0.83 \\
                \midrule
                0.9 & None          & 0.78 & 0.80 & 0.82 & 0.82 & 0.82 & 0.82 & 0.78 & 0.77 & 0.84 & 0.80 \\
                0.9 & Fixed (1)     & 0.82 & 0.79 & 0.84 & 0.76 & 0.82 & 0.82 & 0.80 & 0.80 & 0.80 & 0.79 \\
                0.9 & Adaptive (0.9)& 0.82 & 0.82 & 0.81 & 0.83 & 0.79 & 0.80 & 0.82 & 0.81 & 0.80 & 0.80 \\
                \bottomrule
        \end{tabular}}
    \end{table}

    \subsection{Effective Multiplicity: Monte Carlo Validation of \texorpdfstring{$\widehat{K}_{\mathrm{eff}}$}{K-eff}}
    \label{app:keff_validation}
    The effective multiplicity $K_{\mathrm{eff}}$ defined in \Cref{eq:keff} is a key diagnostic for search‑space dimensionality. In practice, it must be estimated from a finite sample of strategy returns. When the number of strategies $K$ approaches or exceeds the sample size $T$, the sample correlation matrix becomes ill‑conditioned, rendering standard estimators unstable. We therefore apply the shrinkage estimator of \citet{Schaefer2005} as implemented in \texttt{corpcor}. This appendix validates the estimator's finite‑sample performance.

    \paragraph{Design.}
    We conduct a Monte Carlo experiment with $M=1,000$ replications across three regimes:
    \begin{enumerate}
        \renewcommand{\labelenumi}{(\roman{enumi})}
        \item \textbf{Independent:} $K$ uncorrelated strategies ($\Sigma = I$), where the true $K_{\mathrm{eff}} = K$.
        \item \textbf{Factor structure:} $K=100$ strategies driven by $m=3$ common factors plus idiosyncratic noise. The true $K_{\mathrm{eff}}$ is derived from the population covariance matrix implied by fixed factor loadings.
        \item \textbf{High‑dimensional:} Regimes where $K \ge T$, specifically $K=500$ with $T=250$ and $T=500$ under independence.
    \end{enumerate}

    For each replication, we compute $\widehat{K}_{\mathrm{eff}}$ using the shrinkage‑corrected correlation matrix and, as a baseline, the raw sample correlation matrix. We report bias, root mean squared error (RMSE), and the average shrinkage intensity $\lambda$.

    \paragraph{Results.}
    Table~\ref{tab:keff_simulation} reports the performance of the shrinkage estimator. For independent strategies, the estimator is virtually unbiased and exhibits negligible RMSE ($\leq 0.02$). Under the factor model, a small positive bias appears when $T$ is small ($0.29$ at $T=250$), but it decays rapidly as $T$ increases. In the high-dimensional regime ($K > T$), the estimator remains stable with a bias of $-0.03$ and RMSE of $0.05$.

    Table~\ref{tab:keff_comparison} contrasts the shrinkage estimator with the naive sample estimator. The sample estimator is unusable when $K > T$ (RMSE $>300$), and even when $T \gg K$ it exhibits substantial downward bias (e.g., $-9.0$ at $T=1,000$). Shrinkage uniformly dominates the sample estimator across all metrics and scenarios, ensuring that $K_{\mathrm{eff}}$ remains a conservative and robust diagnostic even in high-dimensional search spaces.

    \begin{table}[htbp]
        \centering
        \caption{{Performance of $\widehat{K}_{\mathrm{eff}}$ under shrinkage correlation estimation (Monte Carlo, $M=1000$).}}
        \label{tab:keff_simulation}
        \begin{tabular}{lccccc}
            \toprule
            Scenario & $K$ & $T$ & Bias & RMSE & $\lambda$ \\
            \midrule
            Independent & 100 & 250 & $-0.01$ & 0.02 & 0.99 \\
            Independent & 100 & 500 & 0.00 & 0.01 & 0.99 \\
            Independent & 100 & 1000 & 0.00 & 0.01 & 0.99 \\
            Factor ($m=3$) & 100 & 250 & 0.29 & 0.36 & 0.03 \\
            Factor ($m=3$) & 100 & 500 & 0.14 & 0.20 & 0.02 \\
            Factor ($m=3$) & 100 & 1000 & 0.07 & 0.12 & 0.01 \\
            High-dim ($K>T$) & 500 & 250 & $-0.03$ & 0.05 & 1.00 \\
            High-dim ($K=T$) & 500 & 500 & $-0.01$ & 0.02 & 1.00 \\
            \bottomrule
        \end{tabular}
    \end{table}

    \begin{table}[htbp]
        \centering
        \caption{{Diagnostic Validation: Shrinkage vs. Sample Estimator ($M=1000$).}}
        \label{tab:keff_comparison}
        \begin{tabular}{lcccccc}
            \toprule
            Scenario & $K$ & $T$ & Bias\textsubscript{Shrink} & RMSE\textsubscript{Shrink} & Bias\textsubscript{Sample} & RMSE\textsubscript{Sample} \\
            \midrule
            Independent & 100 & 250 & $-0.01$ & 0.02 & $-28.45$ & 28.46 \\
            Independent & 100 & 500 & 0.00 & 0.01 & $-16.55$ & 16.55 \\
            Independent & 100 & 1000 & 0.00 & 0.01 & $-9.02$ & 9.02 \\
            Factor ($m=3$) & 100 & 250 & 0.29 & 0.36 & $-0.10$ & 0.22 \\
            Factor ($m=3$) & 100 & 500 & 0.14 & 0.20 & $-0.05$ & 0.15 \\
            Factor ($m=3$) & 100 & 1000 & 0.07 & 0.12 & $-0.03$ & 0.10 \\
            High-dim ($K>T$) & 500 & 250 & $-0.03$ & 0.05 & $-333.57$ & 333.57 \\
            High-dim ($K=T$) & 500 & 500 & $-0.01$ & 0.02 & $-250.02$ & 250.02 \\
            \bottomrule
        \end{tabular}
    \end{table}

    \subsection{Robustness Checks}
    \label{app:robustness_checks}

    \subsubsection{Parameter sweeps across null families}
    Table~\ref{tab:robust_param_sweeps_with_workflow} reports results for the null-workflow experiments across five strict null families and two microstructure placebos, each with two parameterizations. Across the strict null families, the qualitative pattern remains stable: correlated tuning (\texttt{HyperparameterTuning}) saturates at low effective multiplicity, whereas independent mining (\texttt{FeatureMining}) continues to inflate in-sample evidence while leaving walk-forward performance near the null benchmark. In the MA(1) microstructure placebos, however, \texttt{HyperparameterTuning} is decisively falsified, with walk-forward failure rates of 100\% and mean walk-forward extremeness exceeding the corresponding in-sample extremeness.

    {\small
        \begin{longtable}{llrrrrrrr}
            \caption{Robustness: parameter sweeps across structural nulls and microstructure placebos. Comparison of saturation (\texttt{HyperparameterTuning}) versus inflation (\texttt{FeatureMining}) across search budgets $K \in \{25,100,200\}$. Based on $N = 1{,}000$ replications per cell.}
            \label{tab:robust_param_sweeps_with_workflow}\\

            \toprule
            \multicolumn{9}{c}{{Panel A: HyperparameterTuning (Correlated Tuning)}} \\
            \cmidrule(lr){1-9}
            Environment & Workflow & $K$ & $\widehat{K}_{\mathrm{eff}}$ & $|Z_{\mathrm{IS}}|$ & $|Z_{\mathrm{WF}}|$ & $\Delta Z$ & Fail \% & BIF \\
            \midrule
            \endfirsthead

            \multicolumn{9}{c}{\tablename\ \thetable{} (continued)} \\
            \toprule
            Environment & Workflow & $K$ & $\widehat{K}_{\mathrm{eff}}$ & $|Z_{\mathrm{IS}}|$ & $|Z_{\mathrm{WF}}|$ & $\Delta Z$ & Fail \% & BIF \\
            \midrule
            \endhead

            \midrule
            \multicolumn{9}{r}{Continued on next page} \\
            \endfoot

            \bottomrule
            \endlastfoot

            GARCH(1,1) Persistent    & HyperparameterTuning & 25  & 1.77 & 1.59 & 0.82 & 0.77  & 5.7   & 1.50 \\
            GARCH(1,1) Persistent    & HyperparameterTuning & 100 & 1.74 & 1.68 & 0.81 & 0.87  & 5.5   & 1.56 \\
            GARCH(1,1) Persistent    & HyperparameterTuning & 200 & 1.74 & 1.71 & 0.82 & 0.89  & 5.8   & 1.60 \\
            GARCH(1,1) Volatile      & HyperparameterTuning & 25  & 1.75 & 1.58 & 0.81 & 0.77  & 4.6   & 1.46 \\
            GARCH(1,1) Volatile      & HyperparameterTuning & 100 & 1.72 & 1.67 & 0.82 & 0.84  & 4.4   & 1.56 \\
            GARCH(1,1) Volatile      & HyperparameterTuning & 200 & 1.72 & 1.69 & 0.82 & 0.87  & 4.5   & 1.59 \\
            MA(1) $\theta = -0.30$   & HyperparameterTuning & 25  & 1.80 & 5.23 & 6.97 & -1.74 & 100.0 & 0.75 \\
            MA(1) $\theta = -0.30$   & HyperparameterTuning & 100 & 1.77 & 5.32 & 6.96 & -1.64 & 100.0 & 0.77 \\
            MA(1) $\theta = -0.30$   & HyperparameterTuning & 200 & 1.76 & 5.34 & 6.95 & -1.61 & 100.0 & 0.77 \\
            MA(1) $\theta = +0.30$   & HyperparameterTuning & 25  & 1.80 & 5.13 & 6.93 & -1.80 & 100.0 & 0.73 \\
            MA(1) $\theta = +0.30$   & HyperparameterTuning & 100 & 1.77 & 5.21 & 6.94 & -1.73 & 100.0 & 0.74 \\
            MA(1) $\theta = +0.30$   & HyperparameterTuning & 200 & 1.76 & 5.24 & 6.93 & -1.70 & 100.0 & 0.75 \\
            Regime Stable            & HyperparameterTuning & 25  & 1.63 & 1.53 & 0.81 & 0.72  & 4.7   & 1.39 \\
            Regime Stable            & HyperparameterTuning & 100 & 1.60 & 1.61 & 0.81 & 0.80  & 4.5   & 1.46 \\
            Regime Stable            & HyperparameterTuning & 200 & 1.59 & 1.63 & 0.81 & 0.83  & 4.3   & 1.48 \\
            Regime Unstable          & HyperparameterTuning & 25  & 1.80 & 1.58 & 0.81 & 0.77  & 5.3   & 1.43 \\
            Regime Unstable          & HyperparameterTuning & 100 & 1.77 & 1.68 & 0.80 & 0.88  & 5.3   & 1.53 \\
            Regime Unstable          & HyperparameterTuning & 200 & 1.77 & 1.70 & 0.80 & 0.90  & 5.6   & 1.56 \\
            Student $t$, df = 5      & HyperparameterTuning & 25  & 1.89 & 1.60 & 0.78 & 0.82  & 3.8   & 1.50 \\
            Student $t$, df = 5      & HyperparameterTuning & 100 & 1.86 & 1.69 & 0.79 & 0.90  & 3.6   & 1.61 \\
            Student $t$, df = 5      & HyperparameterTuning & 200 & 1.86 & 1.71 & 0.79 & 0.92  & 4.2   & 1.61 \\
            Student $t$, df = 10     & HyperparameterTuning & 25  & 1.84 & 1.60 & 0.80 & 0.80  & 4.9   & 1.44 \\
            Student $t$, df = 10     & HyperparameterTuning & 100 & 1.81 & 1.70 & 0.80 & 0.90  & 4.9   & 1.55 \\
            Student $t$, df = 10     & HyperparameterTuning & 200 & 1.80 & 1.73 & 0.80 & 0.92  & 5.0   & 1.55 \\
            White Noise High Vol     & HyperparameterTuning & 25  & 1.82 & 1.56 & 0.79 & 0.77  & 4.8   & 1.47 \\
            White Noise High Vol     & HyperparameterTuning & 100 & 1.78 & 1.66 & 0.78 & 0.88  & 4.5   & 1.54 \\
            White Noise High Vol     & HyperparameterTuning & 200 & 1.78 & 1.69 & 0.78 & 0.91  & 4.6   & 1.63 \\
            White Noise Low Vol      & HyperparameterTuning & 25  & 1.81 & 1.61 & 0.79 & 0.82  & 4.8   & 1.46 \\
            White Noise Low Vol      & HyperparameterTuning & 100 & 1.78 & 1.70 & 0.79 & 0.92  & 4.5   & 1.58 \\
            White Noise Low Vol      & HyperparameterTuning & 200 & 1.78 & 1.74 & 0.79 & 0.95  & 4.3   & 1.63 \\

            \midrule
            \multicolumn{9}{c}{{Panel B: FeatureMining (Independent Search)}} \\
            \cmidrule(lr){1-9}
            Environment & Workflow & $K$ & $\widehat{K}_{\mathrm{eff}}$ & $|Z_{\mathrm{IS}}|$ & $|Z_{\mathrm{WF}}|$ & $\Delta Z$ & Fail \% & BIF \\
            \midrule

            GARCH(1,1) Persistent    & FeatureMining & 25  & 25.00  & 2.29 & 0.78 & 1.51 & 4.6 & 2.21 \\
            GARCH(1,1) Persistent    & FeatureMining & 100 & 100.00 & 2.77 & 0.78 & 1.99 & 5.5 & 2.74 \\
            GARCH(1,1) Persistent    & FeatureMining & 200 & 200.00 & 3.00 & 0.81 & 2.19 & 5.7 & 2.87 \\
            GARCH(1,1) Volatile      & FeatureMining & 25  & 25.00  & 2.28 & 0.83 & 1.46 & 5.7 & 2.11 \\
            GARCH(1,1) Volatile      & FeatureMining & 100 & 100.00 & 2.78 & 0.81 & 1.97 & 5.5 & 2.67 \\
            GARCH(1,1) Volatile      & FeatureMining & 200 & 200.00 & 2.99 & 0.80 & 2.19 & 4.8 & 2.87 \\
            MA(1) $\theta = -0.30$   & FeatureMining & 25  & 25.00  & 2.25 & 0.82 & 1.44 & 5.4 & 2.07 \\
            MA(1) $\theta = -0.30$   & FeatureMining & 100 & 100.00 & 2.79 & 0.79 & 2.00 & 4.3 & 2.66 \\
            MA(1) $\theta = -0.30$   & FeatureMining & 200 & 200.00 & 3.02 & 0.79 & 2.23 & 5.4 & 3.04 \\
            MA(1) $\theta = +0.30$   & FeatureMining & 25  & 25.00  & 2.26 & 0.77 & 1.48 & 4.4 & 2.17 \\
            MA(1) $\theta = +0.30$   & FeatureMining & 100 & 100.00 & 2.77 & 0.84 & 1.94 & 4.7 & 2.68 \\
            MA(1) $\theta = +0.30$   & FeatureMining & 200 & 200.00 & 3.01 & 0.84 & 2.16 & 6.2 & 2.86 \\
            Regime Stable            & FeatureMining & 25  & 25.00  & 2.28 & 0.83 & 1.45 & 5.4 & 2.09 \\
            Regime Stable            & FeatureMining & 100 & 100.00 & 2.79 & 0.80 & 1.99 & 5.0 & 2.71 \\
            Regime Stable            & FeatureMining & 200 & 200.00 & 2.97 & 0.76 & 2.21 & 4.1 & 2.90 \\
            Regime Unstable          & FeatureMining & 25  & 25.00  & 2.27 & 0.82 & 1.45 & 5.5 & 2.13 \\
            Regime Unstable          & FeatureMining & 100 & 100.00 & 2.75 & 0.81 & 1.94 & 4.6 & 2.68 \\
            Regime Unstable          & FeatureMining & 200 & 200.00 & 3.01 & 0.79 & 2.22 & 4.8 & 2.94 \\
            Student $t$, df = 5      & FeatureMining & 25  & 24.99  & 2.27 & 0.81 & 1.46 & 5.4 & 2.17 \\
            Student $t$, df = 5      & FeatureMining & 100 & 100.00 & 2.76 & 0.82 & 1.94 & 5.7 & 2.70 \\
            Student $t$, df = 5      & FeatureMining & 200 & 200.00 & 3.00 & 0.76 & 2.24 & 4.0 & 2.99 \\
            Student $t$, df = 10     & FeatureMining & 25  & 25.00  & 2.29 & 0.80 & 1.49 & 5.2 & 2.18 \\
            Student $t$, df = 10     & FeatureMining & 100 & 100.00 & 2.80 & 0.79 & 2.01 & 4.5 & 2.75 \\
            Student $t$, df = 10     & FeatureMining & 200 & 200.00 & 3.02 & 0.79 & 2.23 & 4.4 & 2.82 \\
            White Noise High Vol     & FeatureMining & 25  & 25.00  & 2.26 & 0.78 & 1.48 & 4.4 & 2.18 \\
            White Noise High Vol     & FeatureMining & 100 & 100.00 & 2.75 & 0.80 & 1.95 & 5.8 & 2.71 \\
            White Noise High Vol     & FeatureMining & 200 & 200.00 & 3.02 & 0.79 & 2.23 & 4.0 & 2.99 \\
            White Noise Low Vol      & FeatureMining & 25  & 25.00  & 2.28 & 0.83 & 1.45 & 6.0 & 2.23 \\
            White Noise Low Vol      & FeatureMining & 100 & 100.00 & 2.76 & 0.81 & 1.96 & 5.3 & 2.71 \\
            White Noise Low Vol      & FeatureMining & 200 & 200.00 & 2.99 & 0.79 & 2.20 & 4.6 & 2.96 \\
        \end{longtable}

        \subsubsection{Split sensitivity, grid density, and seed stability}
        Table~\ref{tab:robust_other_checks} reports robustness to variations in the in-sample/walk-forward split ratio, hyperparameter grid density, and random seed family for the White Noise / \texttt{HyperparameterTuning} configuration at $K=100$. Point estimates remain stable across split and seed perturbations, while denser hyperparameter grids mechanically raise effective multiplicity and in-sample extremeness without altering the qualitative conclusion that walk-forward evidence remains near the null benchmark.

        \begin{table}[htbp]
            \centering
            \caption{Audit parameter sensitivity under White Noise / \texttt{HyperparameterTuning} ($K=100$). Impact of training split size, grid density, and random seed family on effective multiplicity, in-sample and walk-forward extremeness, the absolute gap $\Delta Z$, walk-forward failure rates, and stabilized BIF.}
            \label{tab:robust_other_checks}
            \small
            \resizebox{\textwidth}{!}{%
                \begin{tabular}{lllrrrrrr}
                    \toprule
                    Test & Environment & Variant & $\widehat{K}_{\mathrm{eff}}$ & $|Z_{\mathrm{IS}}|$ & $|Z_{\mathrm{WF}}|$ & $\Delta Z$ & Fail \% & BIF \\
                    \midrule
                    \multicolumn{9}{l}{\textit{Split Sensitivity}} \\
                    Split 0.5 & White Noise & Dev Set & 1.80 & 1.73 & 0.81 & 0.92 & 5.7 & 1.62 \\
                    Split 0.6 & White Noise & Dev Set & 1.78 & 1.70 & 0.79 & 0.92 & 4.5 & 1.58 \\
                    Split 0.7 & White Noise & Dev Set & 1.78 & 1.71 & 0.83 & 0.89 & 5.4 & 1.61 \\
                    \midrule
                    \multicolumn{9}{l}{\textit{Grid Density}} \\
                    Grid & WN Light Grid & Dev Set & 1.78 & 1.70 & 0.80 & 0.89 & 5.2 & 1.55 \\
                    Grid & WN Dense Grid & Dev Set & 2.63 & 2.04 & 0.83 & 1.21 & 6.3 & 1.91 \\
                    \midrule
                    \multicolumn{9}{l}{\textit{Seed Stability}} \\
                    Seed & White Noise & Dev Set & 1.78 & 1.70 & 0.79 & 0.92 & 4.5 & 1.58 \\
                    Seed & White Noise & Holdout Set & 1.79 & 1.70 & 0.82 & 0.87 & 4.7 & 1.59 \\
                    \bottomrule
            \end{tabular}}
        \end{table}

        \subsubsection{Temporal integrity audit}
        Table~\ref{tab:temporal_audit} reports the risk deflation experiment. Shuffled validation underestimates the standard error by approximately $28\%$, inflating the apparent $Z$-statistic from $0.91$ to $1.64$ and increasing the failure rate from $9.2\%$ to $36.4\%$.

        \begin{table}[htbp]
            \centering
            \caption{Temporal integrity audit: walk-forward versus shuffled validation.}
            \label{tab:temporal_audit}
            \begin{tabular}{lcccc}
                \toprule
                Protocol & Mean $|Z|$ & Fail Rate (\%) & Risk Deflation & $p$-value ($H_0$: Ratio $=1$) \\
                \midrule
                Walk-Forward  & 0.91 & 9.2 & 1.00 (ref) & --- \\
                Shuffled Split & 1.64 & 36.4 & 0.72 & $< 10^{-100}$ \\
                \bottomrule
            \end{tabular}
        \end{table}

        \subsubsection{Architecture generalization}
        To assess whether the capacity-driven inflation documented in \Cref{subsec:boosting_effect_sim} is specific to sequential gradient boosting or extends to other non-linear ensembles, we replicate the audit using Random Forests (bagging) alongside XGBoost (boosting). The protocol enforces strict temporal separation between model fitting, adaptive selection, and evaluation. Each candidate model is fit on a training set $T_{\mathrm{train}}$, while the winning configuration is selected by maximizing the absolute HAC statistic on a disjoint development set $T_{\mathrm{dev}}$. The selected winner is then evaluated on a strictly disjoint walk-forward set $T_{\mathrm{WF}}$.

        Table~\ref{tab:rf_vs_xgb} reports the stabilized inflation factor $\mathrm{BIF}_{0.5} = |Z_{\mathrm{dev}}|/\max(|Z_{\mathrm{WF}}|,0.5)$ for search budgets $K=1$ and $K=50$. At $K=1$, the median $\mathrm{BIF}_{0.5}$ remains below unity for both architectures, indicating no systematic inflation when no adaptive search is performed. As the search budget increases to $K=50$, inflation becomes pronounced, with median $\mathrm{BIF}_{0.5}$ rising to 3.31 for
        XGBoost and 2.65 for Random Forest. This confirms that selection bias operates across non-linear ensemble families even when memorization is mitigated by a disjoint development slice: adaptive search alone is sufficient to amplify apparent evidence under the White Noise null, with the effect somewhat stronger for boosting than for bagging.

        \begin{table}[htbp]
            \centering
            \caption{\textbf{Architecture Generalization.} Median stabilized Backtest Inflation Factor ($BIF_{0.5}$) across sequential boosting and parallel bagging algorithms under the global null.}
            \label{tab:rf_vs_xgb}
            \begin{tabular}{lcc}
                \toprule
                Model Architecture & BIF ($K=1$) & BIF ($K=50$) \\
                \midrule
                Gradient Boosting (XGBoost) & 0.87 & 3.31 \\
                Random Forest & 0.85 & 2.65 \\
                \bottomrule
            \end{tabular}
        \end{table}

        \subsubsection{Metric Stabilization: Ratio vs.\ Difference}
        \label{subsubsec:metric_stabilization}

        We evaluate the behavior of inflation metrics in the critical limit where walk-forward evidence vanishes ($Z_{\mathrm{WF}}^\star \to 0$). Figure~\ref{fig:bif_stabilization_panel} contrasts the ratio-based BIF against the difference-based $\Delta Z$.

        The legacy ratio ($\mathrm{BIF}^{\mathrm{raw}}$) exhibits a singularity at zero and extreme instability in the gray zone ($Z_{\mathrm{WF}}^\star < \tau$). Table~\ref{tab:bif_stabilization_comparison} quantifies this divergence, demonstrating the undefined output (NA) of the raw ratio versus the bounded behavior of the stabilized metric as walk-forward evidence approaches zero. The stabilized BIF uses a denominator floor $\tau$ purely to prevent degeneracy when walk-forward evidence is near zero. Consequently, BIF is reported only as an internal ranking heuristic within a study, while our primary inflation diagnostic is the threshold-free absolute gap $\Delta Z$, which is invariant to $\tau$. We report sensitivity to $\tau$ and show that qualitative conclusions are stable over a broad range of values (see Section~\ref{sec:robustness_sensitivity}).

        In contrast, the Absolute Magnitude Gap ($\Delta Z$) maintains linearity throughout the noise floor (Right Panel). Because it avoids division, $\Delta Z$ provides a continuous, parameter-free measure of inflation that remains interpretable even when walk-forward performance is entirely spurious.

        \begin{figure}[htbp]
            \centering
            \includegraphics[width=\textwidth]{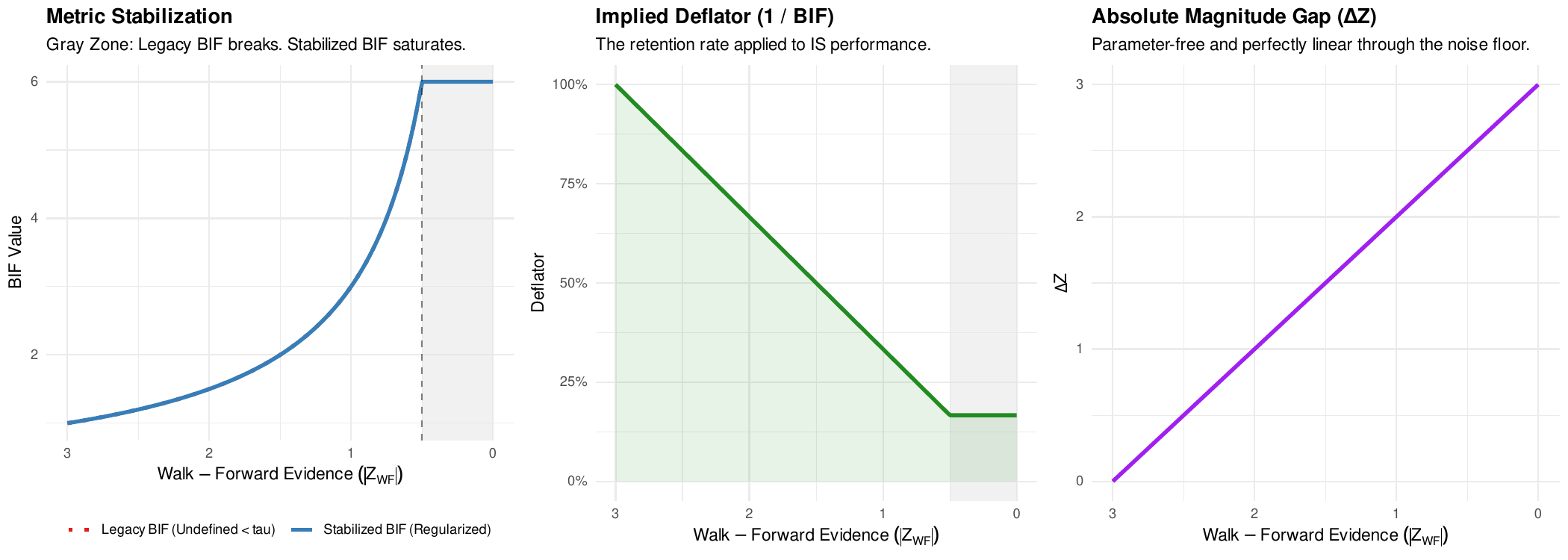}
            \caption{Metric Stabilization Stress Test. Behavior of inflation metrics as walk-forward evidence decays ($Z_{\mathrm{IS}}^\star=3.0$, $\tau=0.5$). Left: The legacy BIF (dotted) is undefined in the gray zone ($Z_{\mathrm{WF}}^\star<\tau$), whereas the stabilized BIF (solid) saturates at the regularization floor. Center: The implied deflator ($1/\mathrm{BIF}^{\mathrm{stab}}_{\tau}$) transitions smoothly into the spurious region, capping the retention rate. Right: The Absolute Magnitude Gap ($\Delta Z$) remains perfectly linear and stable throughout the noise floor, avoiding the singularity issues of ratio-based metrics.}
            \label{fig:bif_stabilization_panel}
        \end{figure}

        \begin{table}[htbp]
            \centering
            \caption{Stress-test values for raw and stabilized inflation metrics as walk-forward evidence decays ($Z_{\mathrm{IS}}^\star=3.0$, $\tau=0.5$). The raw ratio is undefined (NA) when $Z_{\mathrm{WF}}^\star < \tau$. The deflator $d_\tau$ represents the retention rate of in-sample performance.}
            \label{tab:bif_stabilization_comparison}
            \begin{tabular}{ccccccr}
                \toprule
                $Z_{\mathrm{WF}}^\star$ & $\mathrm{BIF}^{\mathrm{raw}}$ & $\mathrm{BIF}^{\mathrm{stab}}_{\tau}$ & Deflator ($d_\tau$) & $\Delta Z$ & Status \\
                \midrule
                3.00 & 1.00 & 1.00 & 1.000 & 0.00 & Agreement \\
                2.00 & 1.50 & 1.50 & 0.667 & 1.00 & Agreement \\
                1.00 & 3.00 & 3.00 & 0.333 & 2.00 & Agreement \\
                0.50 & 6.00 & 6.00 & 0.167 & 2.50 & Boundary \\
                0.25 & NA   & 6.00 & 0.167 & 2.75 & Stabilized \\
                0.00 & NA   & 6.00 & 0.167 & 3.00 & Stabilized \\
                \bottomrule
            \end{tabular}
        \end{table}

        \subsubsection{Parameter Sensitivity and Metric Invariance}
        \label{sec:robustness_sensitivity}

        We assess the structural stability of inflation metrics by sweeping the regularization threshold across a grid $\tau \in [0.1, 1.5]$. Figure~\ref{fig:bif_frontier} contrasts the behavior of the ratio-based $\mathrm{BIF}^{\mathrm{stab}}$ against the difference-based $\Delta Z$.

        The analysis reveals a critical limitation in ratio-based metrics: $\mathrm{BIF}^{\mathrm{stab}}$ exhibits high sensitivity to the choice of floor, with the median inflation factor decaying inversely to $\tau$ (Panel A). While a local plateau exists near $\tau \in [0.3, 0.5]$, the absolute magnitude of the metric remains conditional on an arbitrary parameter choice.

        In contrast, the Absolute Magnitude Gap ($\Delta Z$) demonstrates structural invariance (Panel B). Because $\Delta Z$ operates on the linear difference between evidence scales, it does not require a stability floor to prevent singularities. This parameter independence confirms $\Delta Z$ as the more robust standard for benchmarking falsification audits, while $\tau=0.5$ is retained merely as a standardized convention for legacy BIF comparisons.

        \begin{figure}[htbp]
            \centering
            \includegraphics[width=\textwidth]{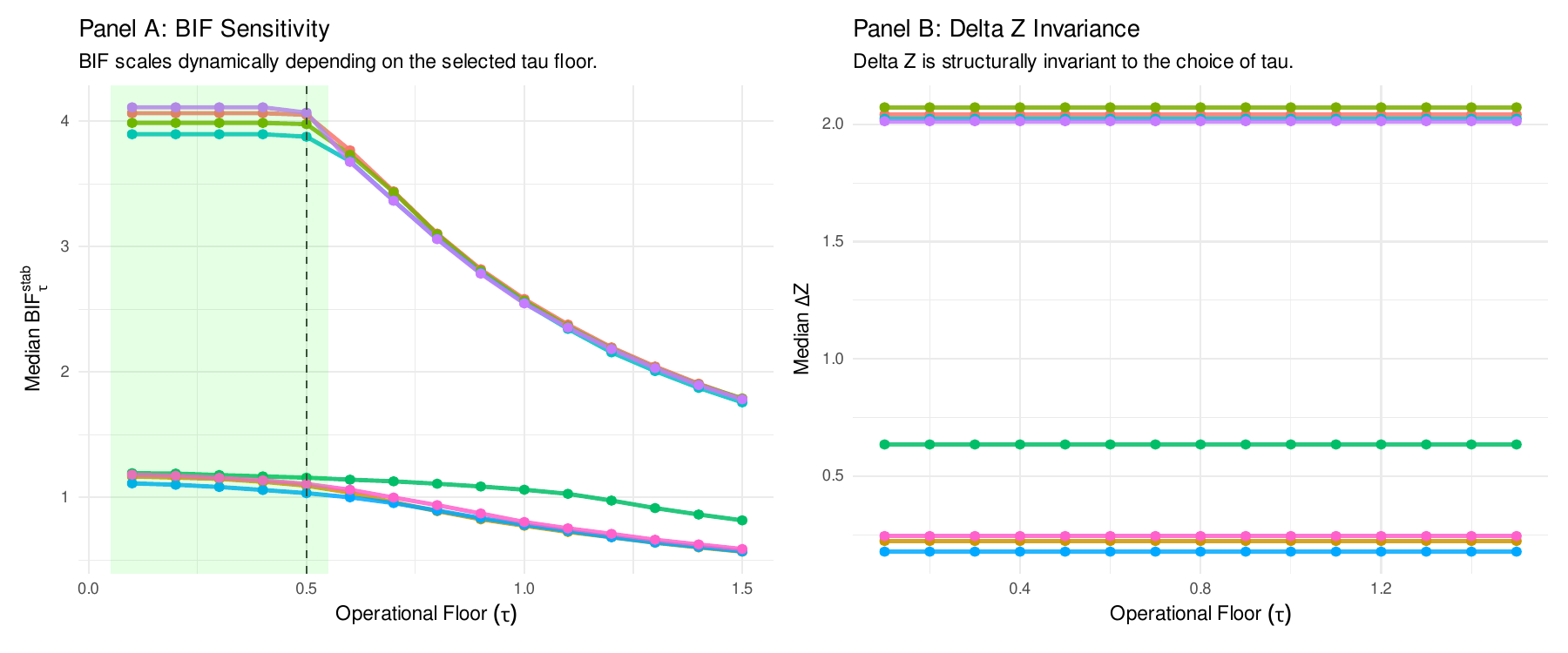}
            \caption{Metric Sensitivity Stress Test. Comparison of inflation metric stability across varying regularization thresholds $\tau$. Panel A demonstrates that ratio-based metrics ($\mathrm{BIF}^{\mathrm{stab}}$) are parameter-sensitive, scaling inversely with the chosen floor even within the stability plateau (green shaded region). Panel B demonstrates that the Absolute Magnitude Gap ($\Delta Z$) is structurally invariant to the regularization parameter, providing a stable benchmark for falsification.}
            \label{fig:bif_frontier}
        \end{figure}

        \subsubsection{Null-calibrated operational thresholds}
        \label{app:calib_thresholds}

        Table~\ref{tab:calib_recommended_thresholds} reports null-calibrated critical values for the absolute magnitude gap ($\Delta Z$) together with legacy ratio-based calibrations derived from the White Noise global null. The reported thresholds correspond to the $95\%$ and $99\%$ empirical quantiles of each metric. Because $\Delta Z$ is parameter-invariant and remains well-defined throughout the noise floor, it serves as the primary threshold for formal inflation classification. By contrast, BIF quantiles are reported only on the stability-gated subset $\{|Z^\star_{\mathrm{WF}}| \ge \tau\}$ with $\tau=0.5$, and should be interpreted strictly as secondary within-study heuristics rather than as parameter-free benchmarks. The table also reports $\zeta_{0.99}$, the $99\%$ quantile of the walk-forward winner magnitude. Across all calibrations, $\zeta_{0.99}$ remains in the vicinity of the standard-normal upper tail, consistent with the bounded null behavior of strictly disjoint walk-forward evaluation.

        \begin{table}[htbp]
            \centering
            \caption{Recommended operational thresholds derived from the empirical distribution of the White Noise global nulls. $\Delta Z$ provides the primary formal evaluation threshold. BIF quantiles are reported on the stability-gated subset $\{|Z^\star_{\mathrm{WF}}| \ge \tau\}$ with $\tau=0.5$. $\zeta$ denotes the $99\%$ quantile of valid walk-forward winner magnitudes.}
            \label{tab:calib_recommended_thresholds}
            \resizebox{\textwidth}{!}{%
                \begin{tabular}{llcccccc}
                    \toprule
                    Source & Setting & $\Delta Z_{0.95}$ & $\Delta Z_{0.99}$ & $\tau$ & $\mathrm{BIF}^{\mathrm{gate}}_{\tau,0.95}$ & $\mathrm{BIF}^{\mathrm{gate}}_{\tau,0.99}$ & $\zeta_{0.99}$\\
                    \midrule
                    Scaling law & K=1000 (Indep.) & 3.68 & 4.12 & 0.5 & 6.48 & 7.29 & 2.77\\
                    Redundancy law & Clusters=1 (Perfect Corr.) & 2.57 & 3.24 & 0.5 & 4.20 & 5.49 & 2.56\\
                    Redundancy law & Clusters=500 (Indep.) & 3.53 & 3.94 & 0.5 & 6.13 & 7.05 & 2.58\\
                    ML Workflows & FeatureMining & 3.44 & 3.85 & 0.5 & 5.88 & 6.76 & 2.69\\
                    ML Workflows & HyperparameterTuning & 3.08 & 3.56 & 0.5 & 4.94 & 6.19 & 2.51\\
                    ML Workflows & TrendFollowing & 2.62 & 3.14 & 0.5 & 4.29 & 5.60 & 2.40\\
                    \bottomrule
            \end{tabular}}
        \end{table}

        \section{Transaction Cost Mechanics: Synthetic Validation}
        \label{app:transaction_costs}
        This appendix validates the mechanical relationship between signal turnover, gross performance, and transaction costs under the audit's cost accounting. The objective is to confirm that the break-even cost metric behaves consistently under strict walk-forward evaluation and a linear per-side cost model. The reported break-even values are not an empirical robustness claim for any specific asset or strategy in the main text.

        We simulate $N=1{,}000$ independent Monte Carlo paths from a regime-switching process calibrated to embed a persistent trend component (annual volatility $15\%$, trend Sharpe $2.0$). The price process is indexed on business days to match the $252$-day annualization convention.

        For each path, the strategy selects its lookback window using the in-sample period $T_{\mathrm{IS}}$ and is then evaluated on the walk-forward period $T_{\mathrm{WF}}$, with parameters frozen out of sample. Let $r^{\mathrm{WF}}_t$ denote the gross walk-forward strategy return and let $s_t \in \{-1,0,1\}$ denote the full continuous signal path. Two-way traded volume on the walk-forward block is defined as
        \begin{equation}
            V_{2}(T_{\mathrm{WF}}) \;=\; \sum_{t \in T_{\mathrm{WF}}} \left| s_t - s_{t-1} \right|,
        \end{equation}
        so that a flip from $+1$ to $-1$ contributes $2$ units, consistent with per-side transaction costs. This accounting includes the transition trade at the $T_{\mathrm{IS}}/T_{\mathrm{WF}}$ boundary whenever the first walk-forward position differs from the final in-sample position. The annualized two-way volume is $\bar V_{2} = V_{2}(T_{\mathrm{WF}}) / (|T_{\mathrm{WF}}|/252)$, and the annualized gross mean return is $\bar\mu_{\mathrm{gross}} = 252 \cdot \mathbb{E}[r^{\mathrm{WF}}_t]$, estimated by the sample mean on $T_{\mathrm{WF}}$.

        Under a linear cost model with cost $c$ per side (in return units), the annualized net mean is $\bar\mu_{\mathrm{net}}(c) = \bar\mu_{\mathrm{gross}} - \bar V_{2}\,c$. The break-even cost is the unique solution to $\bar\mu_{\mathrm{net}}(c)=0$, given by
        \begin{equation}
            c^{\star} \;=\; \frac{\bar\mu_{\mathrm{gross}}}{\bar V_{2}}, \qquad
            c^{\star}_{\mathrm{bps}} \;=\; 10{,}000 \cdot c^{\star}.
        \end{equation}
        We report turnover in one-way units as $\bar V_{1} = \bar V_{2}/2$ for interpretability, while break-even is computed using two-way volume, consistent with per-side costs. To align the economic calculation with the walk-forward significance screen, we compute break-even only for realizations satisfying $Z_{\mathrm{WF}} \ge 1.96$. The resulting $39.5\%$ pass rate should be interpreted as empirical power under the signal-bearing DGP rather than as a null rejection rate; under a no-predictability null, the corresponding one-sided exceedance probability would be $2.5\%$.

        Table~\ref{tab:mc_cost_robustness} summarizes the distribution of annualized one-way turnover and break-even costs across the passing subset. The dispersion in break-even costs reflects path dependence in realized gross performance and trading intensity; the purpose of this validation is to demonstrate that the accounting framework remains mechanically consistent across stochastic realizations.

        \begin{table}[htbp]
            \caption{Transaction Cost Mechanics. Distribution of annualized one-way turnover and break-even transaction costs across 395 Monte Carlo paths ($39.5\%$ of 1000 simulations) that satisfy the one-sided walk-forward significance screen ($Z_{\mathrm{WF}} \ge 1.96$). Break-even cost is defined as the per-side cost that reduces the annualized mean net return to zero.}
            \label{tab:mc_cost_robustness}
            \centering
            \begin{tabular}[t]{lrrrr}
                \toprule
                Metric & Mean & Median & 5\% & 95\%\\
                \midrule
                Ann. Turnover (1-Way) & 13.78 & 12.00 & 6.75 & 28.50\\
                Break-Even Cost (bps) & 100.68 & 86.10 & 33.13 & 204.98\\
                \bottomrule
            \end{tabular}
        \end{table}

        \subsection{Replication Materials}
        \label{app:replication_materials}

        All results in the paper are generated by the accompanying replication scripts, which reproduce the simulation experiments and empirical applications end-to-end, including walk-forward evaluation, induced-null audits, and inflation diagnostics. The authoritative reference for the methodology is the \texttt{QuantAudit} framework and its accompanying code used to generate the tables and figures. These complete replication materials, alongside the documented R package, will be made publicly available upon journal publication.

        \subsection{Simulation Strategy Templates}
        \label{subsec:sim_strategy_templates}

        The simulation suite is organized around a small set of stylized strategy templates that each isolate a distinct failure channel relevant to adaptive financial backtesting. The objective is not to claim realism for any single rule, but to ensure that each diagnostic is exercised against a concrete mechanism: pure noise controls, explicit protocol violations (look-ahead), multiplicity-driven selection, risk-misattribution via systematic exposure, microstructure-style serial dependence, and nonlinear interaction with heteroskedasticity. Table~\ref{tab:sim_specs} provides a compact map from each template to the targeted failure channel, clarifying how the simulation evidence relates to the audit’s interpretive claims.

        \begin{table}[htbp]
            \centering
            \caption{Simulation design map: strategy templates and targeted failure channels.
                Each row defines a stylized trading rule used in the simulation suite and the specific non-signal mechanism it is intended to diagnose under the null and induced-null experiments.}
            \label{tab:sim_specs}
            \small
            \begin{tabularx}{\textwidth}{lX}
                \toprule
                Specification & Targeted failure channel (mechanism tested) \\
                \midrule
                \textit{Panel A: Controls} & \\
                White Noise & Passive control trading on i.i.d.\ noise $S_t \sim \mathcal{N}(0,1)$; baseline Type I error benchmark. \\
                Oracle (Look-ahead) & Oracle with access to $r_{t+1}$; illustrates the maximum impact of look-ahead and serves as an upper bound failure signature. \\
                \midrule
                \textit{Panel B: Spurious and structural modes} & \\
                Spurious Selection & In-sample winner selected from $K=100$ independent noise predictors; isolates multiplicity and selection-induced inflation. \\
                Latent Factor & Linear exposure to a persistent risk premium; isolates the risk-misattribution (beta confound) channel. \\
                Mean Reversion & Linear reversal signal $S_t = -\theta r_{t-1}$; isolates negative serial dependence and microstructure-style artifacts (e.g., bid--ask bounce). \\
                Regime Switching & Threshold rule switching with volatility-state conditioning (e.g., GARCH-based); isolates nonlinear interactions with heteroskedasticity and validation fragility. \\
                \bottomrule
            \end{tabularx}
        \end{table}

    \end{document}